\documentclass[fleqn,usenatbib]{mnras}
\hypersetup{linkcolor=red}          
\usepackage{multirow}
\usepackage[dvipsnames]{xcolor}
\usepackage{newtxtext,newtxmath}
\usepackage{orcidlink}
\usepackage{array}

\def\skirt{{\sc skirt} }
\def\colibre{{\sc colibre} }

\def\cosk{{\sc colibre-skirt} }

\def\equationautorefname~#1\null{Eq.~#1\null}

\title[COLIBRE UVLF at $z\geqslant7$]{The galaxy ultraviolet luminosity function from $z=7$ to $20$ in the COLIBRE simulations}

\author[S. Lu et al.]
{Shengdong Lu\orcidlink{0000-0002-6726-9499}$^{1}$\thanks{E-mail: \url{shengdong.lu@durham.ac.uk}}, 
Carlos S. Frenk\orcidlink{0000-0002-2338-716X}$^{1}$, 
Cedric G. Lacey\orcidlink{0000-0001-9016-5332}$^{1}$, 
Andrea Gebek\orcidlink{0000-0002-0206-8231}$^{2}$,
Joop Schaye\orcidlink{0000-0002-0668-5560}$^{3}$, 
\and
Shaun Cole\orcidlink{0000-0002-5954-7903}$^{1}$, 
Sownak Bose\orcidlink{0000-0002-0974-5266}$^{1}$, 
Anna Durrant\orcidlink{0009-0003-9672-2170}$^{4}$,
Nick Andreadis\orcidlink{0009-0001-9915-6325}$^{2}$,
Maarten Baes\orcidlink{0000-0002-3930-2757}$^{2}$,
\and
Alejandro Benítez-Llambay$^{5}$,
Evgenii Chaikin$^{1,3}$,
Camila Correa$^{3}$,
Robert A. Crain$^{4}$,
Filip Huško$^{3}$,
\and
Robert J. McGibbon$^{3}$,
Sylvia Ploeckinger$^{6}$,
Alexander J. Richings$^{7,8}$,
Matthieu Schaller\orcidlink{0000-0002-2395-4902}$^{3,9}$, 
\and
James W. Trayford$^{10}$
\\
\\
$^{1}$Institute for Computational Cosmology, Department of Physics, University of Durham, South Road, Durham, DH1 3LE, UK\\
$^{2}$Department of Physics and Astronomy, Universiteit Gent, Proeftuinstraat 86 N3, B-9000 Ghent, Belgium\\
$^{3}$Leiden Observatory, Leiden University, PO Box 9513, 2300 RA Leiden, the Netherlands\\
$^{4}$Astrophysics Research Institute, Liverpool John Moores University, 146 Brownlow Hill, Liverpool L3 5RF, UK\\
$^{5}$Dipartimento di Fisica G. Occhialini, Universit\`a degli Studi di Milano Bicocca, Piazza della Scienza, 3 I-20126 Milano MI, Italy\\
$^{6}$Department of Astrophysics, University of Vienna, Türkenschanzstrasse 17, A-1180 Vienna, Austria\\
$^{7}$Centre for Data Science, Artificial Intelligence and Modelling, University of Hull, Cottingham Road, Hull, HU6 7RX, UK\\
$^{8}$E. A. Milne Centre for Astrophysics, University of Hull, Cottingham Road, Hull, HU6 7RX, UK\\
$^{9}$Lorentz Institute for Theoretical Physics, Leiden University, PO Box 9506, 2300 RA Leiden, the Netherlands\\
$^{10}$Institute of Cosmology and Gravitation, University of Portsmouth, Dennis Sciama Building, Burnaby Road, Portsmouth PO1 3FX, UK
}
\date{}
\pubyear{2026}

\begin{document}
\label{firstpage}
\pagerange{\pageref{firstpage}--\pageref{lastpage}}
\maketitle
\begin{abstract}
{\it JWST} has enabled the detection of galaxies in the earliest stages of cosmic history. We compare the ultraviolet luminosity functions (UVLFs) at redshifts $z=7-20$ predicted by a set of new cosmological hydrodynamical simulations, {\sc colibre}, with observations, including those from {\it JWST}. The UV luminosities of {\sc colibre} galaxies are derived using the radiative transfer code {\sc skirt}, which tracks stellar emission and its processing through the multi-phase interstellar medium and dust distribution predicted by {\sc colibre}. We find that although \colibre is consistent with the observed evolution of the stellar mass function up to $z=12$, its dust-attenuated UVLFs fall systematically below the observations at the bright end: at the number density of $10^{-6}\,\mathrm{Mpc^{-3}\,mag^{-1}}$, the brightest galaxies are underluminous by $\approx 1\,\rm mag$ at $z=7$, increasing to $\approx 2.5\,\rm mag$ at $z=15$. Accounting for observational uncertainties brings the {\sc colibre} UVLFs closer to the observational data, but does not fully resolve the discrepancy. Ignoring dust attenuation allows \colibre to produce sufficiently bright galaxies at $7\lesssim z \lesssim 12$, but at $z=15$, \colibre still underpredicts the luminosities of the brightest galaxies, indicating the need for additional physical mechanisms to boost the UV luminosities at the earliest cosmic epochs, such as a ``top-heavy'' stellar initial mass function. We fit the \colibre UVLFs with Schechter functions and calculate the evolution of the best-fit parameters. We find that the galaxy number density decreases, the characteristic luminosity becomes fainter and the faint-end slope becomes steeper towards higher redshifts. The UV luminosity density decreases by a factor of several hundred from $z = 7$ to $z = 15$.
\end{abstract}


\section{Introduction}
\label{sec:introduction}
Galaxies in the early Universe carry key information about the formation and evolution of cosmic structures. They reveal the transition of the Universe from an initially near-uniform state to the one we observe today filled with complex structures, reflecting physical processes such as dark matter halo collapse, gas cooling, star formation and feedback loops. These early galaxies are thought to be the primary source of the ionising photons responsible for cosmic reionisation, which significantly influenced the transparency and subsequent evolution of the Universe (e.g., \citealt{Madau2014,Robertson2015}).

Deep imaging with the {\it Hubble Space Telescope} (HST) and complementary {\it Spitzer}/IRAC observations had already extended galaxy detections to $z\approx 7-11$ \citep[e.g.,][]{Oesch2016,Bouwens2021,Roberts-Borsani2016,Roberts-Borsani2020}, 
providing early constraints on the rest-frame optical 
properties and galaxy stellar masses in the first few hundred million years after the Big Bang. With the launch of the {\it James Webb Space Telescope} ({\it JWST}), our observational horizon expanded dramatically, reaching into the earliest epochs of cosmic history. For example, \citet{Atek2023} found two candidate galaxies at $z\approx 16$ in the first {\it JWST} deep field. More recently, {\it JWST} spectroscopy has pushed the frontier of \emph{spectroscopic confirmations} to higher and higher redshifts, with galaxies now confirmed at $z\approx 14.3$ and $z\approx 14.4$ \citep{Carniani2024,Naidu2026}. \citet{Cullen2023} investigated the rest-frame ultraviolet (UV) continuum slope, $\beta$, of galaxies at $8<z<16$ using {\it JWST} and ground-based imaging, finding that high-redshift galaxies are generally blue, with a $\beta-M_{\rm UV}$ (where $M_{\rm UV}$ is the absolute UV magnitude) relation and other properties consistent with those of young, metal-poor, moderately dust-reddened stellar populations.

Cosmological hydrodynamical simulations, such as EAGLE \citep{Schaye2015,Crain2015}, Horizon-AGN \citep{Dubois2016,Kaviraj2017}, FABLE \citep{Henden2018}, IllustrisTNG \citep{Nelson2019}, SIMBA \citep{Dave2019}, FLAMINGO \citep{Schaye2023} and MillenniumTNG \citep{Pakmor2023}, have made significant progress in reproducing many key observed galaxy properties across cosmic time, including, for example, galaxy stellar mass functions, star formation rates (SFRs), sizes, morphologies and chemical enrichment trends and have thus become crucial tools for interpreting astronomical observations and testing theoretical models of galaxy formation and evolution. 

More recently, the {\sc colibre} collaboration \citep{Schaye2026,Chaikin2026a} introduced a new galaxy formation framework that explicitly models the multiphase interstellar medium (ISM), non-equilibrium H/He chemistry and dust evolution. Spanning a wide range of volumes and resolutions, {\sc colibre} also reproduces key observed galaxy properties, such as the galaxy stellar mass function (GSMF) and its evolution \citep{Chaikin2026b}, the evolution of sizes \citep{Ludlow2026} and resolved star formation laws \citep{Lagos2025}, making it a promising tool for studies of galaxy formation across cosmic time \citep{Schaye2026,Chaikin2026b}.

Most large-volume cosmological simulations (including {\sc colibre}) are calibrated to low-redshift observational constraints, especially the low-$z$ GSMF, since stellar mass is a direct output of the simulations and can be estimated observationally from broad-band spectral energy distribution (SED) fitting (e.g. \citealt{Crain2015,Nelson2019}). However, converting observed luminosities into stellar masses is a form of inverse modelling, which depends on a number of highly uncertain assumptions such as the star formation history, dust attenuation, metallicity and the assumed stellar initial mass function (IMF; e.g., \citealt{Mitchell2013,Lo_Faro2017}). These assumptions enter both into parametric SED-fitting approaches, which adopt specific functional forms for the star formation history (e.g. exponentially declining or delayed-$\tau$ models) and more flexible non-parametric methods, which avoid explicit functional forms but still rely on priors and finite temporal resolution in reconstructing star formation histories (e.g. \citealt{Carnall2019,Leja2019,Pacifici2023}).

An alternative strategy is {\em forward modelling}, in which simulated galaxy properties (such as stellar mass and star formation history) are translated into directly observable quantities, such as luminosities, allowing a closer connection to the data \citep{McAlpine2016}. While this approach is also model-dependent and relies, for example, on stellar population synthesis models with choices made in the simulations for the IMF and dust prescriptions, it is grounded in a single, internally self-consistent galaxy formation model. In particular, the same underlying physical framework simultaneously determines star formation, chemical enrichment, gas content in different phases and dust content, so tensions with observational luminosity functions can provide more direct diagnostics of the underlying physics. As a result, comparisons based on luminosity-related observables, which more naturally account for observational selection effects and survey characteristics, offer a powerful test of simulated galaxy populations, particularly at high redshift where stellar mass estimates remain highly uncertain.

The rest-frame UV luminosity function (UVLF) at high redshift quantifies the abundance and brightness of star-forming galaxies in the early Universe. Since rest-frame UV light traces recent star formation activity, the UVLF offers critical insights into the buildup of stellar mass, the efficiency of star formation and the ionising photon budget during the epoch of reionisation. Accurate measurements of the UVLF are therefore essential for constraining models of galaxy formation and the timeline of cosmic reionisation. Recent deep observations from {\it JWST} have significantly improved the accuracy and depth of UVLF measurements at $z>7$, providing crucial information on early galaxy populations (e.g., \citealt{Harikane2023,Harikane2024,Casey2024,Robertson2024,Naidu2026}). 

Comparing the predictions for UVLFs at high redshift from cosmological simulations to the observations serves to assess the validity of current galaxy formation models and also to identify the key physical processes driving early galaxy evolution. Specific predictions for the UVLFs that {\it JWST} was expected to measure at high redshift in the $\Lambda$CDM cosmology were presented, prior to the launch of {\it JWST}, by \citet{Cowley2018} using the Durham semi-analytic galaxy formation model, {\sc galform} \citep{Cole2000,Lacey2016}. A distinctive feature of this model is that the IMF in starburst episodes is assumed to be top-heavy (a modification introduced to explain the counts of faint submillimeter galaxies; \citealt{Baugh2005}).

Some of the first wave of {\it JWST} papers compared the data to a selection of models (not including that of \citealt{Cowley2018}) and concluded that bright galaxies at the highest redshifts are more abundant than predicted by the theoretical models they considered \citep[e.g.,][]{Finkelstein2024}. However, \citet{Lu2025} compared the pre-existing {\sc galform} predictions of UVLFs from \cite{Cowley2018} to the {\it JWST} and other observations and found remarkable agreement up to $z=12$. An adjustment of the model to take into account the timescale for the growth of dust grains extended this agreement out to the largest redshift for which there are data, $z \approx 14$. 

Several other comparisons with theoretical predictions have been made since the publication of the {\it JWST} data. \citet{Kannan2023} calculated UVLFs for galaxies at $z>10$ in the MillenniumTNG hydrodynamical simulation. This model shows broad agreement with observations up to $z = 10$, but to come close to the data at $z = 10-12$ requires assuming that galaxies are dust free at these redshifts. Even in the absence of dust, the model underpredicts galaxy abundances by approximately an order of magnitude at $z =15$. The authors conjectured that processes not included in the simulation, such as a variable stellar IMF or contributions from Population III (Pop~III) stars, could be necessary to resolve the discrepancy. 

Similarly, \citet{Yung2024}, using the Santa Cruz semi-analytic model \citep{Somerville1999}, presented results for the UVLF at high redshift, finding that the predicted galaxy number densities at $z > 11$ are at least an order of magnitude below the observed values even if galaxies are assumed to be dust free. They too suggested that a top-heavy IMF might eliminate the discrepancy. By contrast, \citet{Somerville2025} explored whether the observed abundance of UV-luminous galaxies at $z \gtrsim 10$ could be explained by invoking higher star formation efficiencies (SFEs) in dense environments. By introducing an {\em ad hoc} density-modulated SFE model into the Santa Cruz semi-analytic model, they showed that increasing the fraction of gas in dense clouds can reproduce the observed UVLFs over the high-redshift range considered, without requiring a top-heavy IMF. However, their analysis focused exclusively on high redshifts and did not present any evidence that the modified model remains consistent with observational constraints at lower redshift. Other recent studies have also compared model predictions with {\it JWST} UVLF measurements; some of these are discussed in \autoref{sec:discussion}.

In a series of three papers, we present a comprehensive study of the luminosity functions predicted by the {\sc colibre} simulations \citep{Schaye2026,Chaikin2026a}, across the full wavelength range from the far-UV (FUV) to the submillimeter and over redshifts from $z=0$ to $z \approx 20$. We post-process the {\sc colibre} simulations with the 3D Monte Carlo radiative transfer code, {\sc skirt}, to calculate the reprocessing of stellar emission by dust in the ISM \citep{Baes2011,Camps2015,Camps2020}. We use for this the calibration-free \cosk pipeline developed by \citet{Gebek2026}, who showed that when processed in this way, \colibre predicts a cosmic spectral energy distribution (CSED) at $z\approx 0$ that is consistent with local observations. In Paper~I \citep{Lu2026a}, using the same \cosk pipeline, we found remarkably good agreement at $z \approx 0$ between the predicted and observed LFs from the FUV to the near-infrared (NIR) and from the far-infrared (FIR) to the submillimeter. This agreement suggests that \colibre successfully captures both the properties of stellar populations and the amount and spatial distribution of interstellar dust in present-day galaxies. In addition, \citet{Chaikin2026b} showed that \colibre roughly reproduces the observed evolution of the galaxy stellar mass function up to $z\approx12$ and the number density of massive quenched galaxies up to $z\approx7$. 

In this second paper in the series, we adopt the same \cosk pipeline that has been shown to be successful at $z\approx 0$ and focus on the high-redshift UVLFs predicted by \colibre at $z=7-20$, comparing them with observational data, particularly those from {\it JWST}. An important feature of this work is that it allows us to quantify the effects of dust attenuation using dust distributions computed self-consistently in \colibre with a live dust model \citep{Trayford2026}. Crucially, the \colibre model has been calculated down to $z=0$ and shown to satisfy key observational constraints in the local Universe, as described above. Readers are referred to Paper~I \citep{Lu2026a} for the study of the local ($z=0$) LFs and to Paper~III (Lu et al. in prep.) for the redshift evolution of broadband LFs, again from the FUV to the submillimetre, up to intermediate redshifts ($z \approx 6$).

This paper is organized as follows: in \autoref{sec:data_method}, we introduce the \colibre simulations (\autoref{sec:colibre}), the galaxy sample selection method (\autoref{sec:sample}), the radiative transfer code \skirt and its setup (\autoref{sec:skirt}) and the compilation of observational data for comparison (\autoref{sec:obs_data}). The main results are presented in \autoref{sec:result}, where we study the correlation between stellar mass and galaxy UV magnitudes (\autoref{sec:correlation}) and compare the UVLFs of \colibre galaxies to observations (\autoref{sec:uvlf}). We discuss our results in \autoref{sec:discussion} and present a summary of our findings in \autoref{sec:conclusion}. Throughout this paper, we adopt the cosmological parameters used in the \colibre simulations (taken from \citealt{Abbott2022}): present-day matter density, $\Omega_{\rm m,0} = 0.306$; present-day baryon density, $\Omega_{\rm b,0} = 0.0486$; linear theory amplitude of matter fluctuations on $8\,\mathrm{Mpc}/h$ scales, $\sigma_8 = 0.807$; and dimensionless Hubble parameter, $h = 0.681$.

\section{Data and method}
\label{sec:data_method}

\subsection{The \colibre simulations}
\label{sec:colibre}

{\colibre} (COLd ISM and Better REsolution\footnote{\url{https://www.colibre-simulations.org/}}; 
\citealt{Schaye2026,Chaikin2026a}) is a new suite of cosmological hydrodynamical simulations of galaxy formation, performed with the open-source code {\sc swift} \citep{Schaller2024}. Below we briefly summarize the key aspects of the \colibre model. Readers are referred to \citet{Schaye2026} for a full description.

\begin{enumerate}
\item {\em Hydrodynamics and gravity}: gas dynamics are solved using the energy-density SPH scheme {\sc Sphenix} \citep{Borrow2022}. Gravitational forces are computed with a hybrid scheme combining a fast multipole method for short-range forces with a particle-mesh approach for long-range forces. Initial conditions are generated with {\sc monofonIC} \citep{Hahn2020,Michaux2021}, starting from $z=63$.

\item {\em Resolved multiphase ISM and non-equilibrium cooling}: radiative cooling is followed down to temperatures of $\approx 10\,\mathrm{K}$ \citep{Ploeckinger2025}, enabling an explicit treatment of the cold, dense ISM phase. Cooling, heating and the time evolution of chemical species are computed with the {\sc chimes} network \citep{Richings2014a,Richings2014b}. Hydrogen and helium are treated in non-equilibrium, while metal-line cooling is calculated element-by-element, accounting for depletion onto dust grains. The model includes molecular formation in the gas phase and on dust, shielding by gas and dust, photoelectric heating, cosmic rays and radiation fields, which together regulate the ionization state, thermal balance and molecular content of the cold ISM. For the metagalactic radiation field, {\colibre} adopts a modified redshift-dependent UV/X-ray background from \citet{Faucher-Giguere2020}, with hydrogen and helium reionization completing at $z\approx7$ and $z\approx3$, respectively. Reionization is therefore treated assuming a spatially uniform background, rather than through explicit radiative transfer, with local attenuation of photoionization, photodissociation and photoheating captured through an approximate treatment of gas and dust self-shielding.

\item {\em Dust evolution}: Dust grain formation, growth and destruction are modelled self-consistently during the simulation using a multi-grain dust evolution model \citep{Trayford2026}. The dust model is fully coupled to the cooling and chemistry calculations, such that the local dust abundance affects molecular formation rates, metal depletion and cooling efficiencies.

\item {\em Star formation and feedback}: Star formation follows a Schmidt law combined with a gravitational instability criterion \citep{Nobels2024}, allowing an effective star formation relation to emerge from the resolved multiphase ISM. Feedback from massive stars includes early stellar feedback (stellar winds, radiation pressure and \ion{H}{ii} regions) and core-collapse supernovae (SN) \citep{Benitez-Llambay2026,Chaikin2023}. Type Ia supernovae and chemical enrichment follow updated nucleosynthetic yields and delay-time distributions \citep{Correa2026}. Most colibre simulations use the purely thermal AGN feedback model of \citet{Schaye2026}, while a subset employ instead the hybrid thermal‑plus‑kinetic-jet model of \citet{Husko2026}, which combines thermally driven winds with kinetic jets.

\item {\em Resolution and box sizes}: {\colibre} is available in three numerical resolutions, m5, m6 and m7, with corresponding gas particle masses of $2.30 \times 10^{5}$, $1.84 \times 10^{6}$ and $1.47 \times 10^{7}\,\mathrm{M_{\odot}}$ and cube sizes ranging from $25$ to $400\,\rm cMpc$. Dark matter is supersampled by a factor of 4 relative to baryonic particles, yielding similar dark matter and baryonic particle masses, in order to suppress spurious transfer of energy from dark matter to stars \citep{Ludlow2019,Ludlow2023}. A redshift-dependent gravitational softening length is adopted to ensure that the spatial resolution increases at high redshift, which is important for resolving the ISM structure of early galaxies.

\item {\em Calibration and galaxy identification}: The model is calibrated at the fiducial m7 resolution to reproduce a small set of key $z=0$ galaxy scaling relations: the galaxy stellar mass function (GSMF), the galaxy size - stellar mass relation (SSMR) and the black hole mass - stellar mass relation (BSMR) \citep{Chaikin2026a}. The calibration to the $z = 0$ GSMF and SSMR is performed using Gaussian process emulators, while the calibration to the $z = 0$ BSMR is carried out manually and independently. After calibrating the m7 model, higher-resolution simulations are not fully recalibrated. Instead, the best-fitting m7 parameters are adopted as a reference point and for the m6 and m5 resolutions, the SN and AGN feedback parameters are manually adjusted starting from this reference to recover a comparable level of agreement with the $z = 0$ GSMF, SSMR and BSMR. Dark matter halos are first identified with a friends-of-friends algorithm \citep{Davis1985} and self-bound subhalos are then found using {\sc hbt-herons} \citep{Han2018,Forouhar_Moreno2025}.

\end{enumerate}

\subsection{Galaxy sample selection}
\label{sec:sample}
\begin{figure*}
\centering
\includegraphics[width=0.95\textwidth]{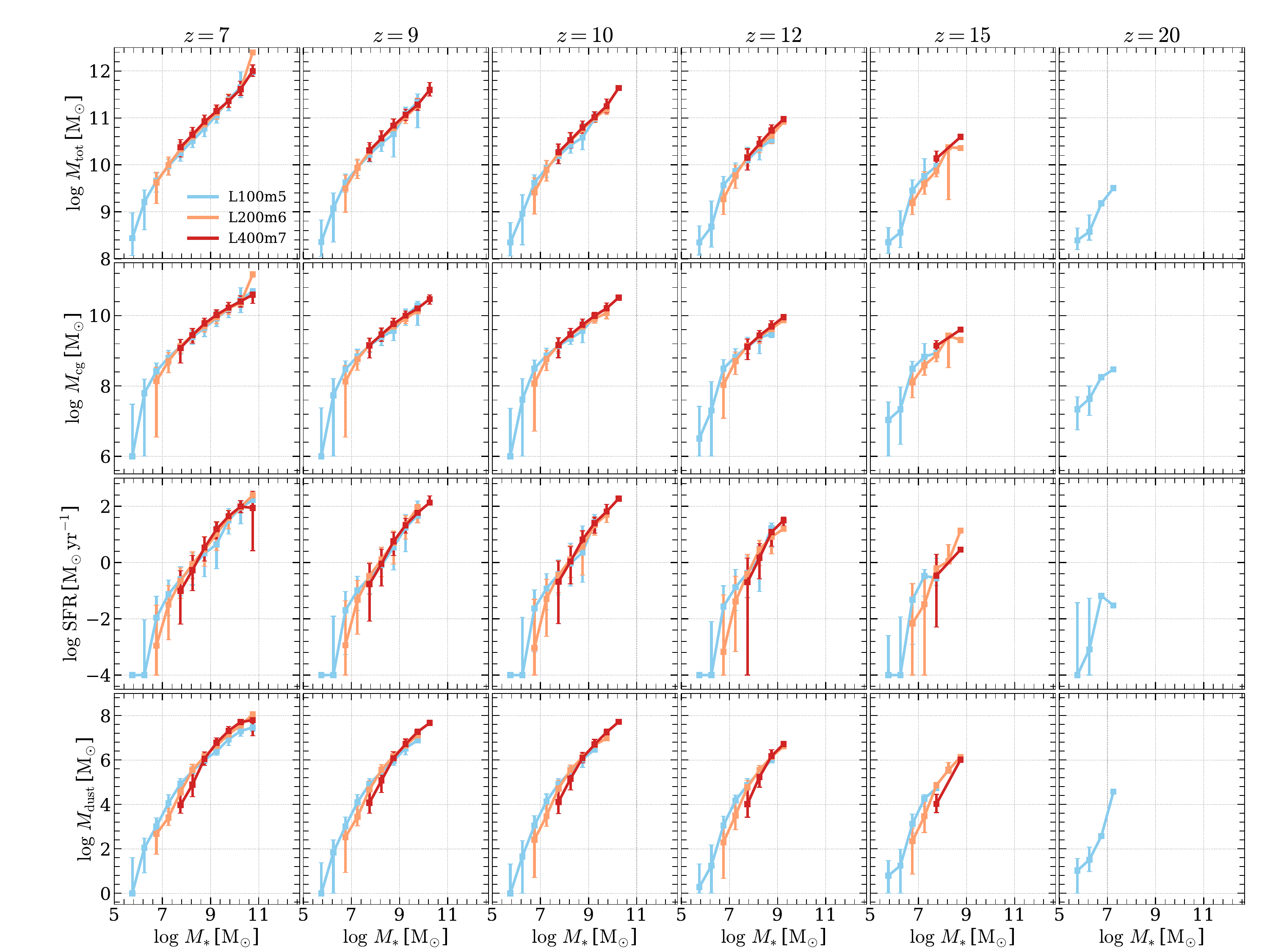}
\caption{Correlations between galaxy stellar mass and other global properties: (1) the total mass (including baryons, dark matter and black holes; $M_{\rm tot}$; top row); (2) the mass of gas in the cool, dense phase (i.e. $T < 10^{4.5}\,\mathrm{K}$ and $\rho_{\rm g}/m_{\mathrm{H}} > 10^{-1}\,\mathrm{cm}^{-3}$), denoted as $M_{\rm cg}$ (second row); (3) the instantaneous SFR (third row) and (4) the total dust mass, denoted as $M_{\rm dust}$ (bottom row) of the galaxies (above certain stellar mass limits; see \autoref{table:simulations} for minimum stellar masses for \colibre simulations with different resolutions) from the three \colibre simulations at the selected redshifts. The correlations at different redshifts are shown in the different columns ($z=7$ to $z=20$ from left to right). The squares represent the median values of the corresponding quantity in the stellar mass bin, with the error bars indicating the 16th-84th percentile range (i.e. the $\pm 1\sigma$ region). Good convergence is seen for all relations among the three \colibre simulations.}
\label{fig:mstar2props}
\end{figure*}

We follow the procedure outlined in Paper~I to select galaxies for which to calculate FUV luminosities including the effects of dust attenuation. The key steps are summarized below:
\begin{enumerate}
\item We select galaxies from three \colibre simulations of different resolutions and volumes (L100m5, L200m6 and L400m7), as listed in \autoref{table:simulations}, in order to derive UVLFs over a broad range of luminosity\footnote{The \colibre galaxy data were accessed using the {\sc python} packages {\sc swiftsimio} \citep{Borrow2020,Borrow2021} and {\sc swiftgalaxy} \citep{Oman2025}.}. 

\item We only analyse simulation outputs corresponding to \colibre ``snapshots'' which contain more detailed information on the particles in galaxies, such as the dust mass fractions of each dust species, required for radiative transfer calculations. From the available snapshots, we select those at $z = 7, 9, 10, 12, 15$ and 20.

In \autoref{fig:mstar2props}, we present the correlations between galaxy stellar mass and several global galactic properties for the three \colibre simulations at the selected redshifts, using galaxies above stellar mass limits of $5\times10^{5}\,\rm M_{\odot}$ for m5, $5\times10^{6}\,\rm M_{\odot}$ for m6 and $5\times10^{7}\,\rm M_{\odot}$ for m7, which correspond to roughly $2-5$ stellar particles\footnote{The approximate number of stellar particles in galaxies at the lower stellar-mass boundary varies slightly with resolution.}. At $z=20$, only three galaxies in L200m6 satisfy this stellar mass limit, while none do so in L400m7. We therefore consider only the results from L100m5 at this redshift.

The properties shown in \autoref{fig:mstar2props} are: (1) the total mass (including baryons, dark matter and black holes), denoted as $M_{\rm tot}$; (2) the mass of gas in the cool and dense phase, denoted as $M_{\rm cg}$ (defined as gas with $T < 10^{4.5}\,\mathrm{K}$ and $\rho_{\rm g}/m_{\mathrm{H}} > 10^{-1}\,\mathrm{cm}^{-3}$, where $T$ and $\rho_{\rm g}$ are the gas temperature and density, respectively and $m_{\mathrm{H}}$ is the mass of a hydrogen atom); (3) the instantaneous SFR; and (4) the dust mass, denoted as $M_{\rm dust}$. The total mass is calculated using all gravitationally bound particles, while the other global properties are measured within a spherical aperture of radius 50~kpc. All these quantities are taken from the Spherical Overdensity and Aperture Processor (SOAP; \citealt{McGibbon2025}) catalogue, which contains a wide range of global properties measured within different apertures. As shown in \autoref{fig:mstar2props}, the three \colibre simulations with different resolutions produce very consistent correlations, which agree down to very low stellar masses (corresponding to about two stellar particles). This convergence test demonstrates the strong consistency of these relations across different \colibre resolutions. 

\item For each \colibre simulation and at each redshift, we remove galaxies with stellar masses below the resolution limits adopted in \autoref{fig:mstar2props}. The remaining galaxies are then grouped in the stellar mass-SFR plane using bins of 0.5 dex in both stellar mass and SFR. From each bin we randomly select 50 galaxies. If a bin contains fewer than 50 galaxies, all galaxies in that bin are included. For each bin we record the sampling fraction, defined as the ratio between the number of selected galaxies and the total number of galaxies in that bin, which is then used to upweight the contribution of the subsampled population to the UVLF (see \autoref{eq:phi}). The number of selected galaxies and their {\it average} fractions relative to the {\em total} galaxy population in the different \colibre simulations are listed in \autoref{table:sample}. Binning galaxies in the stellar mass-SFR plane, rather than by stellar mass alone, ensures that galaxies with different levels of star formation activity are represented within each mass range. This prevents the sample being dominated by either highly star-forming or quiescent systems and therefore better captures the diversity of UV emission that contributes to the UVLF. The target number of galaxies per bin is somewhat arbitrary and is chosen to balance computational cost with adequate sampling. We have verified that this sampling strategy provides sufficiently accurate estimates of the UVLFs across the whole redshift range considered here (see Appendix~\ref{sec:random_sampling}). 
\end{enumerate}

\begin{figure*}
\centering
\includegraphics[width=\textwidth]{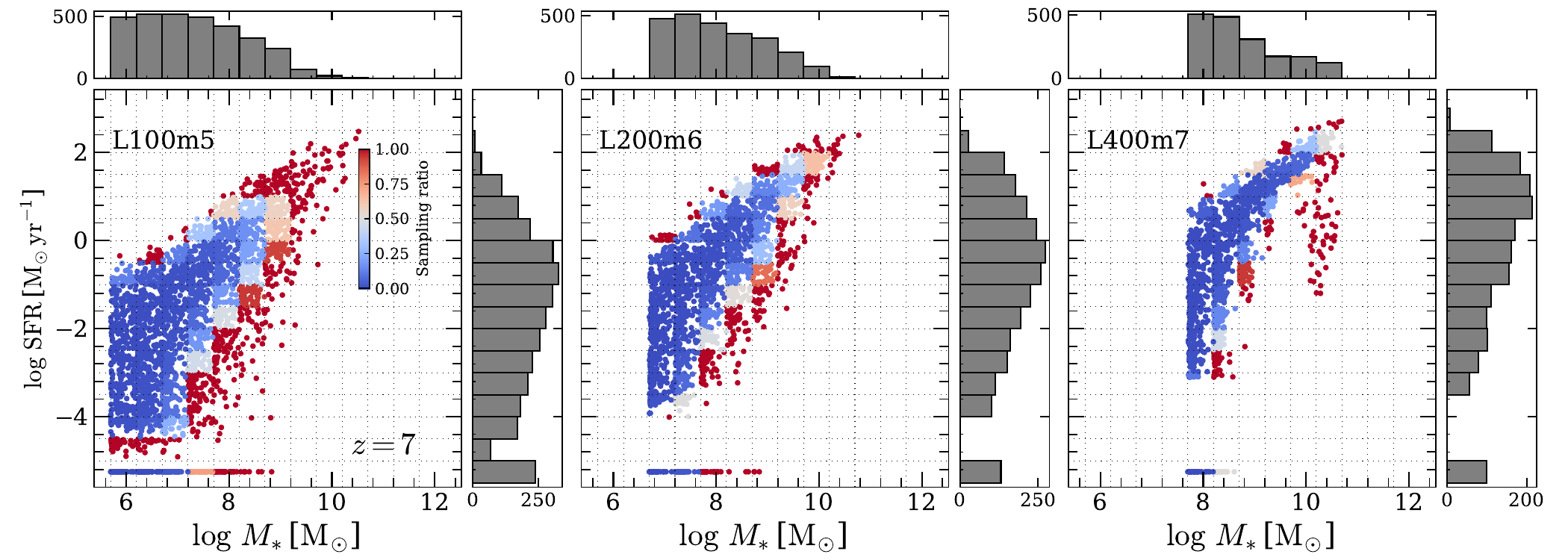}
\caption{The distribution of galaxies selected for post-processing with \skirt on the $\log\,M_{\ast}-\log \,\rm SFR$ plane for L100m5, L200m6 and L400m7 (from left to right). For clarity, SFRs lower than $10^{-5.25}\,\mathrm{M_{\odot}\,yr^{-1}}$ are set to $10^{-5.25}\,\mathrm{M_{\odot}\,yr^{-1}}$. The grid delineated by dotted lines indicates the boundaries of the stellar mass and SFR bins used in the sample selection. The colours represent the sampling fraction of galaxies (i.e., the ratio between the number of galaxies selected from a bin and the total number of galaxies in that bin). The histograms show the distributions of stellar mass and SFR for the selected galaxies. Distributions at other redshifts are shown in \autoref{fig:sample_more}.}
\label{fig:sample}
\end{figure*}

In \autoref{fig:sample}, we show the $\log\,M_{\ast}-\log\,\rm SFR$ distribution of selected galaxy samples at $z=7$ from the three \colibre simulations, colour-coded by their sampling fractions. Distributions of the selected samples at other redshifts 
($z=9, 10, 12, 15$ and 20) are shown in Appendix~\ref{sec:more_sample} (\autoref{fig:sample_more}). As can be seen, rare galaxies with high stellar mass and high or low SFR have higher sampling fractions, indicating a higher probability of being selected. 

We adopt the purely thermal AGN feedback simulations as our default model. We have checked that the choice of AGN feedback model has little impact on the UVLFs at the redshifts of interest (see Appendix~\ref{sec:feedback} for details).

\begin{table*}
\caption{Summary of the \colibre simulations used in this work. Column~(1): simulation name; (2): comoving side of the computational cube; (3): gas particle mass; (4): dark matter (DM) particle mass; (5): number of gas particles in the initial conditions; (6): number of DM particles in the initial conditions; (7): AGN feedback model; (8): lower stellar mass limit of selected galaxies.} \setlength{\tabcolsep}{2mm}
\begin{tabular}{lrcccccc}
\hline
Simulation Name & $L_{\rm box} [\mathrm{cMpc}]$ & $m_{\rm gas} [\mathrm{M_{\odot}}]$ & $m_{\rm DM} [\mathrm{M_{\odot}}]$ & $N_{\rm gas}$ & $N_{\rm DM}$ & AGN feedback model & Minimum stellar mass $[\mathrm{M_{\odot}}]$\\
\hline
L100m5 & 100 & $2.30\times 10^5$ & $3.03\times 10^5$ & $3008^3$ & $4\times 3008^3$ & Thermal & $5\times 10^5$\\
\hline
L100m6 & 100 & $1.84\times 10^6$ & $2.42\times 10^6$ & $1504^3$ & $4\times 1504^3$ & Thermal& $5\times 10^6$\\
\hline
L200m6 & 200 & $1.84\times 10^6$ & $2.42\times 10^6$ & $3008^3$ & $4\times 3008^3$ & Thermal& $5\times 10^6$\\
\hline
L400m7 & 400 & $1.47\times 10^7$ & $1.94\times 10^7$ & $3008^3$ & $4\times 3008^3$ & Thermal& $5\times 10^7$\\
\hline
L100m6h & 100 & $1.84\times 10^6$ & $2.42\times 10^6$ & $1504^3$ & $4\times 1504^3$ & Hybrid& $5\times 10^6$\\
\hline
\end{tabular}
\label{table:simulations}
\vspace{2mm}
\end{table*}

\begin{table}
 \caption{The number of galaxies in the selected samples for calculation of the UVLFs using {\skirt} at different redshifts in the three \colibre simulations. The percentages in parentheses represent the {\it average} sampling fraction (galaxies from different $\log\,M_{\ast}$--$\log\,\rm SFR$ grids have different sampling ratios) among galaxies above the stellar mass thresholds.} \setlength{\tabcolsep}{4mm}
\begin{tabular}{cccc}
\hline
\hline
 & L100m5 & L200m6 & L400m7\\
\hline

$z=7$ & 3100 (0.44\%) & 2415 (0.66\%) & 1759 (0.82\%)\\

\hline
$z=9$ & 2290 (0.56\%) & 1781 (1.55\%) & 1182 (3.30\%)\\

\hline
$z=10$ & 2007 (0.71\%) & 1468 (2.42\%) & 949 (7.22\%)\\

\hline
$z=12$ & 1428 (1.28\%) & 971 (6.20\%) & 521 (35.88\%) \\

\hline
$z=15$ & 793 (4.67\%) & 428 (33.59\%) & 29 (100\%)\\

\hline
$z=20$ & 158 (72.81\%) & - & -\\

\hline
\end{tabular}
\vspace{2mm}
\label{table:sample}
\end{table}

\subsection{Radiative Transfer}
\label{sec:skirt}
We perform dust radiative transfer calculations for the samples of \colibre galaxies described above using the calibration-free \cosk pipeline of \citet{Gebek2026}. Here, ``calibration-free'' means that no {\sc skirt} parameters were calibrated against luminosity-related observational constraints. Following \citet{Gebek2026}, we use a rescaled \citet{Draine2007} dust model because it provides a better match to the local CSED than the other dust models. However, as shown in Paper~I, this choice has only a minor impact on the predicted LFs and is therefore not expected to significantly affect the UVLF results presented here, where dust is used only for attenuation. The \cosk pipeline is built upon the publicly available 3D Monte Carlo radiative transfer code {\sc skirt}\footnote{Version 9; Available at \url{https://skirt.ugent.be}.} \citep{Baes2011,Camps2015,Camps2020}. The code supports complex source and dust geometries derived directly from particle- or grid-based hydrodynamical simulations \citep{Baes2015} and employs adaptive spatial grids to efficiently resolve dust structures \citep{Camps2013,Saftly2013,Saftly2014}. 

\citet{Gebek2026} showed that the local CSED is broadly recovered by the calibration-free \cosk pipeline. In Paper~I, we further showed that {\sc colibre-skirt} produces a galaxy population in remarkable agreement with measured local LFs from the FUV to the submillimetre, with the exception of the MIR, where emission of Polycyclic Aromatic Hydrocarbons is important but not explicitly simulated by {\sc colibre}. Here, we broadly follow the {\sc skirt} setup adopted in Paper~I, which in turn follows the setup of \citet{Gebek2026}, where the fiducial {\sc colibre-skirt} post-processing pipeline is described in full detail, with slight modifications to the SED wavelength range and the size of the dust discretisation box, in order to reduce the computational cost. Below, we briefly summarize the setup.

\begin{enumerate}
\item {\em Dust radiative transfer}: \skirt self-consistently models dust absorption, scattering, heating and thermal re-emission \citep{Camps2015}, supporting various dust species and complex geometries for both sources and dust distributions. Since we focus only on the UVLFs in this work and dust emission in the UV band is negligible, {\sc skirt} is operated in \texttt{ExtinctionOnly} mode. The dust distribution is taken directly from {\sc colibre}, which tracks graphite and silicate grains of two representative sizes (large and small grains). Since \colibre provides only the masses of the small- and large-grain components for each dust particle, whereas \skirt modelling requires a continuous grain size distribution, we reconstruct such a distribution by separately rescaling the small- and large-grain parts of an adopted dust model, divided at $0.03\,\mu\mathrm{m}$, such that their integrated masses match the corresponding dust masses predicted by the simulation. In this work, the continuous grain size distribution and the corresponding optical properties are adopted from \citet{Draine2007}.

\item {\em Luminosity sources}: We decompose the luminosity sources into two components, evolved stars and star-forming regions. Stellar particles older than a threshold age of $t_{\rm R}=10\,\mathrm{Myr}$ are treated as evolved stars and assigned simple stellar population (SSP) templates from BPASS (v2.2.1; \citealt{Eldridge2017,Stanway2018}), consistent with the early stellar feedback module adopted in \colibre \citep{Benitez-Llambay2026} and assuming a \citet{Chabrier2003} IMF, as in {\sc colibre}. Stellar populations younger than $t_{\rm R}$ are resampled using the star-forming gas (including the currently existing gas with non-zero SFR averaged over time $t_{\rm R}$ and the parent gas of the young stellar particles) to better capture the emission from unresolved young stars. The emission from these star-forming regions is computed using the {\sc toddlers} library \citep{Kapoor2023,Kapoor2024}, which models the time-dependent evolution of \ion{H}{ii} regions around young stars and includes the emission lines from the ionised gas in the \ion{H}{ii} regions. The {\sc toddlers} library is also based on the BPASS (v2.2.1) library and is consistent with what we use for evolved stars. In this work, we adopt the dust-free version of {\sc toddlers}, which currently does not include nebular continuum emission (see \citealt{Gebek2026}), such that attenuation by the dust distribution in {\sc colibre} is treated self-consistently in the subsequent {\sc skirt} radiative transfer calculation. The luminosities of star-forming regions are determined using the SFR averaged over the preceding time interval $t_{\rm R}$, assuming a constant star formation history over this period. As mentioned in \autoref{sec:sample}, the lowest-mass galaxies in our sample contain only $2-5$ star particles. However, this limitation is partially removed in the \skirt post-processing because these galaxies usually contain many more cool gas particles (the median ratio of the number of cool gas particles to star particles is $\approx 30$). These cool gas particles enter both the resampling of very young ($<10\,\rm Myr$) stellar populations and the modelling of the dust distribution for radiative transfer.

\item {\em Dust discretisation box}: Given that galaxies at high redshift are much smaller than those at $z=0$ and to optimize computational cost, we adopt a redshift-dependent discretisation box for the dusty medium around each galaxy (unlike in Paper~I, where a fixed-size discretisation box of 100 kpc was used). This volume defines the spatial region within which the dust distribution is discretised onto the \skirt grid. The dust discretisation volume is centred at the location of the most bound particle and has a proper size given by
\begin{equation}
\label{eq:skirt_box}
L_{\rm dust} = 100\,\mathrm{kpc}\times \frac{1.8}{0.7}\times \frac{1}{1+z},
\end{equation}
where $z$ is the redshift. The factors 1.8 and 0.7 correspond to the comoving and maximum physical gravitational softening lengths, respectively, adopted in \colibre at m6 resolution. The factor $(1+z)^{-1}$ converts the comoving softening length into proper units. Although the comoving and maximum physical gravitational softening lengths vary with simulation resolution (see Table~2 of \citealt{Schaye2026}), their ratio is fixed. We therefore adopt \autoref{eq:skirt_box} for all simulations. We have verified that the resulting discretisation box size is sufficiently large to encompass the dust distribution at high redshift in all simulations ($L_{\rm dust}=32.1\,\rm kpc$ at $z=7$ and $12.2\,\rm kpc$ at $z=20$). In {\sc skirt}, dust located outside this discretised medium volume is not included in the radiative transfer calculation. However, the radiation sources (stars and star-forming regions) are treated as gridless entities in {\sc skirt} and are therefore not restricted to this volume, i.e., (gravitationally bound) stars and star-forming regions located outside the dust discretisation box are still included.

\item The wavelength grid used here spans the range 0.09 to $0.2\,\mu\mathrm{m}$, with 40 logarithmically spaced bins.

\item We launch $10^{7.5}$ photon packets for each galaxy in the {\sc skirt} computation. The photon packets are randomly sampled from the source emission spectra at their native spectral resolution \citep{Camps2020}. We confirm that this number of photon packets is sufficient to resolve the SEDs of galaxies in the FUV band.

\item Galaxies are ``observed'' with an idealized mock detector positioned along the $z$-axis of the simulation cube, corresponding to a random viewing orientation relative to the galaxy, at a distance of 10 Mpc. No point-spread function convolution or noise injection is applied. As a result, this procedure does not constitute a full end-to-end forward modelling of the observations, but instead yields idealized galaxy luminosities for direct comparison.

\item Luminosities are computed within a fixed projected circular aperture of proper radius\footnote{Note that only gravitationally bound particles are considered here.} 50~kpc. As shown in \autoref{fig:uvlf_aperture}, the resulting UVLFs are virtually identical to those obtained when including all particles gravitationally bound to each galaxy, indicating that this aperture fully encompasses the galaxies at these redshifts. A detailed assessment of the impact of varying the aperture size on the UVLFs is presented in Appendix~\ref{sec:aperture}.
\end{enumerate}

\subsection{Observational data}
\label{sec:obs_data}
\begin{table*}
 \caption{Observational data on the FUV luminosity functions used at each redshift (see \autoref{sec:obs_data} for our criteria for selecting the observational data). To make fair comparisons, all observational magnitudes and number densities are rescaled as described in the text. All observational UVLFs used in this work for comparison (uncorrected and together with their corresponding $h_{\rm obs}$ values) are available in machine-readable format on both the journal website and \url{https://icc.dur.ac.uk/data/}.}\setlength{\tabcolsep}{4mm}
\begin{tabular}{p{1cm} p{6cm}p{6cm}}
\hline
\hline
Redshift & Pre-\textit{JWST} References & \textit{JWST} References\\
\hline
7 & \citet{Bouwens2011,Bouwens2015,Bouwens2021,Bouwens2022,McLure2013,Schenker2013,Bowler2014,Finkelstein2015,Stefanon2019,Harikane2022} & \citet{Bouwens2023a,Meyer24,Franco2025,Harikane2025,RojasRuiz2025}\\
\hline
9 & \citet{McLure2013,Morishita2018,Stefanon2019,Bowler2020,Bouwens2015,Bouwens2021,Bouwens2022}
& \citet{Bouwens2023a,Bouwens2023b,Donnan2023,Donnan2024,Harikane2023,Harikane2024,Harikane2025,Leung2023,Perez-Gonzalez2023,Adams2024,Finkelstein2024,Meyer24, Willott2024,Chemerynska2026,Franco2025,RojasRuiz2025}\\
\hline
10 & \citet{Oesch2014,Oesch2018,Bouwens2015,Morishita2018,Stefanon2019} & \citet{Bouwens2023a,Bouwens2023b,Castellano2023,Donnan2023,Donnan2024,Perez-Gonzalez2023, Adams2024,Harikane2024,Harikane2025,Willott2024,Chemerynska2026,Franco2025,RojasRuiz2025}\\
\hline
12 & \centering --- & \citet{Bouwens2023a,Bouwens2023b,Harikane2023,Perez-Gonzalez2023,Adams2024,Donnan2024,Robertson2024,Willott2024}\\
\hline
15 & \centering --- &\citet{Casey2024,Donnan2024,Finkelstein2024,Robertson2024,Harikane2025,Naidu2026}\\
\hline
20 & \multicolumn{1}{c}{---} & \multicolumn{1}{c}{---} \\
\hline
\end{tabular}
\vspace{2mm}
\label{table:obs_data}
\end{table*}

The observational UVLFs used in this work for comparison include both new {\it JWST} measurements and earlier estimates from other facilities such as the {\it Hubble Space Telescope} (HST). To compare our UVLFs with observational data at the same redshift, we adopt the following strategy: if an observational UVLF measurement provides a redshift range (i.e., lower and upper redshift limits) with a width no larger than 2, it is included when our target redshift ($z_{\rm target}$) falls within this range, i.e., $z_{\rm obs,min}\leqslant z_{\rm target}\leqslant z_{\rm obs,max}$ and $\Delta z\leqslant 2$, where $\Delta z\equiv z_{\rm obs,max}-z_{\rm obs,min}$ is the redshift bin width. If no redshift range is given and only a mean or median redshift ($z_{\rm obs}$) is reported, the measurement is used only when $|z_{\rm obs} - z_{\rm target}| \leqslant 0.5$. Observational data with redshift bins wider than $\Delta z = 2$ are excluded to avoid contamination from lower-redshift objects. As no observational measurements at $z=20$ satisfy these selection criteria, we present the \colibre results at this redshift as genuine predictions to be tested against future observations. With this selection, we list in \autoref{table:obs_data} the selected observational references at each redshift. 

Differences in the Hubble constant assumed in the observations can introduce systematic biases in the calculation of galaxy distances, affecting both absolute magnitudes and number densities. To ensure a consistent comparison between our simulations and observational data, we rescale all absolute magnitudes in the observational data to the Hubble constant used in {\sc colibre}, using $M+5\log_{10} (h_{\rm sim}/h_{\rm obs})$, where $h_{\rm obs}$ is the Hubble constant stated in the observational study and $h_{\rm sim}$ is the Hubble constant in \colibre ($h_{\rm sim}=0.681$). In addition, all number densities are scaled to the Hubble constant in {\sc colibre}, through $\log_{10} \Phi+3\log_{10} (h_{\rm sim}/h_{\rm obs})$. This standard convention effectively removes the dependence on the Hubble constant in our comparison of simulations with observations and allows for a fair comparison across datasets adopting different cosmologies. We note that variations in other cosmological parameters, such as $\Omega_{\rm m}$, can also influence the inferred LFs by altering the luminosity distance and comoving volume estimates. However, these effects are generally small compared to those from the differences in $h$ and they cannot be readily corrected using a simple analytic rescaling. For this reason, we do not apply an explicit correction for differences in $\Omega_{\rm m}$ in this work.

Collecting observational data from multiple studies can be time-consuming and repetitive. To support reproducibility and make future comparisons easier, we provide all observational datasets used in this series of papers in machine-readable format on both the journal website and \url{https://icc.dur.ac.uk/data/}. For broader discoverability, these datasets are also listed on the DigitizeAstronomy website (\url{https://digastro.org/}), a community platform for sharing machine-readable data extracted from the astronomical literature. We hope this will be a useful resource for the community and encourage similar practices in future work.

\section{Results}
\label{sec:result}

\subsection{Correlations between galaxy properties and UV magnitude}
\label{sec:correlation}

\begin{figure*}
\centering
\includegraphics[width=0.9\textwidth]{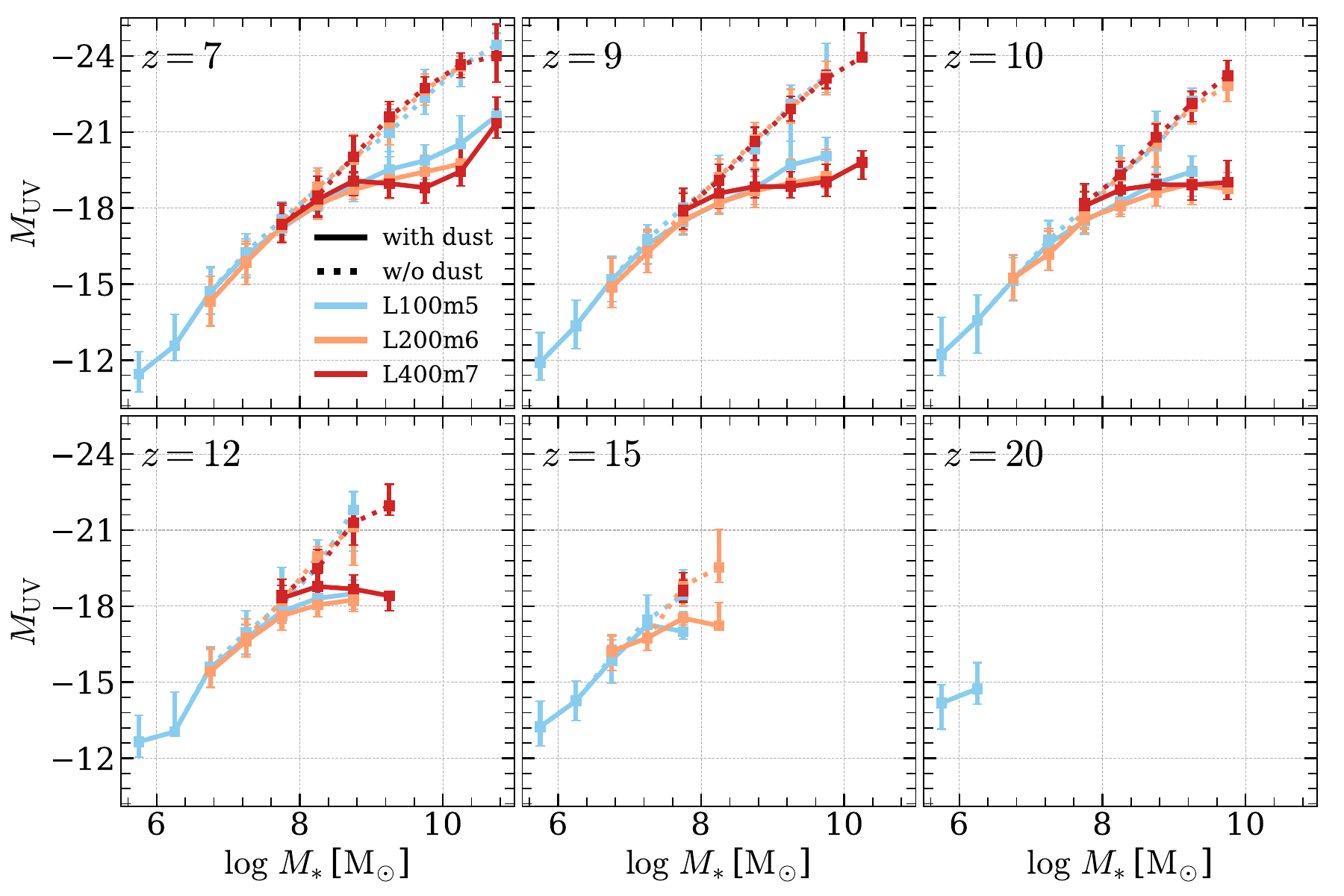}
\caption{Correlations between the stellar mass of galaxies ($\log\,M_{\ast}$) and their UV magnitude, $M_{\rm UV}$ (solid: dust-attenuated; dashed: dust free) at different redshifts (as indicated in each panel). Results for L100m5, L200m6 and L400m7 are indicated by different colours. The squares represent the median $M_{\rm UV}$ in the stellar mass bins, with the error bars indicating the interval from the 16th to 84th percentiles of $M_{\rm UV}$ in each stellar mass bin. \colibre simulations with different resolutions show consistent $M_{\rm UV}-\log\,M_{\ast}$ relations across the redshift range $z=7-15$, especially for dust-free $M_{\rm UV}$.}
\label{fig:correlations}
\end{figure*}

The spectral energy distribution (SED, or spectrum) from \skirt is the sum of emission from evolved stars and star-forming regions processed by the dust integrated over a given projected aperture ($50\,\rm kpc$ radius in this work). Having the output SEDs from {\sc skirt}, we follow the practice adopted in studies using {\it JWST} data (e.g., \citealt{Finkelstein2024}) and obtain the UV magnitude at $\lambda = 1500\,\Angstrom$ by convolving the SED with a top-hat filter spanning $1450-1550\,\Angstrom$. Throughout this work, we denote $M_{\rm UV}$ as the derived rest-frame AB absolute magnitude of galaxies.

In \autoref{fig:correlations}, we present the correlations between stellar mass and $M_{\rm UV}$ (both dust-attenuated and dust-free) for the selected galaxies over $z=7-20$ in the three \colibre simulations listed in \autoref{table:simulations}. At $z=20$, only the L100m5 result is shown, as only three galaxies in L200m6 and none in L400m7 satisfy the stellar mass limit. As expected, at fixed redshift, both dust-attenuated and dust-free $M_{\rm UV}$ on average become brighter with increasing stellar mass. For both dust-attenuated and dust-free results, the three \colibre simulations are in good agreement in terms of the $M_{\rm UV}-M_{\ast}$ relation down to the lower stellar mass limit of the selected samples (see \autoref{table:simulations}), indicating good numerical convergence between m5 and m6 and slightly poorer convergence for m7 (with the dust-free results showing even better consistency). Furthermore, the dust-attenuated and dust-free $M_{\rm UV}$ begin to diverge at $\log\,M_{\ast}/\mathrm{M_{\odot}}\sim 7-8$ (except at $z=20$, where they remain nearly identical), with the difference increasing toward higher stellar masses. This trend is expected because, on average, massive galaxies are more metal-enriched \citep{Tremonti2004} and typically exhibit higher gas surface densities, which together imply larger dust column densities in their ISM and thus stronger attenuation. However, geometry, inclination and the relative star-dust distribution introduce significant scatter. An interesting feature is that the dust-attenuated $M_{\rm UV}-M_{\ast}$ relation in L400m7 at $z=7$ and 9 shows an ``upturn'' at the highest stellar masses. This trend deserves further investigation in future work.

\subsection{The UV Luminosity Functions}
\label{sec:uvlf}

\subsubsection{UVLFs for three \colibre simulations}
\label{sec:uvlfs_individual}

\begin{figure*}
\centering
\includegraphics[width=\textwidth]{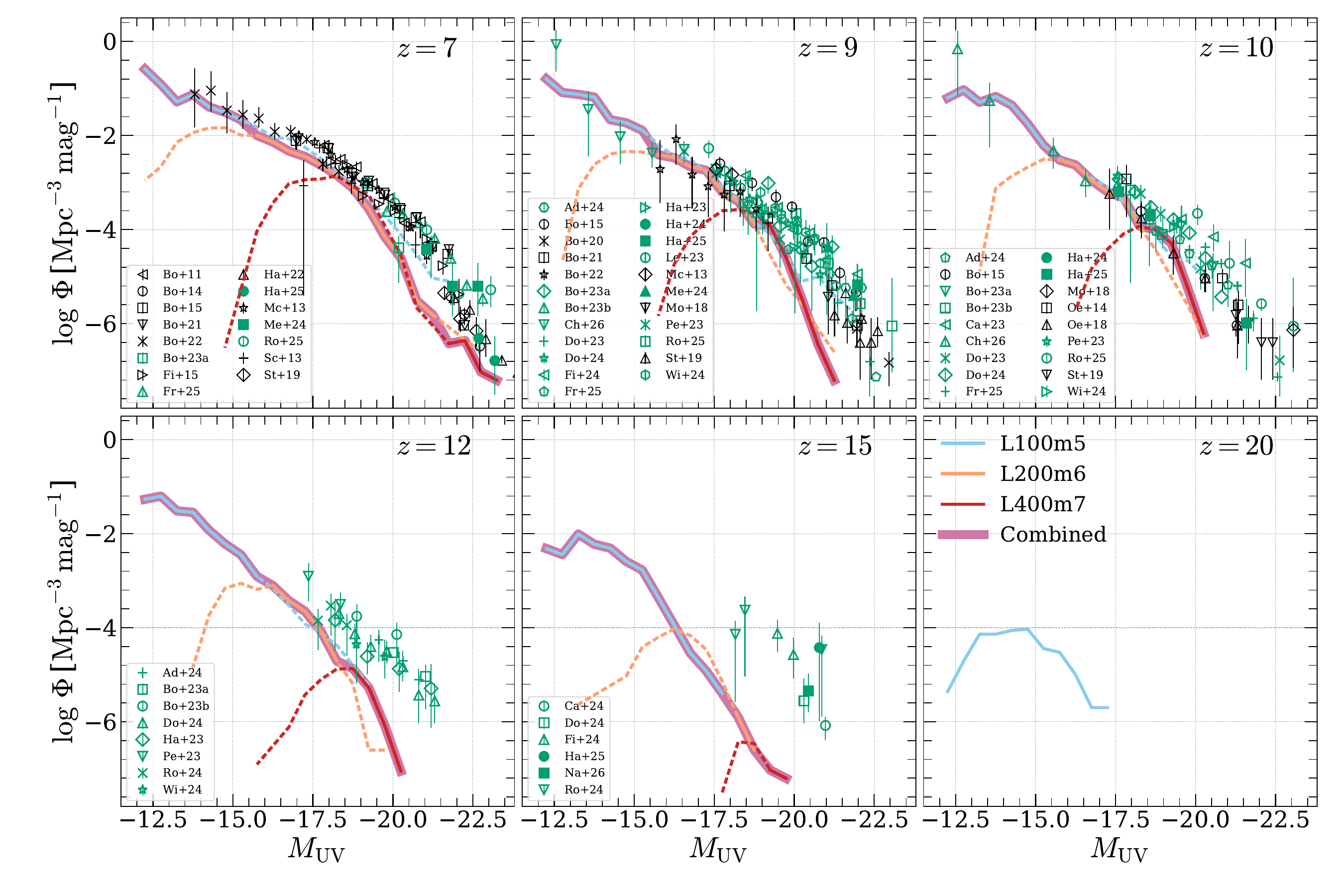}
\caption{Comparison of (dust-attenuated) UVLFs between the predictions of \colibre (consisting of the UVLFs from three individual \colibre simulations and their combined UVLFs; see \autoref{sec:combine} for details) and observational data. We remind readers that the redshifts here correspond to the \colibre snapshots and are {\it not} the average redshifts of the observational data. Please see \autoref{sec:obs_data} for the selection criteria for the observational data and \autoref{table:obs_data} for the list of selected observational datasets at each redshift. For the UVLFs from the three individual \colibre simulations, we use dashed lines to indicate those portions that are not included in the combined UVLFs. In each panel, we overplot observational data, grouped into three categories: (i)~\textit{JWST} data with spectroscopically confirmed redshifts (filled green symbols), (ii)~\textit{JWST} with photometric redshifts only (open green symbols) and (iii)~photometric redshift-based measurements from earlier observations, such as HST (open black symbols). For clarity, we exclude data points labelled as upper or lower limits in the original paper.}
\label{fig:uvlf}
\end{figure*}

\begin{figure*}
\centering
\includegraphics[width=\textwidth]{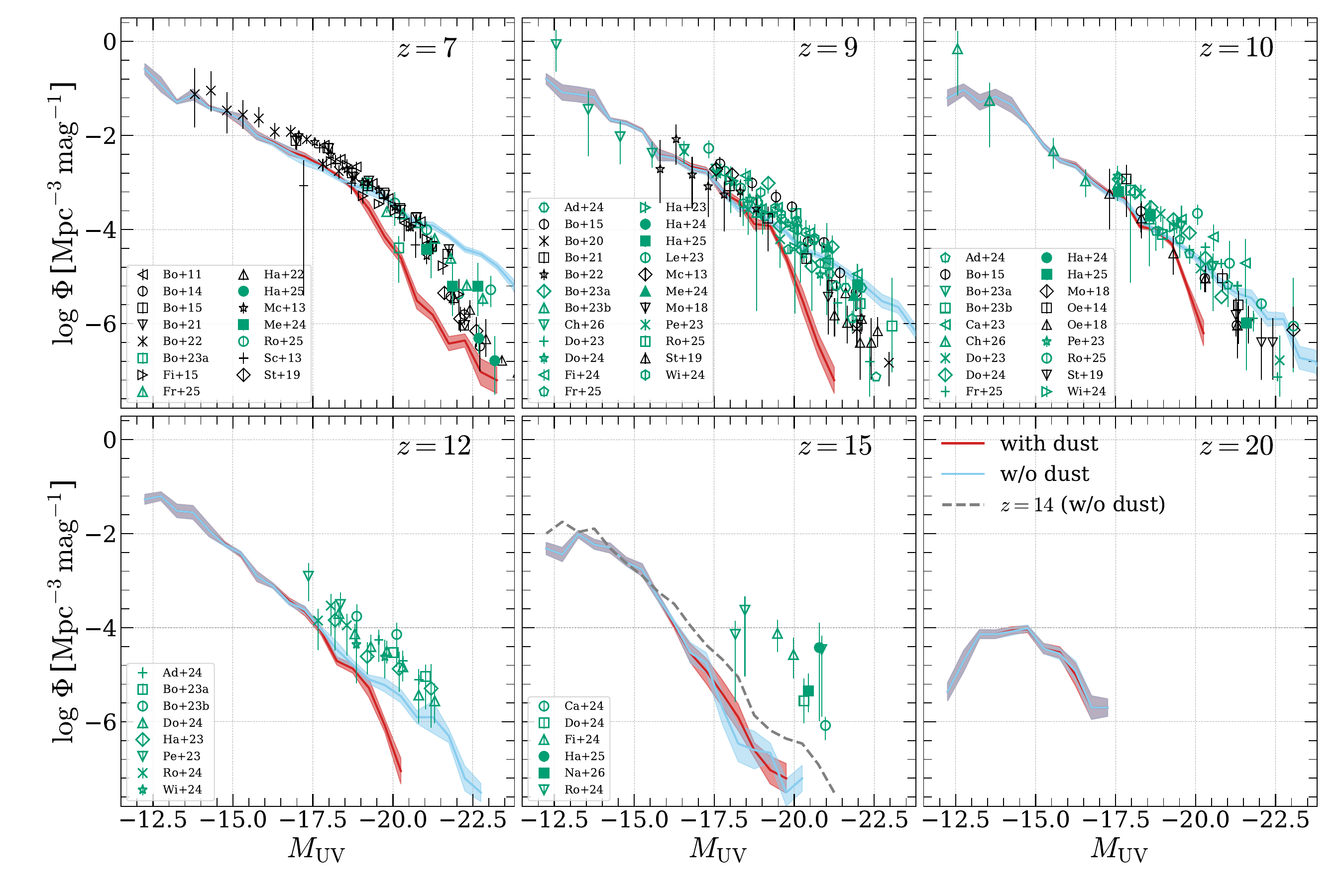}
\caption{Comparison of the combined dust-attenuated and dust-free \colibre UVLFs with observational data. The shaded regions represent the $\pm 1\sigma$ ranges, calculated using the bootstrap method by perturbing both the {\it parent} sample (from which our random sample processed with {\skirt} is drawn) and the selected sample 1000 times (see \autoref{sec:combine} for details). The grey dashed curve in the $z=15$ panel shows the dust-free UVLF of \colibre at $z=14$, providing a more precise comparison with observations that are based on samples with redshifts close to 14. The symbols for the observational data are as in \autoref{fig:uvlf}.}
\label{fig:uvlf_dust}
\end{figure*}

In this work, \colibre UVLFs are computed in bins of width $\delta M =0.5$ magnitudes, while the results are reported as galaxy number densities per magnitude. In each bin, the galaxy abundance is thus calculated as
\begin{equation}
\label{eq:phi}
\Phi_{M} = \frac{\sum_i 1/\eta_i}{\delta M \times L_{\rm box}^3},
\end{equation}
where $\eta_i$ is the sampling ratio (see \autoref{sec:sample}) of the $i$-th galaxy and $L_{\rm box}$ is the side length of the simulation cube. The summation is taken over all galaxies within the given $M_{\rm UV}$ bin. We confirm that the choice of magnitude bin width (e.g., 0.25, 0.5, 0.75 and 1) has only a minor influence on the UVLFs. In \autoref{fig:uvlf}, we present the predicted UVLFs from three \colibre simulations (indicated with different colours) at $z=7-20$. The predicted UVLFs in \autoref{fig:uvlf} are calculated from the dust-attenuated UV magnitudes from {\sc skirt}. Observational data (see \autoref{sec:obs_data} for the selection criteria and \autoref{table:obs_data} for the selected datasets) are overplotted for reference, but a detailed comparison is deferred to \autoref{sec:comparison}.

As shown in \autoref{fig:uvlf}, the UVLFs from different \colibre simulations at varying resolutions exhibit good agreement over certain magnitude ranges. However, at magnitudes fainter than $M_{\rm UV} \approx -18$, the UVLFs of L400m7 begin to deviate from those of the higher-resolution simulations. Similarly, at magnitudes fainter than $M_{\rm UV} \approx -16$, the UVLFs of L200m6 gradually diverge from those of L100m5. This is expected, as we limit our sample to galaxies with stellar mass above a certain threshold (roughly corresponding to $2-5$ stellar particles; see \autoref{table:simulations}), potentially leading to incompleteness at the faint end of each UVLF. When compared to \autoref{fig:correlations}, we find that the $M_{\rm UV}$ thresholds at which different \colibre simulations start to diverge roughly correspond to the median $M_{\rm UV}$ of galaxies near the stellar-mass lower limit. This implies that the UVLFs are already approximately converged even in galaxies containing only a few stellar particles. The good agreement between the \colibre simulations also suggests that the UVLFs are converged with respect to the simulation volume over the magnitude ranges considered.

\subsubsection{Combining UVLFs from three simulations and estimating the errors}
\label{sec:combine}
To construct a representative UVLF from \colibre across a wide magnitude range at each redshift, we combine the UVLFs from different \colibre simulations using the following procedure:
\begin{enumerate}
\item Among the three \colibre simulations, we adopt the UVLFs from L200m6 as the default reference. Starting from $M_{\rm UV} = -17.5$ and moving toward the fainter end, we compare the galaxy abundances from L200m6 and the higher-resolution simulation, L100m5. If the difference between the two exceeds a threshold (0.25 dex) in two consecutive $M_{\rm UV}$ bins, we switch to using the UVLFs from L100m5 for all subsequent (fainter) bins. We note that we adopt $M_{\rm UV} = -17.5$ as the starting point as it ensures sufficient reliability from both the numerical and statistical perspectives. According to \autoref{fig:correlations}, in L200m6, $M_{\rm UV} = -17.5$ corresponds to a stellar mass of $\approx 10^{7.6}\, \mathrm{M_\odot}$, which consists of $\approx 20$ stellar particles. Additionally, as shown in \autoref{fig:uvlf}, $M_{\rm UV} = -17.5$ corresponds to a galaxy abundance of $\sim 10^{-4.4}\,\mathrm{Mpc^{-3}\,mag^{-1}}$ at $z=15$ (and even higher at lower redshifts) in L200m6, which translates to about 320 galaxies per magnitude in the L200 simulation cube. Practically, this choice also facilitates the construction of a smooth combined LF, which would be more difficult to obtain at other magnitudes. Therefore, this starting point guarantees both adequate resolution and robust statistical sampling.

\item Similarly, starting from $M_{\rm UV} = -17.5$ and moving toward the brighter end, we compare the UVLFs from L200m6 and the simulation in the larger volume, L400m7. Again, if the difference exceeds 0.25 dex in two consecutive $M_{\rm UV}$ bins, we switch to using the UVLFs from L400m7 for all subsequent (brighter) bins.
\end{enumerate}
The combined UVLFs are shown as the thick purple curves in \autoref{fig:uvlf}, while measurements from individual \colibre simulations that are {\it not} used in the combination are shown as dashed curves. 

The measurement uncertainties of the UVLFs for individual \colibre simulations, as well as for the combined UVLFs, are estimated using the bootstrap method. Since the UVLFs are computed from a subsample of galaxies (see \autoref{sec:sample}), we assessed the statistical uncertainty arising from this subsampling by performing 1000 bootstrap resamplings of both the parent sample (i.e., the full set of galaxies in \colibre before any selection) and the selected subsample (for \skirt modelling) for each \colibre simulation and each redshift. In each bootstrap iteration, we first randomly draw galaxies from the parent sample with replacement to create a ``mock'' parent sample of the same size. Then, we randomly draw galaxies from the selected subsample (see \autoref{sec:sample}) in each stellar mass-SFR bin, also with replacement, to create a ``mock" subsample (this procedure preserves the number of selected galaxies in each bin). For each bootstrap iteration, we then recalculate the sampling ratio (i.e., the ratio between the number of selected galaxies and the total number of galaxies in the parent sample within each bin) for each (re)selected galaxy, which may appear multiple times in the bootstrap resampling, and then compute the corresponding ``mock'' UVLF. Finally, the uncertainty in the UVLF is quantified as the standard deviation derived from the 1000 mock UVLFs. By bootstrapping both the parent sample and the selected subsample, we naturally capture both the fluctuations in the sampling ratio and the random variations in the detailed properties of the galaxies included in the selected subsample, reflecting statistical uncertainties arising from both the representativeness of the parent population and the stochasticity of the selection. These variations introduce additional fluctuations in the estimated UVLFs, which are used to quantify the measurement uncertainties.

The uncertainty in the combined UVLF is then obtained by applying the same combination strategy described in the previous paragraphs to each of the 1000 individual mock UVLFs. This yields 1000 {\it mock} combined UVLFs and the uncertainty in the combined UVLF is also quantified as the standard deviation of these 1000 realizations.

\subsubsection{Comparison with observations}
\label{sec:comparison}
The observational data used for comparison are selected as described in \autoref{sec:obs_data} and listed in \autoref{table:obs_data}. In \autoref{fig:uvlf_dust}, we compare the \colibre UVLFs with dust attenuation (red), including the measurement uncertainties estimated using the method described above, with the observational data. UVLFs constructed from earlier, pre-\textit{JWST} data, which are all based on photometric redshifts, are indicated by open black symbols. UVLFs derived from \textit{JWST} observations are shown as green symbols, with filled symbols indicating measurements based on fully spectroscopically confirmed high-$z$ candidates, including those from \citet{Harikane2024,Harikane2025,Meyer24,Naidu2026}\footnote{The UVLFs of \citet{Meyer24} are based on FRESCO spectroscopically confirmed H$\beta$+[O~III] emitters. Since this [O~III]-selected sample is incomplete at the faint end owing to line-detection sensitivity and scatter in the [O~III]-to-UV ratio, we follow \citet{Harikane2025} and show only the bright points with $M_{\rm UV}\leqslant -21$.} and open symbols indicating measurements based on photometric redshifts from \textit{JWST}. To avoid confusion in the figure, we exclude data points reported as upper or lower limits. It can be seen that the measurements from \textit{JWST} agree well with pre-\textit{JWST} observations and that the results based on spectroscopically confirmed objects are consistent with those based on photometric redshifts.

As shown in \autoref{fig:uvlf_dust}, the galaxy abundance predicted by \colibre drops by a factor of $\sim 10^{5}$ from $M_{\rm UV} \approx -15$ to $M_{\rm UV} \approx -22$ at $z=7$ when dust attenuation is included (on average, $\Delta\log\,\Phi/\Delta M_{\rm UV}\approx 0.7\,\rm dex$ per magnitude). At higher redshift, e.g., $z=15$, the decreasing trend becomes steeper with the galaxy abundance dropping by a factor of $\sim 10^{4}$ from $M_{\rm UV} \approx -15$ to $M_{\rm UV} \approx -19$ ($\Delta\log\,\Phi/\Delta M_{\rm UV}\approx 1\,\rm dex$ per magnitude). This difference is most clearly seen in the upper left panel of \autoref{fig:uvlf_schechter}, where the combined UVLFs at different redshifts are overplotted for comparison. At $z=20$, the UVLF derived from L100m5 shows a decline similar to that at $z=15$ for $M_{\rm UV}<-15$, while at the faint end it is affected by sample incompleteness.

When compared to the observed UVLFs, the \colibre dust-attenuated UVLFs show faint-end slopes at $z=7-12$ that are broadly consistent with those inferred from observations, while the bright end is noticeably steeper. At $z=7$, the \colibre dust-attenuated UVLF is systematically lower than the observations at the bright end. Equivalently, \cosk underpredicts the luminosities of the most UV-luminous galaxies, with an offset of $\approx 1\,\rm mag$ at a number density of $10^{-6}\,\mathrm{Mpc^{-3}\,mag^{-1}}$. This number density corresponds to 8 galaxies per unit magnitude in the L200m6 simulation box and 64 galaxies per unit magnitude in the L400m7 simulation box. At $z = 9$ and 10, the \colibre UVLFs also agree fairly well with the observations at the faint end, while again underpredicting the luminosities of the most UV-luminous galaxies, with offsets of $\approx 1\,\rm mag$ at $z=9$ and $\approx 1.2\,\rm mag$ at $z=10$ at the same number density as above. The offset increases to $\approx 2.5\,\rm mag$ at $z=15$.

The galaxy abundance at fixed $M_{\rm UV}$ decreases with redshift. For example, in {\sc colibre}, at $M_{\rm UV} = -19.5$, it decreases from $\sim10^{-4.5}\,\mathrm{Mpc^{-3}\,mag^{-1}}$ at $z=9$ to $\sim10^{-6}\,\mathrm{Mpc^{-3}\,mag^{-1}}$ at $z=12$, indicating strong redshift evolution ($\Delta\,\log\,\Phi/\Delta z\approx 0.5$ at $M_{\rm UV}=-19.5$; see also the upper left panel of \autoref{fig:uvlf_schechter}), which is inconsistent with observations (e.g., \citealt{Harikane2023,Finkelstein2024}) for which a milder redshift evolution of the UVLF at high-$z$ is reported ($\Delta\,\log\,\Phi/\Delta z\approx 0.12$ at $M_{\rm UV}\approx-19.5$ from $z=9$ to $z=14$ in \citealt{Finkelstein2024}; $\Delta\,\log\,\Phi/\Delta z\approx 0.05$ at $M_{\rm UV}\approx-19.5$ from $z=9$ to $z=12$ in \citealt{Harikane2023}\footnote{For \citet{Harikane2023}, we only provide a rough estimate of the galaxy abundance at $M_{\rm UV} \approx -19.5$ for $z = 9$ and $z = 12$.}). 

In \autoref{fig:uvlf_dust}, we also show the comparison between dust-free \colibre UVLFs (with measurement uncertainties) and the observations (which automatically include the effects of any dust attenuation present). We note that the dust-free results here are also from \skirt modelling, including the emission from evolved stellar particles and the star-forming regions, but ignoring dust attenuation. We can see from \autoref{fig:uvlf_dust} that the dust-free UVLFs differ from the dust-attenuated results only at the bright end, while the faint end remains nearly unchanged. This trend is consistent with the results shown in \autoref{fig:correlations}, where the difference between dust-attenuated and dust-free $M_{\rm UV}$ clearly increases with stellar mass. 

At $z=7$, the dust-free UVLFs are slightly lower than the observations at the faint end but substantially higher at the bright end. We note, however, that a completely dust-free scenario is inconsistent with observational results at this redshift, as ALMA observations of $z\approx 7$ galaxies (e.g. the REBELS programme) already reveal significant rest-frame infrared emission, implying the presence of dust (\citealt{Barrufet2023}). At $z=9$, the dust-free UVLF of \colibre agrees with the observations at the faint end, as does the dust-attenuated result, but is higher than the observations at the bright end. At $z=10$, the dust-free \colibre UVLF roughly matches the observed UVLF across the whole luminosity range. At $z=12$, the dust-free UVLF lies slightly below the observational data, while at $\Phi = 10^{-6}\,\mathrm{Mpc^{-3}\,mag^{-1}}$, the dust-free UV luminosities of \colibre galaxies roughly match those inferred from the observations. At $z=15$, even the dust-free UVLF remains lower than the observations over the entire luminosity range, meaning that \cosk underpredicts the luminosities of the most UV-luminous galaxies even when dust attenuation is ignored. The luminosity offset is $\approx 2.5\,\rm mag$ at $\Phi = 10^{-6} \, \mathrm{Mpc^{-3}\,mag^{-1}}$, similar to the dust-attenuated case, since the dust content at $z=15$ is negligible.

We note that some of the observational data used for the comparison at $z=15$ are actually based on galaxies at slightly lower redshifts. For example, the spectroscopic constraints from \citet{Naidu2026} rely on sources with $z_{\rm spec} \approx 14$ \citep{Carniani2024,Naidu2026}. Motivated by this, we include the \colibre UVLF at $z=14$, also calculated with {\sc skirt}, in the $z=15$ panel, shown as the grey dashed curve. The \colibre output at $z=14$ is a ``snipshot'' and does not contain the dust-species information required for \skirt modelling. We therefore only present the dust-free UVLF at $z=14$, noting that by $z \approx 15$ the dust-free UVLF is already very similar to the dust-attenuated one. As shown, while the \colibre UVLF at $z=14$ is closer to the observations, the most UV-luminous galaxies are still fainter than observed, with an offset of $\approx 1.5-2\,\rm mag$ at $\Phi = 10^{-6} \, \mathrm{Mpc^{-3}\,mag^{-1}}$.

More recently, \citet{Kreilgaard2026} and \citet{McLeod2026} reported lower UVLFs than many previous {\it JWST} photometric studies and hence results closer to the \colibre predictions. However, their redshift bins are not directly matched to those adopted here. This is particularly important since the UVLF evolves rapidly at these redshifts, especially beyond $z\approx 12$ \citep{McLeod2026}. In the highest-redshift bin of \citet{McLeod2026}, the detected candidates lie well below the upper edge of the adopted redshift interval; the resulting effective volume is therefore large, naturally leading to lower inferred number densities. This illustrates why such broad redshift bins cannot be used for a direct comparison with our target redshifts, even when the target redshift lies within the bin. We therefore do not include these measurements in our observational compilation, following the selection criteria stated in \autoref{sec:obs_data}.

\subsection{Schechter fitting and UV luminosity density}
\label{sec:schechter}

\begin{figure*}
\centering
\includegraphics[width=0.8\textwidth]{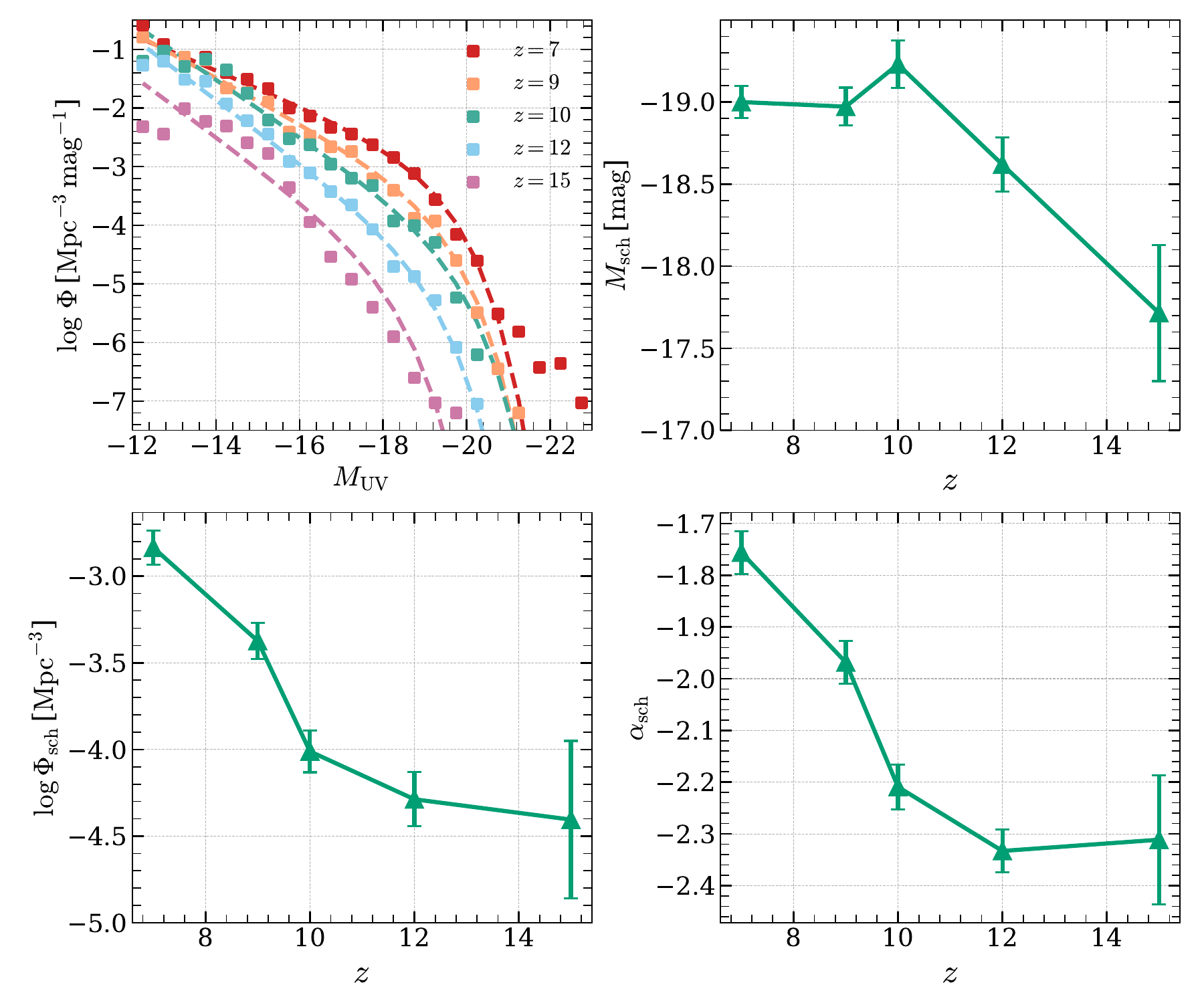}
\caption{Upper left: \citet{Schechter1976} function fits to the combined UVLFs in the \colibre simulations across redshifts $z = 7$ to $z = 15$ (see \autoref{sec:combine} for details on the combination method). Squares indicate galaxy number densities in each magnitude bin and dashed curves the corresponding Schechter fits. Other panels: redshift evolution of the Schechter parameters - characteristic absolute magnitude ($M_{\rm sch}$; upper right), characteristic number density ($\log\,\Phi_{\rm sch}$; lower left) and faint-end slope ($\alpha_{\rm sch}$; lower right). Error bars represent $\pm1\sigma$ uncertainties, estimated with the bootstrap method (see \autoref{sec:schechter} for details).}
\label{fig:uvlf_schechter}
\end{figure*}

\begin{figure}
\centering
\includegraphics[width=\columnwidth]{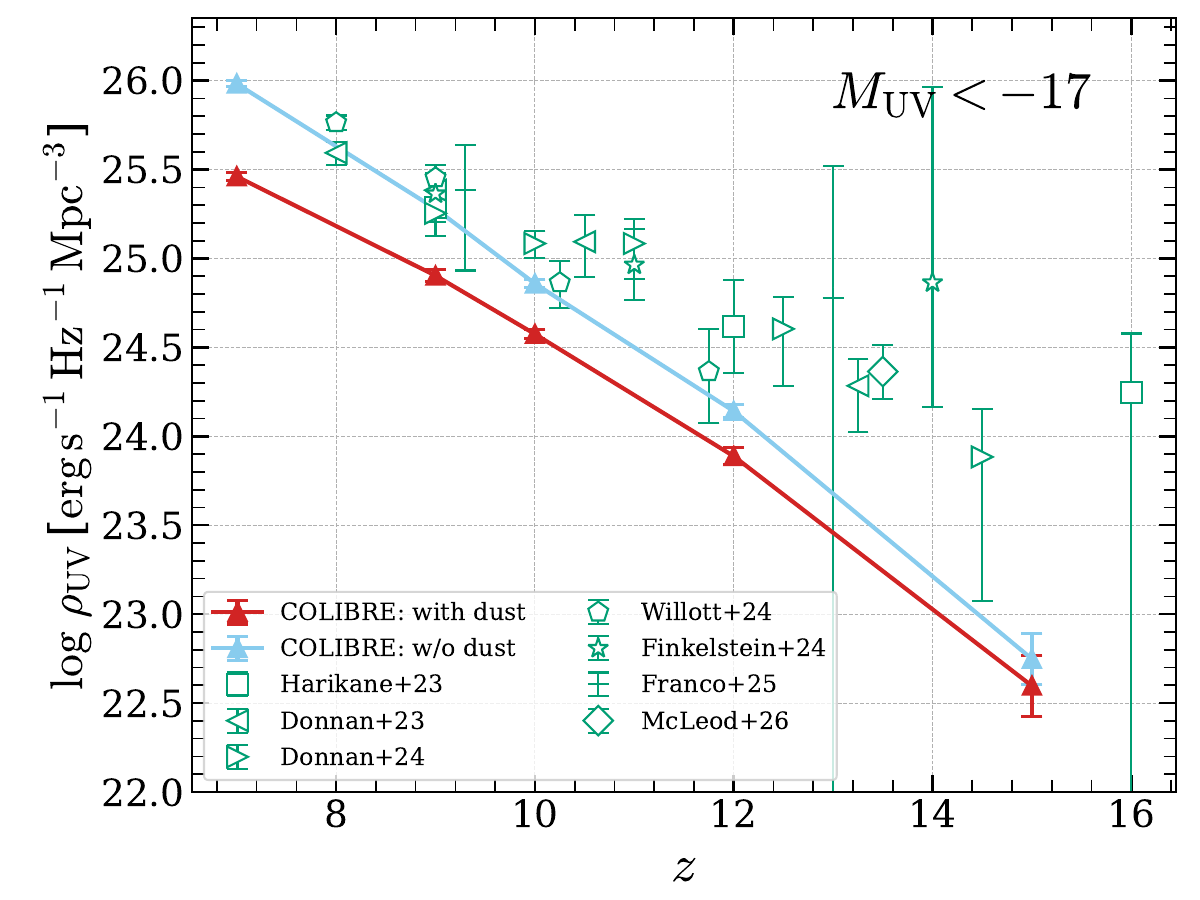}
\caption{Redshift evolution of the UV luminosity density, $\rho_{\rm UV}$, from \colibre\ (including all galaxies brighter than $M_{\rm UV}=-17$) from $z=7$ to $z=15$, compared with {\it JWST} observations (dust-attenuated, including \citealt{Harikane2023,Donnan2023,Donnan2024,Willott2024,Finkelstein2024,Franco2025,McLeod2026}). The red line shows the dust-attenuated \colibre UV luminosity density and the blue line shows the dust-free results. The error bars represent the $\pm1\sigma$ uncertainties, estimated with the bootstrap method by perturbing the parent sample (from which our random sample for \skirt modelling is drawn) and the selected sample 1000 times (see \autoref{sec:combine} and \autoref{sec:schechter} for details).}
\label{fig:uvdensity}
\end{figure}

The \citet{Schechter1976} function is widely used to parametrize the shape of observed LFs at all wavelengths. The evolution of the LF is then often described in terms of the evolution of the Schechter function parameters. In this section, we fit Schechter functions to our combined (dust-attenuated) UVLFs at all the redshifts (except at $z=20$, where the result is only from L100m5 and the bright end is not constrained) and study the redshift evolution of the Schechter parameters. The number density of galaxies within a luminosity range $[L,L+\mathrm{d}L]$ according to the \citet{Schechter1976} function can be written as:
\begin{equation}
\Phi(L)\,\mathrm{d}L = \Phi_{\rm sch} \left( \frac{L}{L_{\rm sch}} \right)^{\alpha_{\rm sch}}
\exp\!\left(-\frac{L}{L_{\rm sch}}\right) \frac{\mathrm{d}L}{L_{\rm sch}},
\end{equation}
where $\Phi(L)$ denotes the number density of galaxies with luminosity $L$ and $\Phi_{\rm sch}$, $L_{\rm sch}$ and $\alpha_{\rm sch}$ represent the characteristic number density, characteristic luminosity and faint-end slope of the UVLF, respectively. In this work, we adopt the magnitude-space formulation of the Schechter function, which can be written as:
\begin{equation}
\begin{aligned}
\Phi(M)\,\mathrm{d}M
&= 0.4\,\ln(10)\,\Phi_{\rm sch}\,
10^{-0.4(\alpha_{\rm sch}+1)(M-M_{\rm sch})} \\
&\quad \times
\exp\!\left[-10^{-0.4(M-M_{\rm sch})}\right]
\,\mathrm{d}M,
\end{aligned}
\end{equation}
where $M_{\rm sch}$ is the characteristic absolute magnitude.

We perform nonlinear least-squares fitting over the $M_{\rm UV}$ range from $-12$ to $-21$. During the fitting, each data point is assigned a weight inversely proportional to the square of its measurement error (in logarithmic space) described in \autoref{sec:combine}, so data points with smaller errors have a larger influence on determining the best-fit parameters. The derived Schechter parameters at different redshifts are listed in \autoref{table:schechter}. In the upper left panel of \autoref{fig:uvlf_schechter}, we show the combined UVLF predictions for different redshifts from \colibre with squares and the best-fitting Schechter functions with dashed curves. As can be seen, at $z=7$, the extreme bright end of the \colibre UVLF cannot be well described by a Schechter function and there appears to be an excess of galaxies in these bins. This is consistent with the ``upturn'' seen in the $M_{\rm UV}-M_{\ast}$ relation of L400m7 at $z=7$ (see \autoref{fig:correlations}).

In the other panels, we show the redshift evolution of the three Schechter parameters: the characteristic magnitude ($M_{\rm sch}$; upper right), the characteristic number density ($\log\,\Phi_{\rm sch}$; lower left) and the faint-end slope ($\alpha_{\rm sch}$; lower right). To obtain the measurement uncertainties of the Schechter parameters, we apply Schechter fitting to the 1000 {\it mock} combined UVLFs at each redshift (see \autoref{sec:combine} for details about obtaining the {\it mock} UVLFs) and calculate the standard deviation of the 1000 sets of Schechter parameters. As can be seen, all three parameters exhibit clear evolutionary trends with redshift, which are generally monotonic. Specifically, the characteristic absolute magnitude generally becomes fainter with increasing redshift, the faint-end slope tends to steepen and the characteristic number density generally declines.

We note that we did not compare with observational estimates of the Schechter parameters here, as many observational studies adopt different methods when applying Schechter fitting, for example, fixing the faint-end slope (e.g., \citealt{Harikane2023}) or fixing both the faint-end slope and the characteristic magnitude (e.g., \citealt{Bouwens2023b}). Comparing Schechter parameters is, in any case, not a straightforward way to compare predictions with observations. Firstly, both the predicted and observed LFs may not be well described by a Schechter function. Secondly, observational Schechter-function fits may account for effects such as redshift evolution across the sample and photo-$z$ uncertainties, whereas these effects are not included in our fits to the \colibre UVLFs. We therefore focus primarily on direct comparisons between the predicted and observed LFs.

For a further comparison with observations, we show, in \autoref{fig:uvdensity}, the redshift evolution of the UV luminosity density ($\rho_{\rm UV}$; both dust-attenuated and dust-free), together with recent observational results based on \textit{JWST} observations (dust-attenuated; including \citealt{Harikane2023,Donnan2023,Donnan2024,Willott2024,Finkelstein2024,Franco2025,McLeod2026}). The observational results are selected using criteria similar to those described in \autoref{sec:obs_data}: we only include measurements obtained over redshift intervals narrower than 2 and we ensure that all references use the same faint-end limit $M_{\rm UV}<-17$ for calculating the UV luminosity density. To compute the UV luminosity density in {\sc colibre}, we calculate the contributions from galaxies of different luminosities separately in different simulation volumes (L100m5, L200m6 and L400m7). The boundaries that determine which volume is used for a given luminosity range are the same as those adopted in the construction of the combined UVLF (see \autoref{sec:combine}). For each luminosity interval, we sum the UV luminosities of galaxies in the corresponding simulation, divide by the volume to obtain the luminosity-density contribution and then sum over all intervals brighter than the chosen threshold ($M_{\rm UV} = -17$, following the practice in the above-mentioned references) to derive the global UV luminosity density. 

As can be seen in \autoref{fig:uvdensity}, the UV luminosity density decreases towards higher redshifts, by a factor of $\approx 700$ for the dust-attenuated case and by a factor of $\approx 1700$ for the dust-free case from $z=7$. Compared to observations, as expected, the dust-attenuated UV luminosity densities from \colibre are systematically lower at all redshifts, by $0.5-1$~dex, consistent with the lower galaxy abundances in \colibre relative to observations (see \autoref{fig:uvlf} and \autoref{fig:uvlf_dust}). The dust-free results, as expected, yield higher UV luminosity densities than the dust-attenuated ones. The dust-free \colibre UV luminosity density agrees reasonably well with the observational data up to $z=10$, but the observational measurements plotted here have not been corrected for dust attenuation. At higher redshifts, most observationally inferred UV luminosity densities lie above the \colibre predictions. The main exception is the measurement from \citet{Willott2024} at $z\approx 12$, which is broadly consistent with the dust-free \colibre result. In addition, the measurements from \citet{Franco2025} at $z\approx 13$ (and also that from \citealt{Harikane2023} at $z\approx 16$) have very large uncertainties, such that the dust-free \colibre result, or its extrapolation in the case of the $z\approx 16$ point, remains within their $1\sigma$ ranges.

The difference between dust-free and dust-attenuated luminosity densities decreases toward higher redshifts, indicating weaker dust attenuation at these earlier times, consistent with \autoref{fig:uvlf_dust}. The derived UV luminosity densities and their uncertainties (estimated using 1000 bootstrap iterations) are listed in \autoref{table:schechter}. We note, however, that the observational estimates of $\rho_{\rm UV}$ rely on fitting Schechter functions to the measured UVLFs and integrating over the fitted functions, which introduces additional uncertainty. For this reason, direct comparison of the UVLFs themselves provides a more direct discrimination between the simulations and the observations.

\begin{table*}
 \caption{The best-fitting Schechter parameters and the derived (dust-attenuated and dust-free) UV luminosity densities of the combined UVLFs of \colibre from $z = 7$ to $z = 15$. The fitting is performed over the range $-21 \leqslant M_{\rm UV} \leqslant -12$. The measurement uncertainties of the Schechter parameters and UV luminosity density are evaluated using the bootstrap method (1000 iterations). The UV luminosity density is calculated by summing the UV luminosities of all eligible galaxies in the \colibre simulations, rather than by integrating the Schechter-function fits to the UVLFs (see \autoref{sec:schechter} for more details).}\setlength{\tabcolsep}{3mm}
\begin{tabular}{cccccc}
\hline
\hline
 $z$ & $M_{\rm sch}$ & $\log\,\Phi_{\rm sch}$ & $\alpha_{\rm sch}$ & $\log\,\rho_{\rm UV} \mathrm{(with\,dust)}$ & $\log\,\rho_{\rm UV} \mathrm{(w/o\,dust)}$\\
 & $[\mathrm{mag}]$ & $[\mathrm{Mpc^{-3}}]$ & & $[\mathrm{erg\,s^{-1}\,Hz^{-1}\,Mpc^{-3}}]$ & $[\mathrm{erg\,s^{-1}\,Hz^{-1}\,Mpc^{-3}}]$\\
 
\hline

7 & $-19.00\pm 0.10$ & $-2.84\pm 0.10$ & $-1.76\pm 0.04$ & $25.46\pm 0.02$ & $25.98\pm 0.02$\\

\hline
9 & $-18.97\pm 0.11$ & $-3.37\pm 0.10$ & $-1.97\pm 0.04$ & $24.90\pm 0.03$ & $25.28\pm 0.02$\\

\hline
10 & $-19.23\pm 0.15$ & $-4.01\pm 0.12$ & $-2.21\pm 0.04$ & $24.57\pm 0.03$ & $24.86\pm 0.02$\\

\hline
12 & $-18.62\pm 0.16$ & $-4.28\pm 0.16$ & $-2.33\pm 0.04$ & $23.89\pm 0.05$ & $24.14\pm 0.04$\\

\hline
15 & $-17.71\pm 0.41$ & $-4.40\pm 0.45$ & $-2.31\pm 0.12$ & $22.60\pm 0.17$ & $22.75\pm 0.14$\\

\hline
\end{tabular}
\vspace{2mm}
\label{table:schechter}
\end{table*}

\section{Discussion}
\label{sec:discussion}

\subsection{Caveats for \colibre model and \skirt post-processing}
\label{sec:caveats}

As seen in \autoref{sec:result}, the dust-attenuated UVLFs in \colibre appear to be systematically lower than the observations, especially at the highest redshifts ($z=12$ and $15$), where even the dust-free UVLFs in \colibre remain lower than the observations (see \autoref{fig:uvlf_dust}). These differences may arise from physical processes that are not included in the current simulations as well as from numerical limitations. In the following, we discuss several caveats that may affect the predicted UVLFs.

First, \colibre does not include Population~III (Pop~III) stars. These (almost) metal-free stars are expected to form in the first galaxies and may contribute to the UV luminosity budget at very high redshift. Although Pop~III stars are expected to form primarily in the earliest minihaloes (e.g., \citealt{Abel2002}; see \citealt{Bromm2004} for a review) and therefore directly affect only the faintest galaxies, their feedback can indirectly influence the evolution of more massive systems. In particular, Pop~III stars can enrich the surrounding medium with metals and produce strong radiative and supernova feedback, which affects the transition to Population~II star formation and the early star formation histories of galaxies. As a result, their absence may influence the overall evolution of the galaxy population. For example, \citet{Ventura2024} made use of a semi-analytic model to study the impact of Pop~III stars on UVLFs and reported that Pop~III stars can have a significant effect on the total UVLF at $z = 12-16$. In particular, they found that Pop~III-dominated systems can contribute significantly to the bright end of the UVLF at these redshifts, especially for a top-heavy Pop~III IMF and may help explain the abundance of bright galaxies observed by {\it JWST} at $z \gtrsim 12$ without requiring an increased star formation efficiency.

Second, \colibre assumes a redshift-dependent, spatially uniform metagalactic ionizing and photodissociating background. This is likely to be a poor approximation during hydrogen reionization (which is completed at $z\approx 7$ in the UV-background model adopted by {\sc colibre}) when the radiation field is expected to be highly inhomogeneous. In particular, galaxies in overdense regions may experience a different ionizing background from those in more isolated environments, which could affect gas cooling, star formation histories and hence the predicted UV luminosities of high-redshift galaxies.

Third, \colibre assumes that the interstellar radiation field depends only on the local gas density \citep{Ploeckinger2025}. This means that dense gas clouds that have not yet formed stars are nevertheless exposed to an interstellar radiation field, which may suppress star formation and delay the formation of the first stars. This approximation could therefore suppress the early build-up of the galaxy population and lead to a moderate underestimate of the predicted UVLFs at the highest redshifts considered here. 

Fourth, the UV luminosities considered in this work do not include any contribution from AGN. While the UV emission of high-redshift galaxies is generally expected to be dominated by stellar light, AGN may contribute to the luminosities of the most massive systems (e.g., \citealt{Giallongo2015, Matsuoka2018, Harikane2022}), potentially affecting the brightest end of the UVLF. We note, however, that the highest-redshift AGN currently known is at $z\approx 10.6$ \citep{Maiolino2024a,Maiolino2024b} and so AGN are unlikely to explain the difference between modelled and observed UVLFs at $z\geqslant12$.

In addition, nebular continuum emission is not included in the {\sc toddlers} library used in {\sc skirt}. Nebular continuum can modestly enhance the UV luminosities of very young stellar populations (e.g., \citealt{Schaerer2003, Raiter2010}) and may therefore slightly modify the UVLF, particularly at the bright end and at very high redshift where star formation episodes are dominated by young stellar populations. A test of the impact of nebular continuum on the UVLF can be found in \citet{Durrant2026}, where nebular continuum is found to affect the UVLF at $z \geqslant 9$, with the effect reaching up to $\approx 0.5\,\mathrm{mag}$ at the brightest end of the UVLF at $z=15$. This is too small to reconcile the simulations with the observations in \autoref{fig:uvlf_dust}. The choices made in the {\sc colibre}-{\sc skirt} pipeline may also introduce additional uncertainties in the predicted UVLFs. We refer the reader to \citet{Gebek2026} and Paper I of this series \citep{Lu2026a} for tests exploring the impact of different pipeline choices.

Other caveats include:
\begin{enumerate}
\item The dust model in {\sc colibre}. Although the \colibre dust model has been shown to reproduce the $z=0$ galaxy dust mass function and dust scaling relations well (see \citealt{Trayford2026}), it has not yet been tested over the redshift range studied in this work ($7\leqslant z \leqslant 20$). Thus, the dust-attenuated results should be interpreted with caution. This is why we also present the dust-free results in \autoref{fig:uvlf_dust}. We discuss this point further in \autoref{sec:discussion_dust}.

\item Finite numerical resolution. The finite mass and spatial resolution of the simulations may influence both the formation of low-mass galaxies and the internal structure of massive galaxies. In particular, the star formation and feedback processes in galaxies may not be fully resolved, which could affect the predicted UV luminosities and therefore the UVLF. However, the convergence of \colibre simulations with resolution (see \autoref{fig:uvlf}) suggests that this may not be a significant shortcoming. 

\item Limited simulation volume. Since the simulations have a limited volume, the abundance of extremely rare and very luminous galaxies may not be fully sampled. This effect primarily impacts the brightest end of the UVLF, although the impact is expected to be modest given the relatively large volume of the simulations at m6 and m7 resolutions in which the brightest observed galaxies are well sampled.
\end{enumerate}

\subsection{Testing observational biases on the measured LFs}
\label{sec:test_on_obs_effect}

\begin{figure*}
\centering
\includegraphics[width=\textwidth]{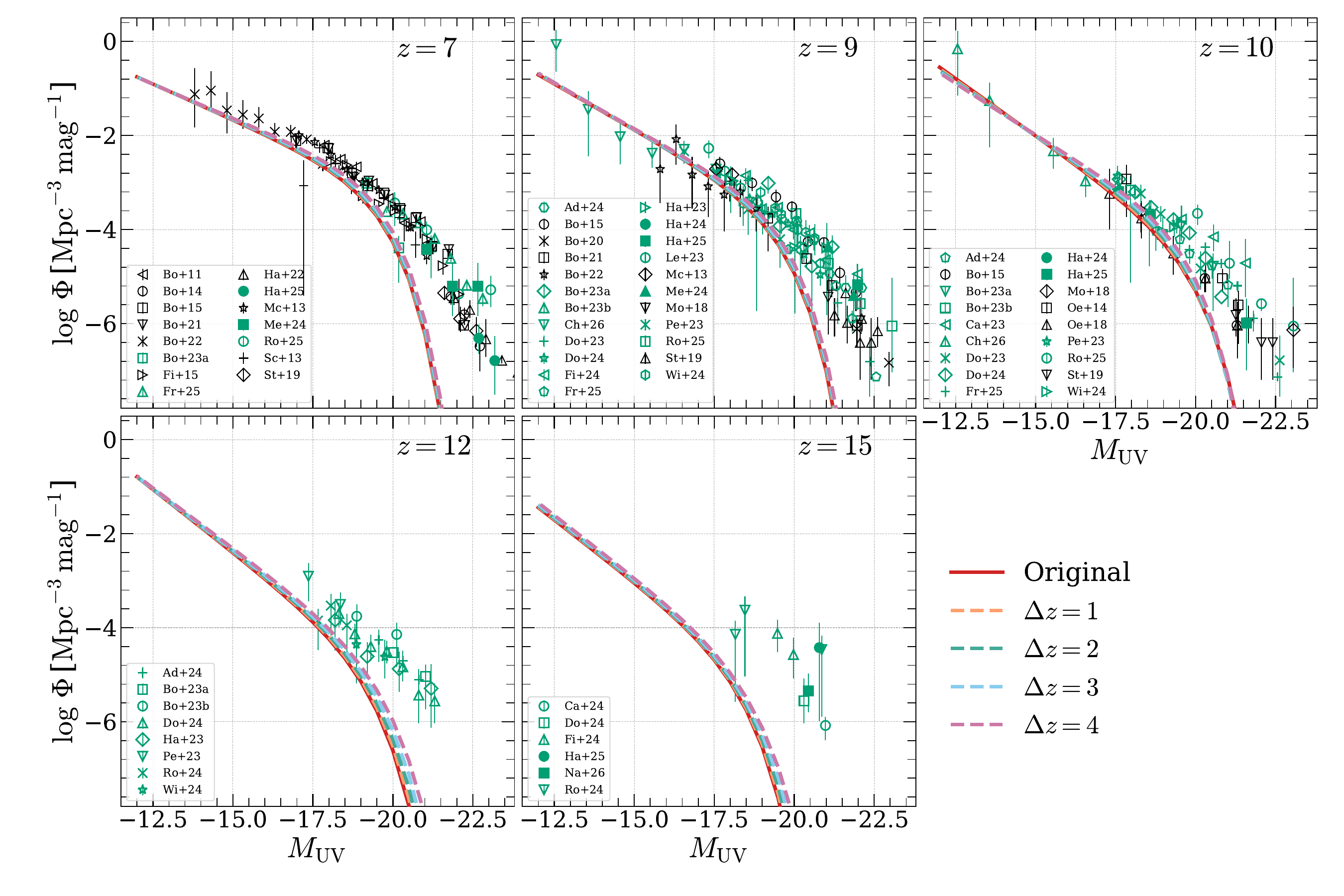}\vspace{-0.6cm}
\caption{The effect of redshift binning on the predicted UVLFs. In each panel, the Schechter fit to the original (dust-attenuated) UVLF from \colibre (without the redshift binning effect) is shown with a red solid curve, while those including the binning effect are shown with dashed curves of different colour, corresponding to different values of $\Delta z$ (i.e. the width of the redshift bin: $\Delta z=1,2,3,4$; see \autoref{sec:binning} for details). The symbols for the observational data are the same as in \autoref{fig:uvlf}.}
\label{fig:uvlf_binning}
\end{figure*}

\begin{figure*}
\centering
\includegraphics[width=\textwidth]{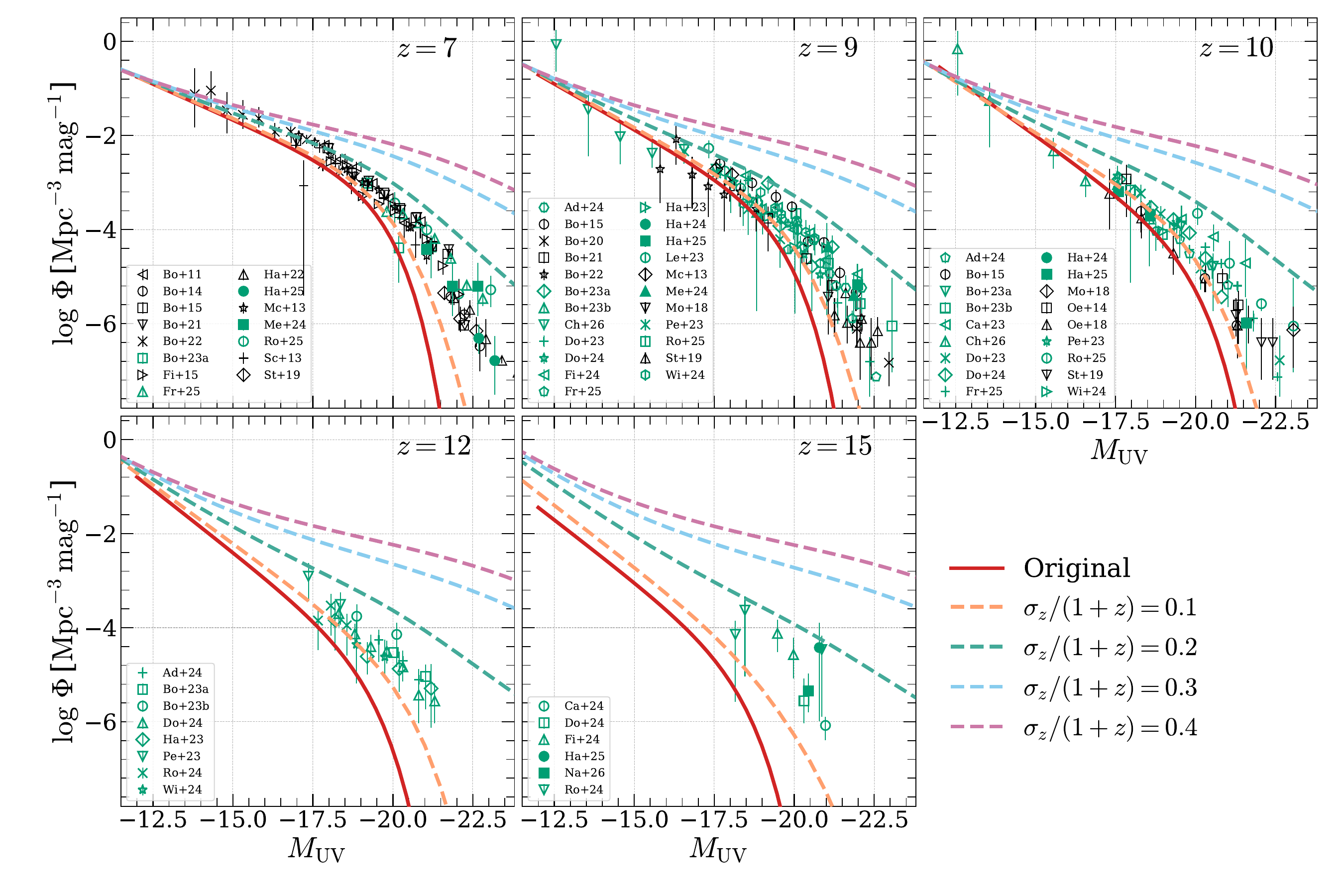}\vspace{-0.6cm}
\caption{The effect of photo-$z$ errors on the UVLFs. In each panel, the Schechter fit to the original (dust-attenuated) UVLF from \colibre (without the effect of photo-$z$ errors) is shown with a red solid curve, while those including the photo-$z$ error effect are shown with dashed curves of varying colour, corresponding to different values of $\sigma_z$, the measurement uncertainty of photo-$z$: $\sigma_z/(1+z)=0.1,\,0.2,\,0.3$ and $0.4$ (see \autoref{sec:phtoz} for details). We note that the larger values, $\sigma_z/(1+z)=0.2,\,0.3$ and $0.4$, are unlikely to represent realistic root-mean-square uncertainties and are included only to illustrate the potential effects of extreme photo-$z$ errors. Our treatment assumes symmetric photo-$z$ errors and does not include catastrophic failures. The symbols for the observational data are the same as in \autoref{fig:uvlf}.}
\label{fig:uvlf_photoz}
\end{figure*}

When comparing simulated LFs with observational data, several observational uncertainties and limitations should be kept in mind. Firstly, observational LFs are typically calculated over finite redshift bins, whereas we use simulation snapshots at a single redshift. If the LF evolves rapidly with redshift, this mismatch may lead to systematic differences, as the observed galaxy population is often biased toward the lower-redshift end of the bin \citep{Page2000}.

Secondly, photometric-redshift errors can introduce systematic differences in the inferred high-redshift UVLFs. Since the bright end of the UVLF is steep and evolves rapidly with redshift, scatter in the estimated photo-$z$s can cause more galaxies to be scattered to higher rather than lower redshifts, thereby causing their inferred absolute magnitudes to appear brighter than their true values and boosting the inferred number density at the bright end \citep{Subbarao1996,Sheth2007}. Such effects may be particularly relevant at high redshift, where photometric redshift uncertainties are significant \citep{Serjeant2023,Arrabal_Haro2023,Bouwens2023a,Fujimoto2023} and most LFs rely entirely on photo-$z$ estimates.

Thirdly, measurement uncertainties in luminosities can also introduce \citet{Eddington1913} bias in which galaxies are scattered due to measurement errors across a steeply declining LF. In this case, the larger number of intrinsically faint galaxies scattered into brighter bins than vice versa by magnitude errors artificially enhances the inferred number density at the bright end. 

Below, we investigate the impact of these observational effects on the UVLF by implementing them in the \colibre UVLFs. As these tests rely on the Schechter form of the \colibre UVLFs and aim to explain the difference between \colibre and observations, we only apply these tests to the \colibre UVLFs at $z\leq 15$.

\subsubsection{The effect of redshift binning}
\label{sec:binning}
In observations, UVLFs are typically estimated within finite redshift bins, rather than at exact redshifts as in simulations. Consequently, the observed UVLFs at a given redshift represent an average over a redshift range. If the number of galaxies decreases rapidly with increasing redshift, this averaging tends to overestimate the galaxy abundance compared to the true value at the centre of the redshift bin (\citealt{Page2000}). The magnitude of this difference depends on the width of the redshift bin, to which we refer here as the ``redshift binning effect''.

To investigate this effect, we recalculate the UVLF at a given redshift, $z$, as the average value from $z-\Delta z/2$ to $z+\Delta z/2$, using:
\begin{equation}
\label{eq:weighted_phi}
\Phi(M,z) = \frac{\sum \Phi(M,z_i) \times \frac{\mathrm{d}V}{\mathrm{d}z}(z_i) \times \delta z}{\sum \frac{\mathrm{d}V}{\mathrm{d}z}(z_i) \times \delta z},
\end{equation}
where $\Phi(M,z_i)$ is the galaxy abundance at absolute magnitude $M$ and redshift $z_i$; $\delta z$ is the thickness of the redshift slice (set to 0.1 here); and $\frac{\mathrm{d}V}{\mathrm{d}z}(z_i)$ is the comoving volume element per unit redshift at $z_i$, given by
\begin{equation}
\label{eq:dvdz}
\frac{\mathrm{d}V}{\mathrm{d}z}(z_i) = \frac{c}{H(z_i)} D_M^2(z_i) \, \Delta \Omega,
\end{equation}
where $c$ is the speed of light; $H(z_i)$ is the Hubble parameter at redshift $z_i$; $D_M(z_i)$ is the transverse comoving distance at redshift $z_i$; and $\Delta \Omega$ is the solid angle subtended by the survey area. The summation extends over the range $z-\Delta z/2$ to $z+\Delta z/2$. Since we compute the ``real'' UVLFs of {\sc colibre} only at a subset of redshifts (i.e., $z = 7, 9, 10, 12$ and 15), we use the Schechter functions fitted to the UVLFs to interpolate or extrapolate them to other redshifts.

In \autoref{fig:uvlf_binning}, we show the comparison between the original \colibre UVLFs as represented by the best-fitting Schechter functions (dust-attenuated with a projected aperture size of 50 kpc) and those including the redshift binning effect (for different bin widths: $\Delta z = 1, 2, 3,$ and $4$, indicated by dashed curves of different colour). As shown, larger $\Delta z$ (i.e., wider redshift bins) only results in slightly higher galaxy abundances and even with an extremely large redshift bin ($\Delta z = 4$), the \colibre UVLFs still fall short of the observations. As mentioned in \autoref{sec:obs_data}, the largest observational redshift bin size allowed for comparison in this work is $\Delta z = 2$. Thus, accounting for the binning effect alone cannot explain away more than a very small part of the discrepancy between simulations and observations.

We note that some of the observational studies used for comparison have already tried to correct for the redshift binning effect. For example, \citet{Donnan2024} use the full redshift probability distribution for each galaxy, combined with a UV magnitude-based prior, to distribute each galaxy's redshift probability across bins, thereby reducing the effect of redshift bin sizes in calculating the UVLF. Therefore, even after accounting for binning effects either in the simulations or in the corrected observational data, the discrepancy between the \colibre UVLFs and the observed abundances remains significant.

\subsubsection{The effect of photo-$z$ errors}
\label{sec:phtoz}

\begin{figure*}
\centering
\includegraphics[width=\textwidth]{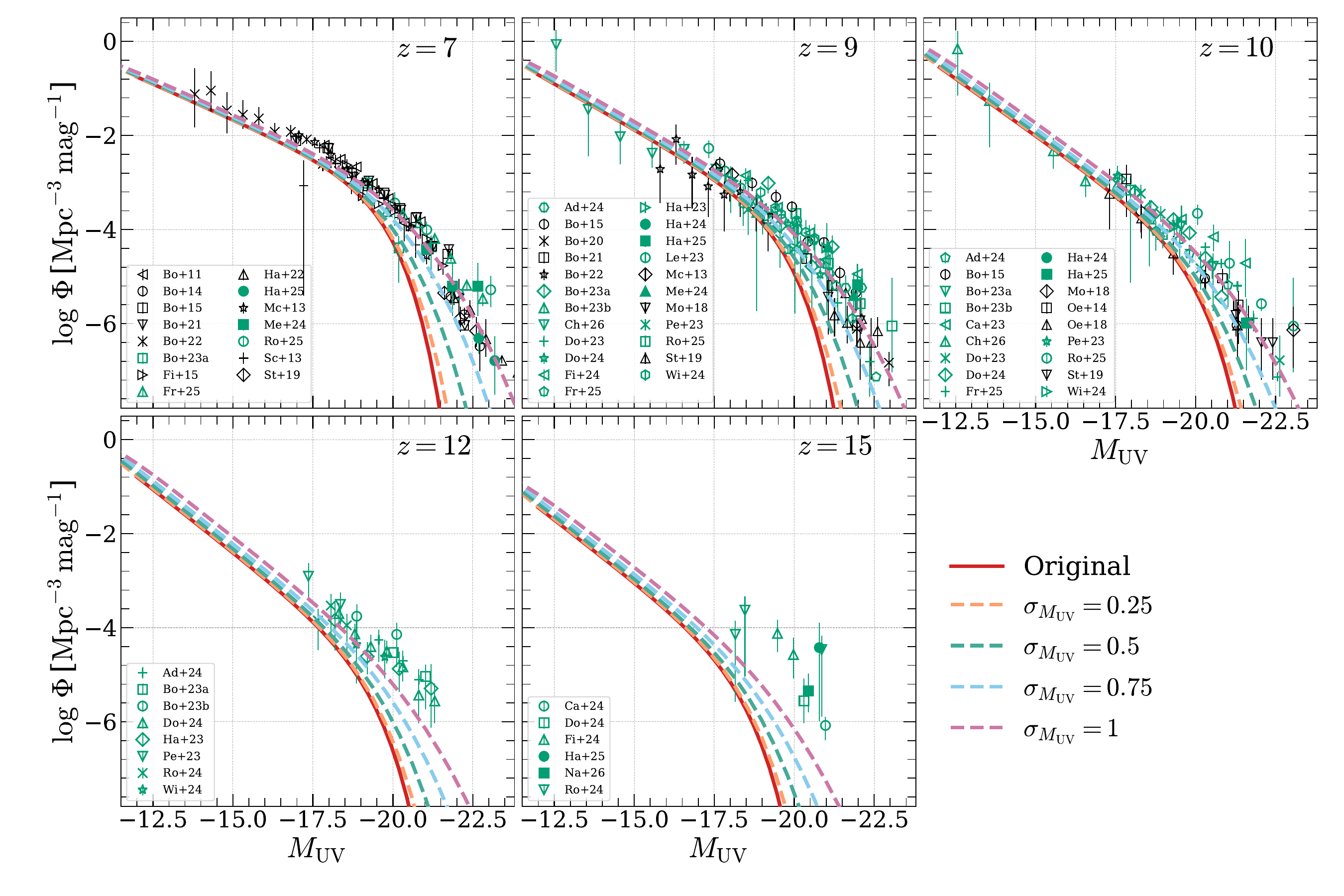}\vspace{-0.6cm}
\caption{Effect of Eddington bias on the UVLFs (dust-attenuated). In each panel, the red solid curve shows the Schechter fit to the original dust-attenuated UVLF from \colibre without Eddington bias (i.e., without measurement uncertainty in $M_{\rm UV}$), while the dashed curves show the results with Eddington bias (convolved with measurement uncertainties in $M_{\rm UV}$ of $\sigma_{M_{\rm UV}} = 0.25,\,0.5,\,0.75$ and $1$). The observational data points are the same as in \autoref{fig:uvlf}.}
\label{fig:uvlf_eddington}
\end{figure*}

\begin{figure*}
\centering
\includegraphics[width=\textwidth]{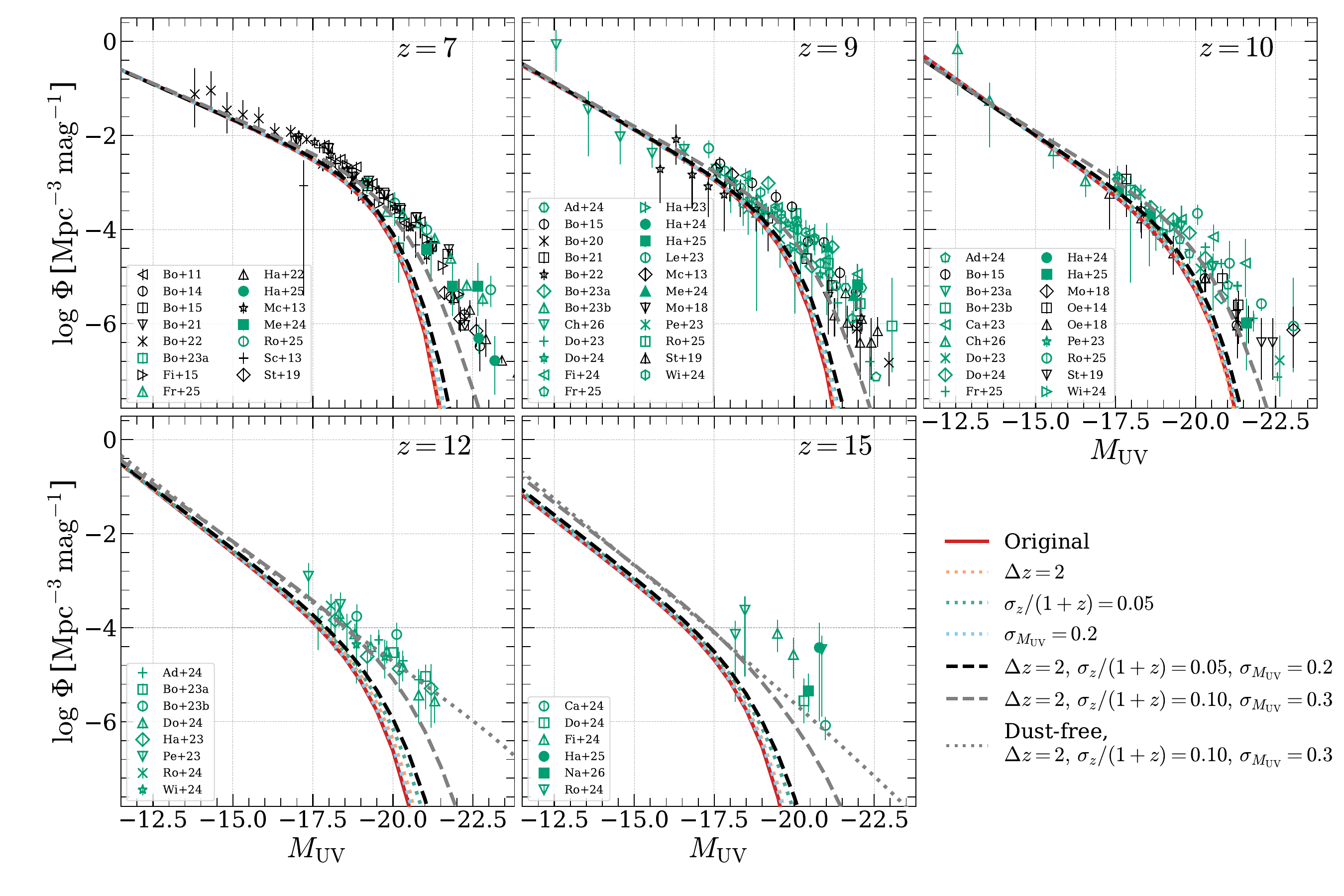}\vspace{-0.6cm}
\caption{The combined effect of observational uncertainties on the dust-attenuated UVLFs (redshift binning, photometric redshift errors and Eddington bias). In each panel, the original dust-attenuated UVLF from \colibre is shown by the red solid curve. The dotted curves in different colours show the UVLFs obtained when including a {\it single} observational effect (assuming median observational values or redshift bin size of 2). The black dashed curve shows the result including all three observational effects, assuming a redshift bin size of 2, a photometric redshift error of $\sigma_{z}/(1+z)=0.05$ (the median value in observations) and an uncertainty in $M_{\rm UV}$ of 0.2 (also the median observational value). The grey dashed curve shows the result when adopting the 84th percentiles of the photometric redshift and $M_{\rm UV}$ uncertainties ($\sigma_{z}/(1+z)=0.10$ and $\sigma_{M_{\rm UV}}=0.3$). The grey dotted curves in the $z=12$ and $z=15$ panels show the dust-free \colibre UVLFs combined with a redshift bin size of 2, a photo-$z$ uncertainty of $\sigma_{z}/(1+z)=0.10$ and an $M_{\rm UV}$ uncertainty of $\sigma_{M_{\rm UV}}=0.3$. The symbols for the observational data are the same as in \autoref{fig:uvlf}.}
\label{fig:uvlf_obs_combine}
\end{figure*}

For a galaxy with apparent UV magnitude, $m_{\rm UV}$, its inferred absolute UV magnitude will be brighter than the true value if it is deemed to be at a higher redshift than its actual redshift. Because (1) low-redshift galaxies are intrinsically more numerous than high-redshift galaxies and (2) a fixed $m_{\rm UV}$ corresponds to a fainter $M_{\rm UV}$ at lower redshift (where the galaxy abundance is also higher), symmetric measurement uncertainties in the photo-$z$ scatter more low-redshift galaxies into higher-redshift bins than the reverse. As a result, the number of galaxies in a given $M_{\rm UV}$ bin at high redshift is artificially increased, leading to an overestimation of the galaxy abundance relative to the true value.

Here we investigate this effect using the following steps, where we again use the Schechter fit to the \colibre UVLF at each snapshot redshift:
\begin{enumerate}
\item For a given redshift bin, [$z$, $z+\delta z$], where $\delta z$ is set to 0.1 here, we calculate the intrinsic number of galaxies within an absolute magnitude bin, [$M$, $M-\Delta M$], as:
\begin{equation}
\label{eq:n_int}
N_{\rm int}(z) = \Phi(M,z) \times \frac{\mathrm{d}V}{\mathrm{d}z}(z)\times \delta z \times \Delta M.
\end{equation}

\item Due to measurement uncertainties in photometric redshifts, the ``observed'' number of galaxies in the given redshift and magnitude bin is lower than the intrinsic number (ignoring, at this stage, contributions from other redshift bins), due to photo-$z$ errors scattering galaxies into other redshift bins. This can be written as:
\begin{equation}
N_{\rm obs}^{\prime} = N_{\rm int}\times P_{\rm stay}, 
\end{equation}
where $P_{\rm stay}$ is the probability that galaxies within [$z$, $z+\delta z$] have observed photo-$z$ values that remain in the same redshift bin.

\item Some galaxies in other redshift bins, [$z_i$, $z_i+\delta z$], may be wrongly inferred to lie within [$z$, $z+\delta z$] due to photo-$z$ errors, with the number given by:
\begin{equation}
\Delta N_{i}= N_{\rm int}(z_i)\times P_{\mathrm{scatter},i},
\end{equation}
where $N_{\rm int}(z_i)$ is the intrinsic number of galaxies in [$z_i$, $z_i+\delta z$], calculated using \autoref{eq:n_int}. Note that here we refer to galaxies with the same {\em apparent} UV magnitude, rather than the same absolute UV magnitude, since apparent magnitudes are what we observe. $P_{\mathrm{scatter},i}$ is the probability of galaxies in [$z_i$, $z_i+\delta z$] being scattered into [$z$, $z+\delta z$].

\item The final ``observed'' number of galaxies in [$z$, $z+\delta z$] within the absolute magnitude bin [$M$, $M-\Delta M$] is:
\begin{equation}
N_{\rm obs}^{\rm final} = N_{\rm obs}^{\prime} + \sum_{i} \Delta N_{i}.
\end{equation}
The summation extends over all redshift bins except the original bin, [$z$, $z+\delta z$]. The new ``observed'' galaxy abundance can then be calculated accordingly.

\item The intrinsic galaxy abundances used to compute $N_{\rm int}$ are inferred from our Schechter fits to the simulated UVLFs. We adopt the real (i.e. measured from \colibre galaxies) Schechter parameters of the UVLFs at $z = 7, 9, 10, 12,$ and $15$, while those at other redshifts are obtained by interpolation (and extrapolation where necessary, for $z<7$ and $z>15$).

\item The probabilities (i.e. the probability of galaxies staying in the same redshift bin, $P_{\rm stay}$ and the probabilities of galaxies being scattered in from other redshift bins, $P_{\mathrm{scatter},i}$) are derived under the assumption that the measured photo-$z$ distribution follows a Gaussian centred on the true redshift, with a fixed photo-$z$ error, ($\sigma_z/(1+z)$), that does not vary with redshift\footnote{This assumption is broadly consistent with the observations reported by \citet{Bouwens2023a,Bouwens2023b,Donnan2023,Finkelstein2023,Harikane2023,McLeod2024}.}. We ignore the effect of $k$-corrections in this simplified test.
\end{enumerate}

Larger $\sigma_z$ corresponds to a higher probability of galaxies being assigned an incorrect redshift, resulting in lower $P_{\rm stay}$ and higher $P_{\mathrm{scatter},i}$ (for a given [$z_i$, $z_i+\delta z$]). In \autoref{fig:uvlf_photoz}, we present the original dust-attenuated \colibre UVLFs (in Schechter form) and those including the photo-$z$ bias (for different $\sigma_z/(1+z)$, indicated by dashed curves of different colour), together with observational results. As can be seen, larger $\sigma_z/(1+z)$ leads to higher abundances at all $M_{\rm UV}$. At all redshifts, $z=7-15$, $\sigma_z/(1+z)\approx 0.1-0.2$ is sufficient to explain the difference between the original \colibre predictions and the observations. However, from a compilation of observational studies of galaxies at $z>6$ \citep{Bouwens2023a,Bouwens2023b,Donnan2023,Finkelstein2023,Harikane2023,McLeod2024}, we find that the median reported value of $\sigma_z/(1+z)$ is only 0.05, with an upper limit given by the 84th percentile of about 0.1. This is smaller than the value required to account for the discrepancy found here. We note, however, that the quoted photo-$z$ errors here are in fact uncertainties derived from the photometric-redshift estimation method, rather than true errors relative to the actual redshifts and may therefore be underestimated. In addition, catastrophic photo-$z$ errors, which may scatter low-redshift interlopers into high-redshift samples or move galaxies between widely separated redshift bins (e.g. \citealt{Bradac2026}), are not taken into account.

\citet{Robertson2024} (labelled Ro+24 in our figures and appearing only in the $z=12$ and $z=15$ panels) state that they correct for the effect of photo-$z$ uncertainties. Their estimates are still higher than our original UVLFs (red solid curves, without accounting for photo-$z$ errors). Furthermore, estimates based on spectroscopically confirmed galaxies (\citealt{Harikane2024} at $z=9$ and 10; \citealt{Harikane2025} at $z=7,9,10,$ and 15; \citealt{Naidu2026} at $z=15$), which are not affected by photo-$z$ uncertainties, are also higher than our original UVLFs. Together, these results indicate that photo-$z$ errors alone cannot explain the discrepancy between the \colibre predictions and the observations.

We note that the test presented here is intended only as a simplified experiment. In particular, extrapolating the Schechter fits obtained at $z = 7-15$ to substantially lower redshifts is inherently uncertain and should not be interpreted as a physically motivated prediction of the UVLF evolution. The shape and normalization of the UVLFs are expected to evolve non-linearly with cosmic time and this evolution may flatten towards lower redshift. As such, our extrapolation likely overestimates the number density of low-redshift galaxies that could be scattered into high-redshift bins due to large photometric redshift uncertainties. 

In addition, in this test we assume that the probability distribution of photometric redshift measurements is Gaussian and centred on the true redshift. We also ignore catastrophic photo-$z$ errors. In practice, however, photometric redshift probability distributions are often non-Gaussian and can be significantly asymmetric, with the upper uncertainty typically larger than the lower one (e.g., \citealt{Bouwens2023a,Bouwens2023b,Donnan2023,Finkelstein2023,Harikane2023,McLeod2024}) and real photo-$z$ errors may be substantially larger than assumed here. 

\subsubsection{The effects of Eddington bias}
\label{sec:eddington}

Due to the decreasing abundance of galaxies from the faint to the bright end of the LF, errors in measurements of magnitudes can cause the ``observed'' number of galaxies in bright bins to be overestimated: more intrinsically faint galaxies are scattered into brighter bins than intrinsically bright galaxies into fainter bins. This effect is known as the Eddington bias \citep{Eddington1913}.

To investigate the impact of Eddington bias, we assume that the measured $M_{\rm UV}$ follows a Gaussian distribution centred on the true value with a dispersion of $\sigma_{M_{\rm UV}}$. The \colibre UVLFs (in the Schechter form) are then convolved with a Gaussian kernel of width $\sigma_{M_{\rm UV}}$ to account for the observational uncertainty in $M_{\rm UV}$. 

In \autoref{fig:uvlf_eddington}, we compare the original dust-attenuated \colibre UVLFs (in the Schechter form, without Eddington bias) to those including Eddington bias (with $\sigma_{M_{\rm UV}}=0.25,\,0.5,\,0.75$ and 1), alongside observational data. As can be seen, accounting for Eddington bias indeed brings the \colibre predictions closer to observations, but even $\sigma_{M_{\rm UV}}=1$ cannot fully explain the discrepancy at $z=12$ and $z=15$. The estimated $M_{\rm UV}$ measurement uncertainties in observations are much smaller than this (with a median value of 0.2 mag and an 84th percentile value of 0.3 mag according to \citealt{Bouwens2023a,Bouwens2023b,Finkelstein2023,Harikane2023}) and thus cannot reconcile the \colibre predictions with the data.

\subsubsection{Combined observational effects}
As shown in the previous sections, all three observational effects bring the \colibre UVLF predictions closer to the observations. We now test the combined effect of these three observational biases. Since we consider only observational studies with a redshift range no larger than 2 (see \autoref{sec:obs_data}), we adopt a redshift bin size of 2. For the photo-$z$ and $M_{\rm UV}$ errors, we consider two sets of values, one corresponding to the median values reported in observational studies ($\sigma_z/(1+z)=0.05$ and $\sigma_{M_{\rm UV}}=0.2$; \citealt{Bouwens2023a,Bouwens2023b,Donnan2023,Finkelstein2023,Harikane2023,McLeod2024}) and another corresponding to their 84th percentiles ($\sigma_z/(1+z)=0.1$ and $\sigma_{M_{\rm UV}}=0.3$).

In \autoref{fig:uvlf_obs_combine}, we show (1) the original dust-attenuated \colibre UVLFs (in Schechter form), (2) the dust-attenuated \colibre UVLFs including a {\it single} observational effect (either the median values of the photo-$z$ or $M_{\rm UV}$ errors, or a redshift bin size of 2), (3) the dust-attenuated \colibre UVLFs including the combination of all observational effects with the median photo-$z$ and $M_{\rm UV}$ errors, (4) the dust-attenuated \colibre UVLFs including the combination of all observational effects with the 84th percentiles of the photo-$z$ and $M_{\rm UV}$ errors and (5) the dust-free \colibre UVLFs including the combination of all observational effects with the 84th percentiles of the photo-$z$ and $M_{\rm UV}$ errors (only for $z=12$ and $z=15$ panels). In cases (3), (4) and (5), a redshift bin size of 2 is adopted. 

As can be seen in the figure, combining all three uncertainties, assuming the median values of the photo-$z$ and $M_{\rm UV}$ errors, shifts the dust-attenuated UVLF horizontally by $\lesssim 0.5\,\rm mag$ at $\Phi = 10^{-6}\,\mathrm{Mpc^{-3}\,mag^{-1}}$. However, the UV luminosities of galaxies at this number density remain fainter than observed. The combination assuming the 84th-percentile values of the photo-$z$ and $M_{\rm UV}$ errors for the dust-attenuated \colibre UVLFs leads to rough agreement with observations at $z=9$ and 10, but still underestimates the galaxy luminosities at a given number density at $z=12$ and 15. Only when the effects of redshift binning and the 84th-percentile photo-$z$ and $M_{\rm UV}$ errors are combined with the dust-free assumption at $z=12$ and 15 (see the grey dotted curves in \autoref{fig:uvlf_obs_combine}) do the \colibre UVLFs come close to the observations at $z=12$ although they are still well below the data at $z=15$. Considering that some studies, such as \citet{Robertson2024} and \citet{Naidu2026}, may not be strongly affected by photo-$z$ errors (which make the largest contribution to the shift of the UVLFs), as mentioned earlier in \autoref{sec:phtoz}, additional physical mechanisms may still be required to boost the UV luminosities of high-redshift galaxies in order to explain the discrepancy between the \colibre UVLFs and the observations.

Throughout this section, we do not explicitly model the effects of cosmic variance. The \colibre UVLFs used at the bright end are measured from large simulation volumes, mainly L200m6 and L400m7, with volumes of $8\times10^{6}$ and $6.4\times10^{7}\,{\rm cMpc}^{3}$, respectively. These volumes are several hundred to several thousand times larger than the effective volumes of current high-redshift {\it JWST} surveys, which are typically of order $10^{3}-10^{5}\,{\rm cMpc}^{3}$ (e.g., \citealt{Castellano2023,Finkelstein2024,Chemerynska2026}). In addition, the \colibre simulations are periodic cosmological volumes whose mean density is fixed by the adopted cosmology, rather than randomly selected over or underdense survey fields (see Section 2.2 of \citealt{Schaye2026}). We therefore regard the \colibre UVLFs as being close to the cosmic mean predictions.

On the observational side, the constraints compiled in \autoref{table:obs_data} are drawn from multiple studies and from several independent or partially independent survey fields. The scatter among these measurements should therefore already reflect, at least partly, the effect of cosmic variance, together with differences in sample selection, photometric-redshift modelling and completeness corrections. In addition, some published studies (e.g., \citealt{Donnan2024,Chemerynska2026}) attempt to account for the effects of cosmic variance in their error bars. We therefore do not further account for the effect of cosmic variance in our comparison.

\subsection{Possible physical explanations}

\label{sec:solutions}
We have seen that observational uncertainties are not large enough to resolve the discrepancy between the \colibre UVLFs and observations. In this section, we discuss possible physical reasons that could explain the discrepancy.

\subsubsection{Impact of dust content}
\label{sec:discussion_dust}
Following the first wave of {\it JWST} papers claiming that the data were inconsistent with theoretical models within a $\Lambda$CDM framework \citep[e.g.][]{Finkelstein2023,Labbe2023}, several studies have tried to explain the difference between predicted and observed UVLFs without challenging the entire $\Lambda$CDM paradigm. An early explanation was reduced dust attenuation at high redshift. For example, \citet{Ferrara2023}, using an empirical model, argued that the measured UVLFs could be reproduced up to $z \approx 14$ if galaxies at $z \gtrsim 11$ contained negligible amounts of dust. More recently, \citet{Prada2026} also reported, using the empirical model {\sc UniverseMachine}, that the predicted dust-free UVLFs show good agreement with the observational data at $z=12$. Moreover, \citet{Vijayan2026} found that observational upper limits on dust masses at high redshift are only marginally consistent with the median dust mass-stellar mass relation predicted by {\sc colibre}, suggesting that dust grain growth in \colibre at $z\gtrsim 10$ may be more efficient than in the real Universe. Nevertheless, even in the limiting case of completely neglecting dust attenuation, our results in \autoref{fig:uvlf_dust} show that the observed abundance of UV-bright galaxies at $z \gtrsim 10$ cannot be reproduced. Consistent with this finding, \citet{Iocco2024} argued that, although dust attenuation may be minimal at the highest redshifts, an enhanced star formation efficiency is still required to reproduce the observed UVLFs and generate sufficiently luminous galaxies.

\cite{Lu2025} noted that the small age of the Universe at $z=14$ --only 300 million years-- limits the amount of dust that can form from metals by grain growth in the ISM. This assumption is supported by recent observational constraints indicating very low dust content in galaxies at $z \gtrsim 12$ \citep{Mitsuhashi2026}. Using the Durham semi-analytic model, {\sc galform}, they showed that a reduced dust-to-metal ratio (relative to the Milky Way value) at $z = 12$ and $z=14$ (due to the non-zero dust growth timescale) brings the model predictions closer to observations. Additional factors beyond dust attenuation must be considered. \citet{Lu2025} proposed that the missing factor is a top-heavy IMF in starbursts and they showed explicitly that the pre-JWST-launch \cite{Cowley2018} {\sc galform} predictions --which assume such an IMF-- together with a realistic model for dust growth at very high redshifts, provide a good match to the {\it JWST} data over the entire redshift range (see below for further details).

\subsubsection{Impact of the initial mass function}
\label{sec:IMF}
\citet{Chaikin2026b} compared the GSMFs in the \colibre simulations as a function of redshift with those inferred from observations. They found better agreement with the data than the level of agreement we find here between the predicted UVLFs and observations. This difference may reflect the fact that stellar masses are dominated by long-lived low-mass stars, whereas the UV luminosity is primarily produced by short-lived massive stars. A possible explanation of the difference is therefore that the simulations (with a universal \citealt{Chabrier2003} IMF) underestimate the relative abundance of massive stars that dominate the UV emission. Variations in the IMF that increase the fraction of massive stars would boost the UV luminosity without strongly altering the stellar mass function, making the top-heavy IMF advocated by \citet{Cowley2018} and others a plausible way to reconcile the predicted UVLFs with observations.

Observational studies have provided substantial evidence for the existence of a top-heavy IMF. For example, using C, N and O isotopic abundance ratios in starburst galaxies, \citet{Romano2017} reported compelling evidence for a top-heavy IMF, suggesting that intensely star-forming environments may preferentially produce a larger fraction of massive stars. \citet{Sneppen2022} fitted a temperature-dependent IMF to galaxies in the COSMOS2015 catalogue \citep{Laigle2016} and found that the IMF becomes increasingly top-heavy at higher redshifts. \citet{Cameron2024} reported two Lyman-$\alpha$ emitting galaxies at redshifts 5.9 and 7.9 that show signatures of exceptionally top-heavy IMFs. \citet{Liang2021} argued that the IMF tends to be more top-heavy at lower metallicities (see also \citealt{Yan2017}), further supporting the likelihood of a top-heavy IMF at high redshift.
 
A dual IMF - a solar neighbourhood-like form for quiescent/disc star formation and a top-heavy power law with slope $x=1$ (compared with $x=1.35$ for the high-mass end of a \citealt{Chabrier2003} IMF) in starbursts - has been adopted in the Durham semi-analytic model \textsc{galform} \citep{Cole2000,Lacey2016} since 2005 to match the observed number counts and redshift distribution of submillimetre galaxies (\citealt{Baugh2005}, who adopted an even more top-heavy IMF with $x=0$). \citet{Lu2025} compared the {\sc galform} predictions of \citet{Cowley2018} for high-$z$ galaxies, published before the launch of {\it JWST}, with the data and found that a combination of a top-heavy IMF and reduced dust content at high redshift yields good agreement with the observed UVLFs out to $z=14$. However, the effect of a top-heavy IMF on stellar feedback, including the increased number of supernovae expected for a top-heavy IMF, was not explicitly modelled and, although the effect of the IMF on metal enrichment was included, the model neglects other possible effects of the IMF on grain growth from metals. More recently, \citet{Fontanot2026} showed, using the semi-analytic model {\sc gaea} \citep{DeLucia2024}, that a variable IMF can improve agreement with observed UVLFs up to $z\approx 13$. However, although their increasingly top-heavy IMF affects the stellar populations, returned fraction, SNe fraction and chemical enrichment, it is not explicitly coupled to the stellar-feedback prescription. They also did not recalibrate the {\sc gaea} parameters after introducing the variable IMF.

\citet{Cueto2024}, using the {\sc Astraeus} framework that couples a semi-analytic galaxy evolution model with a semi-numerical reionisation scheme, argued that a top-heavy IMF alone is insufficient to account for the abundance of bright galaxies at high redshift. While such an IMF does enhance UV photon production per unit stellar mass, it also increases the strength of supernova feedback, which suppresses further star formation and limits the net gain in UV luminosity. They argued that, as a result, the UVLFs remain largely consistent with those predicted under a \citet{Salpeter1955} IMF. In contrast, \citet{Hutter2025}, using the same {\sc Astraeus} framework but adopting an IMF that becomes increasingly top-heavy in gas-rich, highly star-forming galaxies, found that such an evolving IMF can reproduce the observed UVLFs over $z=5-15$. They showed that the luminosity boost is largest in massive galaxies, where it is less efficiently offset by supernova feedback, thereby enhancing the abundance of UV-bright systems at $z>10$.

Thus, assessing the effect of a top-heavy IMF on the high-redshift UVLF requires a self-consistent model that includes its impact on supernova feedback, metal production, dust evolution and stellar luminosities. This has recently been investigated by \citet{Durrant2026}, who used \colibre to explore the effects of a variable IMF on the high-redshift UVLFs within a self-consistent forward-modelling framework. They find that a variable IMF that becomes increasingly top-heavy in high gas density regions goes a long way towards explaining the high-redshift UVLFs.

\subsubsection{Other possibilities}
\citet{Somerville2025} incorporated a density-modulated star-formation efficiency (SFE) into a semi-analytic galaxy formation model and showed that a higher effective SFE can reproduce or even exceed the observed number densities of UV-bright galaxies at $z\approx 6-17$. They further argued that this effect, together with decreasing dust attenuation and enhanced stochasticity in UV luminosities at earlier cosmic times, could help reproduce the observed evolution. However, the reduction in dust attenuation was implemented through a simplified post-processing prescription and the increased burstiness was likewise introduced in post-processing rather than being fully self-consistently predicted by the model. In addition, this updated model was not explicitly recalibrated to reproduce galaxy constraints at $z=0$ so its validity is difficult to assess.

Another possibility is that bursty star formation, or more generally, increased scatter in UV luminosity at fixed halo mass, could boost the bright end of the high-redshift UVLF. \citet{Mason2023}, using a simple empirical model based on halo formation timescales derived from extended Press-Schechter theory, explored how variations in halo assembly histories, together with variations in SFE and dust attenuation, could affect the bright end of the high-redshift UVLF. They showed that scatter in halo formation timescales can produce substantial scatter in the SFR at fixed halo mass, such that the youngest and most rapidly star-forming systems are scattered to brighter UV luminosities, thereby boosting the bright end of the UVLF. 

\citet{Sun2023} used the high-redshift version of the FIRE-2 cosmological zoom-in simulations \citep{Ma2018a,Ma2018b} to study the bright end of the UVLFs at $8\leqslant z\leqslant 12$. They showed that, with a multichannel implementation of standard stellar feedback processes (including, e.g., supernovae, stellar winds and radiation from young stars), bursty star formation histories arise self-consistently in high-redshift galaxies and can reproduce the observed UVLFs over this redshift range without invoking a top-heavy IMF, non-standard cosmology, or strongly enhanced star formation efficiency. Similarly, using a minimal empirical framework, \citet{Shen2023} argued that variability in galaxy UV luminosities at fixed halo mass, potentially driven by bursty star formation and dust clearance during strong feedback cycles, can significantly increase the abundance of UV-bright galaxies and thereby help reconcile model predictions with {\it JWST} observations at high redshift. However, their analysis was based on a highly simplified empirical model, in which the UV-luminosity variability was introduced phenomenologically rather than being predicted self-consistently from an explicit model of galaxy formation.

While {\em ad hoc} models such as those discussed in this subsection, which introduce physical mechanisms explicitly designed to match the high-redshift observations, are useful guides to the sort of mechanisms that could be important, they cannot be regarded as fully successful models of galaxy formation, since there is no guarantee that they would produce a realistic galaxy population at later times including at the present day.

\section{Conclusions}
\label{sec:conclusion}
This paper is the second in a series of three in which we aim to study comprehensively the multi-wavelength galaxy luminosity functions and their redshift evolution predicted by the new state-of-the-art hydrodynamical cosmological simulation set, {\sc colibre} \citep{Schaye2026,Chaikin2026a}, which resolves the cooling of the interstellar medium down to $10\,{\rm K}$, combined with the \skirt radiative transfer code \citep{Baes2011,Camps2015,Camps2020}. Specifically, in this paper we study the ultraviolet luminosity functions (UVLFs) at high redshifts, $z=7$ to $z=20$ and compare them with observations, especially those from {\it JWST}. Our key conclusions are:

\begin{enumerate}
\item \colibre simulations that span two orders of magnitude in numerical resolution show excellent convergence in the relations between stellar mass and other galaxy properties, such as total mass, mass of gas in the cool and dense phase, dust mass, SFRs, etc. down to very low stellar masses (corresponding to $2-5$ stellar particles; \autoref{fig:mstar2props}). 

\item We carried out \skirt\ radiative transfer calculations of both dust-free and dust-attenuated UV luminosities for a representative subsample of galaxies (see \autoref{fig:sampling_effect} for a test of the galaxy subsampling), in order to derive the UVLFs in three \colibre simulations of different volumes and numerical resolutions (L100m5, L200m6 and L400m7). The three \colibre simulations, when processed using {\sc skirt}, also show excellent convergence in the $M_{\rm UV}-M_{\ast}$ relation, both with and without dust attenuation (\autoref{fig:correlations}). The UVLFs derived from the different simulations show good convergence over their overlapping UV-luminosity ranges (\autoref{fig:uvlf}). To obtain UVLFs over the widest accessible luminosity range, we combine the results from all three simulations (see \autoref{sec:combine}).

\item The \colibre UVLFs show stronger redshift evolution than the observational data (e.g., \citealt{Harikane2023,Finkelstein2024}), with the abundance at fixed $M_{\rm UV}$ decreasing more rapidly towards higher redshift. The individual, as well as the combined, dust-attenuated UVLFs in the three \colibre simulations lie systematically below the observational determinations, particularly at the bright end and this discrepancy increases rapidly towards higher redshift (\autoref{fig:uvlf}). At $z=7$ and $z=9$, the dust-free UVLFs in \colibre have a higher amplitude than the observed UVLFs at the bright end; at $z=10$, they agree well with observations. At $z=12$, although the dust-free UVLF in \colibre still lies slightly below the observational data, the UV magnitude of galaxies at $\Phi = 10^{-6}\,\rm mag^{-1}\,Mpc^{-3}$ roughly matches the observed value. However, at $z=15$, even the dust-free \colibre galaxies are $\approx 2.5\,\rm mag$ fainter than observed at this same number density. At $z=20$, where only L100m5 contains a sufficient number of resolved galaxies, the dust-attenuated and dust-free UVLFs are nearly identical. In the absence of observational measurements satisfying our selection criteria, we present this result as a prediction for future observations (\autoref{fig:uvlf_dust}).

\item We fit Schechter functions to the combined, dust-attenuated UVLFs from {\sc colibre} and find that the Schechter parameters exhibit strong evolution with redshift: the characteristic magnitude, $M_{\rm sch}$, becomes fainter; the faint-end slope, $\alpha_{\rm sch}$, becomes more negative (i.e. steeper); and the characteristic number density, $\log\,\Phi_{\rm sch}$, decreases towards higher redshift (\autoref{fig:uvlf_schechter}).

\item The derived dust-attenuated and dust-free UV luminosity densities both decrease towards higher redshifts, by factors of $\approx 700$ and $\approx 1700$, respectively. At $z > 10$, even the dust-free result is systematically lower than most observational estimates by $\approx 0.5-1$~dex, with larger discrepancies at higher redshifts, consistent with the lower galaxy abundance across all redshifts (\autoref{fig:uvdensity}).

\item Accounting for observational uncertainties in the measured UVLFs, including those resulting from finite redshift bin sizes, measurement errors in photometric redshifts (assumed to be symmetric and ignoring catastrophic errors) and errors in $M_{\rm UV}$ (i.e. Eddington bias), brings the \colibre UVLFs closer to the observations. However, even the combination of all three effects does not fully resolve the discrepancy between the modelled and observed UVLFs when observationally reasonable values are assumed, indicating that additional physical mechanisms are required to boost the UV luminosities of high-redshift galaxies in the simulations (Figs.~\ref{fig:uvlf_binning} to \ref{fig:uvlf_obs_combine}).

There may be multiple mechanisms capable of boosting the UV luminosities of high-redshift galaxies, as discussed in \autoref{sec:solutions}. However, their overall impact on galaxy formation cannot be assessed in isolation, nor can it be robustly inferred from {\em ad hoc} modifications introduced solely to enhance the UV output, because such changes are also expected to affect other aspects of galaxy evolution. For example, a higher star formation efficiency (SFE) may lead not only to higher UV luminosities, but also to stronger stellar feedback, more rapid gas consumption, enhanced metal enrichment and potentially different dust production and attenuation histories. These coupled effects may, in turn, regulate subsequent star formation and partly offset the initial increase in UV emission (see also the discussion of the top-heavy IMF in \autoref{sec:IMF}). It is therefore necessary to test the effects of the proposed mechanisms within a self-consistent framework that simultaneously follows their impact on star formation, feedback, chemical enrichment, dust evolution and the emergent luminosities of galaxies at all times.
\end{enumerate}

\section*{Acknowledgements}
We thank the referee for their constructive comments, which improved the paper. We thank Prof. Cheng Li for helpful comments on Eddington bias and Dr. Fengwu Sun for comments on {\it JWST} aperture sizes. We acknowledge support from STFC (ST/X001075/1). CSF acknowledges support from the European Research Council through Advanced Investigator grant DMIDAS (GA 786910). SB is supported by a UK Research and Innovation Future Leaders Fellowship (grant numbers MR/V023381/1 and UKRI2044). AD is supported by an STFC doctoral studentship. NA acknowledges financial support by the Flemish Fund for Scientific Research (FWO-Vlaanderen) through the research grant G0C4723N. ABL acknowledges support by the Italian Ministry for Universities (MUR) program “Dipartimenti di Eccellenza 2023-2027” within the Centro Bicocca di Cosmologia Quantitativa (BiCoQ) and support by UNIMIB’s Fondo Di Ateneo Quota Competitiva (project 2024-ATEQC-0050). EC acknowledges support from the Netherlands Organization for Scientific Research (NWO) through research programme Athena 184.034.002 and STFC consolidated grant ST/X001075/1. FH acknowledges funding from the Netherlands Organization for Scientific Research (NWO) through research programme Athena 184.034.002. SP acknowledges support by the Austrian Science Fund (FWF) through grant-DOI: 10.55776/V982. JT acknowledges support from an STFC Early Stage Research and Development grant (ST/X004651/1). This work used the DiRAC@Durham facility managed by the Institute for Computational Cosmology on behalf of the STFC DiRAC HPC Facility (\url{www.dirac.ac.uk}). The equipment was funded by BEIS capital funding via STFC capital grants ST/K00042X/1, ST/P002293/1, ST/R002371/1 and ST/S002502/1, Durham University and STFC operations grant ST/R000832/1. DiRAC is part of the UK National e-Infrastructure.

\section*{Data availability}
The predicted UV luminosity functions from {\sc colibre}, together with the compiled observational UVLFs used for comparison in this work, are included as supplementary material and can be obtained from both the journal website and \url{https://icc.dur.ac.uk/data/}. More datasets in a machine-readable format are available from the DigitizeAstronomy website (\url{https://digastro.org}).

\bibliographystyle{mnras}
\bibliography{ref} 

@ARTICLE{Tremonti2004,
       author = {{Tremonti}, Christy A. and {Heckman}, Timothy M. and {Kauffmann}, Guinevere and {Brinchmann}, Jarle and {Charlot}, St{\'e}phane and {White}, Simon D.~M. and {Seibert}, Mark and {Peng}, Eric W. and {Schlegel}, David J. and {Uomoto}, Alan and {Fukugita}, Masataka and {Brinkmann}, Jon},
        title = "{The Origin of the Mass-Metallicity Relation: Insights from 53,000 Star-forming Galaxies in the Sloan Digital Sky Survey}",
      journal = {\apj},
         year = 2004,
        month = oct,
       volume = {613},
       number = {2},
        pages = {898-913},
          doi = {10.1086/423264},
archivePrefix = {arXiv},
       eprint = {astro-ph/0405537},
 primaryClass = {astro-ph},
       adsurl = {https://ui.adsabs.harvard.edu/abs/2004ApJ...613..898T}
}

@ARTICLE{McAlpine2016,
       author = {{McAlpine}, S. and {Helly}, J.~C. and {Schaller}, M. and {Trayford}, J.~W. and {Qu}, Y. and {Furlong}, M. and {Bower}, R.~G. and {Crain}, R.~A. and {Schaye}, J. and {Theuns}, T. and {Dalla Vecchia}, C. and {Frenk}, C.~S. and {McCarthy}, I.~G. and {Jenkins}, A. and {Rosas-Guevara}, Y. and {White}, S.~D.~M. and {Baes}, M. and {Camps}, P. and {Lemson}, G.},
        title = "{The EAGLE simulations of galaxy formation: Public release of halo and galaxy catalogues}",
      journal = {Astronomy and Computing},
         year = 2016,
        month = apr,
       volume = {15},
        pages = {72-89},
          doi = {10.1016/j.ascom.2016.02.004},
archivePrefix = {arXiv},
       eprint = {1510.01320},
 primaryClass = {astro-ph.GA},
       adsurl = {https://ui.adsabs.harvard.edu/abs/2016A&C....15...72M}
}

@ARTICLE{Vijayan2026,
       author = {{Vijayan}, Aswin P. and {Trayford}, James W. and {Schaye}, Joop and {Ploeckinger}, Sylvia and {Gebek}, Andrea and {Andreadis}, Nick and {Baes}, Maarten and {Ben{\'\i}tez-Llambay}, Alejandro and {Chaikin}, Evgenii and {Frenk}, Carlos S. and {Hu{\v{s}}ko}, Filip and {McGibbon}, Robert J. and {Richings}, Alexander J. and {Schaller}, Matthieu},
        title = "{The evolution of galaxy dust scaling relations in the COLIBRE simulations}",
      journal = {arXiv e-prints},
         year = 2026,
        month = jul,
          eid = {arXiv:2607.26058},
        pages = {arXiv:2607.26058},
archivePrefix = {arXiv},
       eprint = {2607.26058},
 primaryClass = {astro-ph.GA},
       adsurl = {https://ui.adsabs.harvard.edu/abs/2026arXiv260726058V}
}

@ARTICLE{Labbe2023,
       author = {{Labb{\'e}}, Ivo and {van Dokkum}, Pieter and {Nelson}, Erica and {Bezanson}, Rachel and {Suess}, Katherine A. and {Leja}, Joel and {Brammer}, Gabriel and {Whitaker}, Katherine and {Mathews}, Elijah and {Stefanon}, Mauro and {Wang}, Bingjie},
        title = "{A population of red candidate massive galaxies  600 Myr after the Big Bang}",
      journal = {\nat},
         year = 2023,
        month = apr,
       volume = {616},
       number = {7956},
        pages = {266-269},
          doi = {10.1038/s41586-023-05786-2},
archivePrefix = {arXiv},
       eprint = {2207.12446},
 primaryClass = {astro-ph.GA},
       adsurl = {https://ui.adsabs.harvard.edu/abs/2023Natur.616..266L}
}

@article{Oman2025, doi = {10.21105/joss.09278}, url = {https://doi.org/10.21105/joss.09278}, year = {2025}, publisher = {The Open Journal}, volume = {10}, number = {114}, pages = {9278}, author = {Oman, Kyle A.}, title = {SWIFTGalaxy: a Python package to work with particle groups from SWIFT simulations}, journal = {Journal of Open Source Software} }

@ARTICLE{Ploeckinger2025,
       author = {{Ploeckinger}, Sylvia and {Richings}, Alexander J. and {Schaye}, Joop and {Trayford}, James W. and {Schaller}, Matthieu and {Chaikin}, Evgenii},
        title = "{HYBRID-CHIMES: a model for radiative cooling and the abundances of ions and molecules in simulations of galaxy formation}",
      journal = {\mnras},
         year = 2025,
        month = oct,
       volume = {543},
       number = {2},
        pages = {891-916},
          doi = {10.1093/mnras/staf1402},
archivePrefix = {arXiv},
       eprint = {2506.15773},
 primaryClass = {astro-ph.GA},
       adsurl = {https://ui.adsabs.harvard.edu/abs/2025MNRAS.543..891P}
}

@ARTICLE{Richings2014a,
       author = {{Richings}, A.~J. and {Schaye}, J. and {Oppenheimer}, B.~D.},
        title = "{Non-equilibrium chemistry and cooling in the diffuse interstellar medium - I. Optically thin regime}",
      journal = {\mnras},
         year = 2014,
        month = jun,
       volume = {440},
       number = {4},
        pages = {3349-3369},
          doi = {10.1093/mnras/stu525},
archivePrefix = {arXiv},
       eprint = {1401.4719},
 primaryClass = {astro-ph.GA},
       adsurl = {https://ui.adsabs.harvard.edu/abs/2014MNRAS.440.3349R}
}

@ARTICLE{Richings2014b,
       author = {{Richings}, A.~J. and {Schaye}, J. and {Oppenheimer}, B.~D.},
        title = "{Non-equilibrium chemistry and cooling in the diffuse interstellar medium - II. Shielded gas}",
      journal = {\mnras},
         year = 2014,
        month = aug,
       volume = {442},
       number = {3},
        pages = {2780-2796},
          doi = {10.1093/mnras/stu1046},
archivePrefix = {arXiv},
       eprint = {1403.6155},
 primaryClass = {astro-ph.GA},
       adsurl = {https://ui.adsabs.harvard.edu/abs/2014MNRAS.442.2780R}
}

@ARTICLE{Schaye2026,
       author = {{Schaye}, Joop and {Chaikin}, Evgenii and {Schaller}, Matthieu and {Ploeckinger}, Sylvia and {Hu{\v{s}}ko}, Filip and {McGibbon}, Robert J. and {Trayford}, James W. and {Ben{\'\i}tez-Llambay}, Alejandro and {Correa}, Camila and {Frenk}, Carlos S. and {Richings}, Alexander J. and {Forouhar Moreno}, Victor J. and {Bah{\'e}}, Yannick M. and {Borrow}, Josh and {Durrant}, Anna and {Gebek}, Andrea and {Helly}, John C. and {Jenkins}, Adrian and {Lacey}, Cedric G. and {Ludlow}, Aaron and {Nobels}, Folkert S.~J.},
        title = "{The COLIBRE project: cosmological hydrodynamical simulations of galaxy formation and evolution}",
      journal = {\mnras},
         year = 2026,
        month = may,
       volume = {548},
       number = {1},
          eid = {stag375},
        pages = {stag375},
          doi = {10.1093/mnras/stag375},
archivePrefix = {arXiv},
       eprint = {2508.21126},
 primaryClass = {astro-ph.GA},
       adsurl = {https://ui.adsabs.harvard.edu/abs/2026MNRAS.548ag375S}
}

@ARTICLE{Husko2026,
       author = {{Hu{\v{s}}ko}, Filip and {Lacey}, Cedric G. and {Schaye}, Joop and {Schaller}, Matthieu and {Chaikin}, Evgenii and {Ploeckinger}, Sylvia and {Ben{\'\i}tez Llambay}, Alejandro and {Richings}, Alexander J. and {Trayford}, James W.},
        title = "{A hybrid active galactic nucleus feedback model with spinning black holes, winds and jets}",
      journal = {\mnras},
         year = 2026,
        month = apr,
       volume = {547},
       number = {2},
          eid = {stag324},
        pages = {stag324},
          doi = {10.1093/mnras/stag324},
archivePrefix = {arXiv},
       eprint = {2509.05179},
 primaryClass = {astro-ph.GA},
       adsurl = {https://ui.adsabs.harvard.edu/abs/2026MNRAS.547ag324H}
}

@ARTICLE{Durrant2026,
       author = {{Durrant}, Anna and {Crain}, Robert A. and {Lacey}, Cedric G. and {Schaye}, Joop and {Smit}, Renske and {Gebek}, Andrea and {Schaller}, Matthieu and {Lu}, Shengdong and {Chaikin}, Evgenii and {Andreadis}, Nick and {Baes}, Maarten and {Bate}, Matthew R. and {Ben{\'\i}tez-Llambay}, Alejandro and {Frenk}, Carlos S. and {Hu{\v{s}}ko}, Filip and {McGibbon}, Robert J. and {Ploeckinger}, Sylvia and {Richings}, Alexander J.},
        title = "{Cosmological simulations of the high-redshift galaxy population adopting a variable stellar initial mass function}",
      journal = {arXiv e-prints},
         year = 2026,
        month = jul,
          eid = {arXiv:2607.16404},
        pages = {arXiv:2607.16404},
          doi = {10.48550/arXiv.2607.16404},
archivePrefix = {arXiv},
       eprint = {2607.16404},
 primaryClass = {astro-ph.GA},
       adsurl = {https://ui.adsabs.harvard.edu/abs/2026arXiv260716404D}
}

@ARTICLE{Gebek2026,
       author = {{Gebek}, Andrea and {Baes}, Maarten and {Andreadis}, Nick and {Schaye}, Joop and {Kapoor}, Anand Utsav and {Bottrell}, Connor and {Lu}, Shengdong and {Lacey}, Cedric G. and {Ben{\'\i}tez-Llambay}, Alejandro and {Camps}, Peter and {Chaikin}, Evgenii and {Durrant}, Anna and {Frenk}, Carlos S. and {Hu{\v{s}}ko}, Filip and {Ploeckinger}, Sylvia and {Richings}, Alexander J. and {Schaller}, Matthieu and {Trayford}, James W. and {Vijayan}, Aswin P.},
        title = "{The COLIBRE-SKIRT pipeline: Calibration-free dust radiative transfer postprocessing for cosmological simulations}",
      journal = {arXiv e-prints},
         year = 2026,
        month = jul,
          eid = {arXiv:2607.14901},
        pages = {arXiv:2607.14901},
          doi = {10.48550/arXiv.2607.14901},
archivePrefix = {arXiv},
       eprint = {2607.14901},
 primaryClass = {astro-ph.GA},
       adsurl = {https://ui.adsabs.harvard.edu/abs/2026arXiv260714901G}
}

@ARTICLE{Lu2026a,
       author = {{Lu}, Shengdong and {Frenk}, Carlos S. and {Lacey}, Cedric G. and {Gebek}, Andrea and {Schaye}, Joop and {Cole}, Shaun and {Bose}, Sownak and {Andreadis}, Nick and {Baes}, Maarten and {Ben{\'\i}tez-Llambay}, Alejandro and {Chaikin}, Evgenii and {Crain}, Robert A. and {Durrant}, Anna and {Hu{\v{s}}ko}, Filip and {Ploeckinger}, Sylvia and {Richings}, Alexander J. and {Schaller}, Matthieu and {Trayford}, James W.},
        title = "{Galaxy luminosity functions from far-UV to submillimetre at $z=0$ in the COLIBRE simulations}",
      journal = {arXiv e-prints},
         year = 2026,
        month = may,
          eid = {arXiv:2605.02022},
        pages = {arXiv:2605.02022},
archivePrefix = {arXiv},
       eprint = {2605.02022},
 primaryClass = {astro-ph.GA},
       adsurl = {https://ui.adsabs.harvard.edu/abs/2026arXiv260502022L}
}

@ARTICLE{Benitez-Llambay2026,
       author = {{Ben{\'\i}tez-Llambay}, Alejandro and {Ploeckinger}, Sylvia and {Schaye}, Joop and {Richings}, Alexander J. and {Chaikin}, Evgenii and {Schaller}, Matthieu and {Trayford}, James W. and {Frenk}, Carlos S. and {Hu{\v{s}}ko}, Filip and {Correa}, Camila},
        title = "{Non-explosive pre-supernova feedback in the COLIBRE model of galaxy formation}",
      journal = {\mnras},
         year = 2026,
        month = mar,
       volume = {546},
       number = {4},
          eid = {stag268},
        pages = {stag268},
          doi = {10.1093/mnras/stag268},
archivePrefix = {arXiv},
       eprint = {2509.25309},
 primaryClass = {astro-ph.GA},
       adsurl = {https://ui.adsabs.harvard.edu/abs/2026MNRAS.546ag268B}
}

@ARTICLE{Correa2026,
       author = {{Correa}, Camila A. and {Schaye}, Joop and {Schaller}, Matthieu and {Trayford}, James W. and {Chaikin}, Evgenii and {Ben{\'\i}tez-Llambay}, Alejandro and {Frenk}, Carlos S. and {Ploeckinger}, Sylvia and {Richings}, Alexander J.},
        title = "{A subgrid model for chemical enrichment in cosmological simulations}",
      journal = {\mnras},
         year = 2026,
        month = may,
       volume = {548},
       number = {3},
          eid = {stag645},
        pages = {stag645},
          doi = {10.1093/mnras/stag645},
archivePrefix = {arXiv},
       eprint = {2604.00980},
 primaryClass = {astro-ph.GA},
       adsurl = {https://ui.adsabs.harvard.edu/abs/2026MNRAS.548ag645C}
}

@ARTICLE{Abel2002,
       author = {{Abel}, Tom and {Bryan}, Greg L. and {Norman}, Michael L.},
        title = "{The Formation of the First Star in the Universe}",
      journal = {Science},
         year = 2002,
        month = jan,
       volume = {295},
       number = {5552},
        pages = {93-98},
          doi = {10.1126/science.1063991},
archivePrefix = {arXiv},
       eprint = {astro-ph/0112088},
 primaryClass = {astro-ph},
       adsurl = {https://ui.adsabs.harvard.edu/abs/2002Sci...295...93A}
}

@ARTICLE{Chaikin2023,
       author = {{Chaikin}, Evgenii and {Schaye}, Joop and {Schaller}, Matthieu and {Ben{\'\i}tez-Llambay}, Alejandro and {Nobels}, Folkert S.~J. and {Ploeckinger}, Sylvia},
        title = "{A thermal-kinetic subgrid model for supernova feedback in simulations of galaxy formation}",
      journal = {\mnras},
         year = 2023,
        month = aug,
       volume = {523},
       number = {3},
        pages = {3709-3731},
          doi = {10.1093/mnras/stad1626},
archivePrefix = {arXiv},
       eprint = {2211.04619},
 primaryClass = {astro-ph.GA},
       adsurl = {https://ui.adsabs.harvard.edu/abs/2023MNRAS.523.3709C}
}

@ARTICLE{Chaikin2026a,
       author = {{Chaikin}, Evgenii and {Schaye}, Joop and {Schaller}, Matthieu and {Ploeckinger}, Sylvia and {Bah{\'e}}, Yannick M. and {Ben{\'\i}tez-Llambay}, Alejandro and {Correa}, Camila and {Forouhar Moreno}, Victor J. and {Frenk}, Carlos S. and {Hu{\v{s}}ko}, Filip and {Kugel}, Roi and {McGibbon}, Robert and {Richings}, Alexander J. and {Trayford}, James W. and {Borrow}, Josh and {Crain}, Robert A. and {Helly}, John C. and {Lacey}, Cedric G. and {Ludlow}, Aaron and {Nobels}, Folkert S.~J.},
        title = "{COLIBRE: calibrating subgrid feedback in cosmological simulations that include a cold gas phase}",
      journal = {\mnras},
         year = 2026,
        month = may,
       volume = {548},
       number = {1},
          eid = {stag300},
        pages = {stag300},
          doi = {10.1093/mnras/stag300},
archivePrefix = {arXiv},
       eprint = {2509.04067},
 primaryClass = {astro-ph.GA},
       adsurl = {https://ui.adsabs.harvard.edu/abs/2026MNRAS.548ag300C}
}

@ARTICLE{Chaikin2026b,
       author = {{Chaikin}, Evgenii and {Schaye}, Joop and {Schaller}, Matthieu and {Ploeckinger}, Sylvia and {Ben{\'\i}tez-Llambay}, Alejandro and {Frenk}, Carlos S. and {Hu{\v{s}}ko}, Filip and {McGibbon}, Robert J. and {Richings}, Alexander J. and {Trayford}, James W.},
        title = "{The evolution of the galaxy stellar mass function and star formation rates in the COLIBRE simulations from redshift 17 to 0}",
      journal = {\mnras},
         year = 2026,
        month = jun,
       volume = {548},
       number = {4},
          eid = {stag740},
        pages = {stag740},
          doi = {10.1093/mnras/stag740},
archivePrefix = {arXiv},
       eprint = {2509.07960},
 primaryClass = {astro-ph.GA},
       adsurl = {https://ui.adsabs.harvard.edu/abs/2026MNRAS.548ag740C}
}

@ARTICLE{Abbott2022,
       author = {{Abbott}, T.~M.~C. and {Aguena}, M. and {Alarcon}, A. and {Allam}, S. and {Alves}, O. and {Amon}, A. and {Andrade-Oliveira}, F. and {Annis}, J. and {Avila}, S. and {Bacon}, D. and {Baxter}, E. and {Bechtol}, K. and {Becker}, M.~R. and {Bernstein}, G.~M. and {Bhargava}, S. and {Birrer}, S. and {Blazek}, J. and {Brandao-Souza}, A. and {Bridle}, S.~L. and {Brooks}, D. and {Buckley-Geer}, E. and {Burke}, D.~L. and {Camacho}, H. and {Campos}, A. and {Carnero Rosell}, A. and {Carrasco Kind}, M. and {Carretero}, J. and {Castander}, F.~J. and {Cawthon}, R. and {Chang}, C. and {Chen}, A. and {Chen}, R. and {Choi}, A. and {Conselice}, C. and {Cordero}, J. and {Costanzi}, M. and {Crocce}, M. and {da Costa}, L.~N. and {da Silva Pereira}, M.~E. and {Davis}, C. and {Davis}, T.~M. and {De Vicente}, J. and {DeRose}, J. and {Desai}, S. and {Di Valentino}, E. and {Diehl}, H.~T. and {Dietrich}, J.~P. and {Dodelson}, S. and {Doel}, P. and {Doux}, C. and {Drlica-Wagner}, A. and {Eckert}, K. and {Eifler}, T.~F. and {Elsner}, F. and {Elvin-Poole}, J. and {Everett}, S. and {Evrard}, A.~E. and {Fang}, X. and {Farahi}, A. and {Fernandez}, E. and {Ferrero}, I. and {Fert{\'e}}, A. and {Fosalba}, P. and {Friedrich}, O. and {Frieman}, J. and {Garc{\'\i}a-Bellido}, J. and {Gatti}, M. and {Gaztanaga}, E. and {Gerdes}, D.~W. and {Giannantonio}, T. and {Giannini}, G. and {Gruen}, D. and {Gruendl}, R.~A. and {Gschwend}, J. and {Gutierrez}, G. and {Harrison}, I. and {Hartley}, W.~G. and {Herner}, K. and {Hinton}, S.~R. and {Hollowood}, D.~L. and {Honscheid}, K. and {Hoyle}, B. and {Huff}, E.~M. and {Huterer}, D. and {Jain}, B. and {James}, D.~J. and {Jarvis}, M. and {Jeffrey}, N. and {Jeltema}, T. and {Kovacs}, A. and {Krause}, E. and {Kron}, R. and {Kuehn}, K. and {Kuropatkin}, N. and {Lahav}, O. and {Leget}, P. -F. and {Lemos}, P. and {Liddle}, A.~R. and {Lidman}, C. and {Lima}, M. and {Lin}, H. and {MacCrann}, N. and {Maia}, M.~A.~G. and {Marshall}, J.~L. and {Martini}, P. and {McCullough}, J. and {Melchior}, P. and {Mena-Fern{\'a}ndez}, J. and {Menanteau}, F. and {Miquel}, R. and {Mohr}, J.~J. and {Morgan}, R. and {Muir}, J. and {Myles}, J. and {Nadathur}, S. and {Navarro-Alsina}, A. and {Nichol}, R.~C. and {Ogando}, R.~L.~C. and {Omori}, Y. and {Palmese}, A. and {Pandey}, S. and {Park}, Y. and {Paz-Chinch{\'o}n}, F. and {Petravick}, D. and {Pieres}, A. and {Plazas Malag{\'o}n}, A.~A. and {Porredon}, A. and {Prat}, J. and {Raveri}, M. and {Rodriguez-Monroy}, M. and {Rollins}, R.~P. and {Romer}, A.~K. and {Roodman}, A. and {Rosenfeld}, R. and {Ross}, A.~J. and {Rykoff}, E.~S. and {Samuroff}, S. and {S{\'a}nchez}, C. and {Sanchez}, E. and {Sanchez}, J. and {Sanchez Cid}, D. and {Scarpine}, V. and {Schubnell}, M. and {Scolnic}, D. and {Secco}, L.~F. and {Serrano}, S. and {Sevilla-Noarbe}, I. and {Sheldon}, E. and {Shin}, T. and {Smith}, M. and {Soares-Santos}, M. and {Suchyta}, E. and {Swanson}, M.~E.~C. and {Tabbutt}, M. and {Tarle}, G. and {Thomas}, D. and {To}, C. and {Troja}, A. and {Troxel}, M.~A. and {Tucker}, D.~L. and {Tutusaus}, I. and {Varga}, T.~N. and {Walker}, A.~R. and {Weaverdyck}, N. and {Wechsler}, R. and {Weller}, J. and {Yanny}, B. and {Yin}, B. and {Zhang}, Y. and {Zuntz}, J. and {DES Collaboration}},
        title = "{Dark Energy Survey Year 3 results: Cosmological constraints from galaxy clustering and weak lensing}",
      journal = {\prd},
         year = 2022,
        month = jan,
       volume = {105},
       number = {2},
          eid = {023520},
        pages = {023520},
          doi = {10.1103/PhysRevD.105.023520},
archivePrefix = {arXiv},
       eprint = {2105.13549},
 primaryClass = {astro-ph.CO},
       adsurl = {https://ui.adsabs.harvard.edu/abs/2022PhRvD.105b3520A}
}

@ARTICLE{Adams2024,
       author = {{Adams}, Nathan J. and {Conselice}, Christopher J. and {Austin}, Duncan and {Harvey}, Thomas and {Ferreira}, Leonardo and {Trussler}, James and {Juod{\v{z}}balis}, Ignas and {Li}, Qiong and {Windhorst}, Rogier and {Cohen}, Seth H. and {Jansen}, Rolf A. and {Summers}, Jake and {Tompkins}, Scott and {Driver}, Simon P. and {Robotham}, Aaron and {D'Silva}, Jordan C.~J. and {Yan}, Haojing and {Coe}, Dan and {Frye}, Brenda and {Grogin}, Norman A. and {Koekemoer}, Anton M. and {Marshall}, Madeline A. and {Pirzkal}, Nor and {Ryan}, Russell E. and {Maksym}, W. Peter and {Rutkowski}, Michael J. and {Willmer}, Christopher N.~A. and {Hammel}, Heidi B. and {Nonino}, Mario and {Bhatawdekar}, Rachana and {Wilkins}, Stephen M. and {Bradley}, Larry D. and {Broadhurst}, Tom and {Cheng}, Cheng and {Dole}, Herv{\'e} and {Hathi}, Nimish P. and {Zitrin}, Adi},
        title = "{EPOCHS. II. The Ultraviolet Luminosity Function from 7.5 < z < 13.5 Using 180 arcmin$^{2}$ of Deep, Blank Fields from the PEARLS Survey and Public JWST Data}",
      journal = {\apj},
         year = 2024,
        month = apr,
       volume = {965},
       number = {2},
          eid = {169},
        pages = {169},
          doi = {10.3847/1538-4357/ad2a7b},
archivePrefix = {arXiv},
       eprint = {2304.13721},
 primaryClass = {astro-ph.GA},
       adsurl = {https://ui.adsabs.harvard.edu/abs/2024ApJ...965..169A}
}

@ARTICLE{Arrabal_Haro2023,
       author = {{Arrabal Haro}, Pablo and {Dickinson}, Mark and {Finkelstein}, Steven L. and {Fujimoto}, Seiji and {Fern{\'a}ndez}, Vital and {Kartaltepe}, Jeyhan S. and {Jung}, Intae and {Cole}, Justin W. and {Burgarella}, Denis and {Chworowsky}, Katherine and {Hutchison}, Taylor A. and {Morales}, Alexa M. and {Papovich}, Casey and {Simons}, Raymond C. and {Amor{\'\i}n}, Ricardo O. and {Backhaus}, Bren E. and {Bagley}, Micaela B. and {Bisigello}, Laura and {Calabr{\`o}}, Antonello and {Castellano}, Marco and {Cleri}, Nikko J. and {Dav{\'e}}, Romeel and {Dekel}, Avishai and {Ferguson}, Henry C. and {Fontana}, Adriano and {Gawiser}, Eric and {Giavalisco}, Mauro and {Harish}, Santosh and {Hathi}, Nimish P. and {Hirschmann}, Michaela and {Holwerda}, Benne W. and {Huertas-Company}, Marc and {Koekemoer}, Anton M. and {Larson}, Rebecca L. and {Lucas}, Ray A. and {Mobasher}, Bahram and {P{\'e}rez-Gonz{\'a}lez}, Pablo G. and {Pirzkal}, Nor and {Rose}, Caitlin and {Santini}, Paola and {Trump}, Jonathan R. and {de la Vega}, Alexander and {Wang}, Xin and {Weiner}, Benjamin J. and {Wilkins}, Stephen M. and {Yang}, Guang and {Yung}, L.~Y. Aaron and {Zavala}, Jorge A.},
        title = "{Spectroscopic Confirmation of CEERS NIRCam-selected Galaxies at z ≃ 8-10}",
      journal = {\apjl},
         year = 2023,
        month = jul,
       volume = {951},
       number = {1},
          eid = {L22},
        pages = {L22},
          doi = {10.3847/2041-8213/acdd54},
archivePrefix = {arXiv},
       eprint = {2304.05378},
 primaryClass = {astro-ph.GA},
       adsurl = {https://ui.adsabs.harvard.edu/abs/2023ApJ...951L..22A}
}

@ARTICLE{Atek2023,
       author = {{Atek}, Hakim and {Shuntov}, Marko and {Furtak}, Lukas J. and {Richard}, Johan and {Kneib}, Jean-Paul and {Mahler}, Guillaume and {Zitrin}, Adi and {McCracken}, H.~J. and {Charlot}, St{\'e}phane and {Chevallard}, Jacopo and {Chemerynska}, Iryna},
        title = "{Revealing galaxy candidates out to z   16 with JWST observations of the lensing cluster SMACS0723}",
      journal = {\mnras},
         year = 2023,
        month = feb,
       volume = {519},
       number = {1},
        pages = {1201-1220},
          doi = {10.1093/mnras/stac3144},
archivePrefix = {arXiv},
       eprint = {2207.12338},
 primaryClass = {astro-ph.GA},
       adsurl = {https://ui.adsabs.harvard.edu/abs/2023MNRAS.519.1201A}
}

@ARTICLE{Baes2011,
       author = {{Baes}, Maarten and {Verstappen}, Joris and {De Looze}, Ilse and {Fritz}, Jacopo and {Saftly}, Waad and {Vidal P{\'e}rez}, Edgardo and {Stalevski}, Marko and {Valcke}, Sander},
        title = "{Efficient Three-dimensional NLTE Dust Radiative Transfer with SKIRT}",
      journal = {\apjs},
         year = 2011,
        month = oct,
       volume = {196},
       number = {2},
          eid = {22},
        pages = {22},
          doi = {10.1088/0067-0049/196/2/22},
archivePrefix = {arXiv},
       eprint = {1108.5056},
 primaryClass = {astro-ph.CO},
       adsurl = {https://ui.adsabs.harvard.edu/abs/2011ApJS..196...22B}
}

@ARTICLE{Baes2015,
       author = {{Baes}, M. and {Camps}, P.},
        title = "{SKIRT: The design of a suite of input models for Monte Carlo radiative transfer simulations}",
      journal = {Astronomy and Computing},
         year = 2015,
        month = sep,
       volume = {12},
        pages = {33-44},
          doi = {10.1016/j.ascom.2015.05.006},
archivePrefix = {arXiv},
       eprint = {1505.07708},
 primaryClass = {astro-ph.IM},
       adsurl = {https://ui.adsabs.harvard.edu/abs/2015A&C....12...33B}
}

@ARTICLE{Barrufet2023,
       author = {{Barrufet}, L. and {Oesch}, P.~A. and {Bouwens}, R. and {Inami}, H. and {Sommovigo}, L. and {Algera}, H. and {da Cunha}, E. and {Aravena}, M. and {Dayal}, P. and {Ferrara}, A. and {Fudamoto}, Y. and {Gonzalez}, V. and {Graziani}, L. and {Hygate}, A.~P.~S. and {de Looze}, I. and {Nanayakkara}, T. and {Pallottini}, A. and {Schneider}, R. and {Stefanon}, M. and {Topping}, M. and {van der Werf}, P.},
        title = "{The ALMA REBELS Survey: the first infrared luminosity function measurement at z {\ensuremath{\sim}} 7}",
      journal = {\mnras},
         year = 2023,
        month = jul,
       volume = {522},
       number = {3},
        pages = {3926-3934},
          doi = {10.1093/mnras/stad1259},
archivePrefix = {arXiv},
       eprint = {2303.11321},
 primaryClass = {astro-ph.GA},
       adsurl = {https://ui.adsabs.harvard.edu/abs/2023MNRAS.522.3926B}
}

@article{Bradac2026,
  title={Two Exciting High-redshift Galaxy Candidates Turn Out to Be Two Exciting Ultra-cool Brown Dwarfs},
  author={Brada{\v{c}}, Maru{\v{s}}a and Willott, Chris and Asada, Yoshihisa and Albert, Lo{\"\i}c and Rihtar{\v{s}}i{\v{c}}, Gregor and Harshan, Anishya and Jude{\v{z}}, Jon and Martis, Nicholas S and Ferrara, Andrea and Allingham, Joseph FV and others},
  journal={arXiv preprint arXiv:2604.23668},
  year={2026}
}

@ARTICLE{Baugh2005,
       author = {{Baugh}, C.~M. and {Lacey}, C.~G. and {Frenk}, C.~S. and {Granato}, G.~L. and {Silva}, L. and {Bressan}, A. and {Benson}, A.~J. and {Cole}, S.},
        title = "{Can the faint submillimetre galaxies be explained in the {\ensuremath{\Lambda}} cold dark matter model?}",
      journal = {\mnras},
         year = 2005,
        month = jan,
       volume = {356},
       number = {3},
        pages = {1191-1200},
          doi = {10.1111/j.1365-2966.2004.08553.x},
archivePrefix = {arXiv},
       eprint = {astro-ph/0406069},
 primaryClass = {astro-ph},
       adsurl = {https://ui.adsabs.harvard.edu/abs/2005MNRAS.356.1191B}
}

@article{Borrow2020,
  doi = {10.21105/joss.02430},
  url = {https://doi.org/10.21105/joss.02430},
  year = {2020},
  publisher = {The Open Journal},
  volume = {5},
  number = {52},
  pages = {2430},
  author = {Josh Borrow and Alexei Borrisov},
  title = {swiftsimio: A Python library for reading SWIFT data},
  journal = {Journal of Open Source Software}
}

@ARTICLE{Borrow2021,
       author = {{Borrow}, Josh and {Kelly}, Ashley J.},
        title = "{Projecting SPH Particles in Adaptive Environments}",
      journal = {arXiv e-prints},
         year = 2021,
        month = jun,
          eid = {arXiv:2106.05281},
        pages = {arXiv:2106.05281},
          doi = {10.48550/arXiv.2106.05281},
archivePrefix = {arXiv},
       eprint = {2106.05281},
 primaryClass = {astro-ph.GA},
       adsurl = {https://ui.adsabs.harvard.edu/abs/2021arXiv210605281B}
}

@ARTICLE{Borrow2022,
       author = {{Borrow}, Josh and {Schaller}, Matthieu and {Bower}, Richard G. and {Schaye}, Joop},
        title = "{SPHENIX: smoothed particle hydrodynamics for the next generation of galaxy formation simulations}",
      journal = {\mnras},
         year = 2022,
        month = apr,
       volume = {511},
       number = {2},
        pages = {2367-2389},
          doi = {10.1093/mnras/stab3166},
archivePrefix = {arXiv},
       eprint = {2012.03974},
 primaryClass = {astro-ph.GA},
       adsurl = {https://ui.adsabs.harvard.edu/abs/2022MNRAS.511.2367B}
}

@ARTICLE{Bouwens2011,
       author = {{Bouwens}, R.~J. and {Illingworth}, G.~D. and {Oesch}, P.~A. and {Labb{\'e}}, I. and {Trenti}, M. and {van Dokkum}, P. and {Franx}, M. and {Stiavelli}, M. and {Carollo}, C.~M. and {Magee}, D. and {Gonzalez}, V.},
        title = "{Ultraviolet Luminosity Functions from 132 z \raisebox{-0.5ex}\textasciitilde 7 and z \raisebox{-0.5ex}\textasciitilde 8 Lyman-break Galaxies in the Ultra-deep HUDF09 and Wide-area Early Release Science WFC3/IR Observations}",
      journal = {\apj},
         year = 2011,
        month = aug,
       volume = {737},
       number = {2},
          eid = {90},
        pages = {90},
          doi = {10.1088/0004-637X/737/2/90},
archivePrefix = {arXiv},
       eprint = {1006.4360},
 primaryClass = {astro-ph.CO},
       adsurl = {https://ui.adsabs.harvard.edu/abs/2011ApJ...737...90B}
}

@ARTICLE{Bouwens2015,
       author = {{Bouwens}, R.~J. and {Illingworth}, G.~D. and {Oesch}, P.~A. and {Trenti}, M. and {Labb{\'e}}, I. and {Bradley}, L. and {Carollo}, M. and {van Dokkum}, P.~G. and {Gonzalez}, V. and {Holwerda}, B. and {Franx}, M. and {Spitler}, L. and {Smit}, R. and {Magee}, D.},
        title = "{UV Luminosity Functions at Redshifts z {\ensuremath{\sim}} 4 to z {\ensuremath{\sim}} 10: 10,000 Galaxies from HST Legacy Fields}",
      journal = {\apj},
         year = 2015,
        month = apr,
       volume = {803},
       number = {1},
          eid = {34},
        pages = {34},
          doi = {10.1088/0004-637X/803/1/34},
archivePrefix = {arXiv},
       eprint = {1403.4295},
 primaryClass = {astro-ph.CO},
       adsurl = {https://ui.adsabs.harvard.edu/abs/2015ApJ...803...34B}
}

@ARTICLE{Bouwens2021,
       author = {{Bouwens}, R.~J. and {Oesch}, P.~A. and {Stefanon}, M. and {Illingworth}, G. and {Labb{\'e}}, I. and {Reddy}, N. and {Atek}, H. and {Montes}, M. and {Naidu}, R. and {Nanayakkara}, T. and {Nelson}, E. and {Wilkins}, S.},
        title = "{New Determinations of the UV Luminosity Functions from z   9 to 2 Show a Remarkable Consistency with Halo Growth and a Constant Star Formation Efficiency}",
      journal = {\aj},
         year = 2021,
        month = aug,
       volume = {162},
       number = {2},
          eid = {47},
        pages = {47},
          doi = {10.3847/1538-3881/abf83e},
archivePrefix = {arXiv},
       eprint = {2102.07775},
 primaryClass = {astro-ph.GA},
       adsurl = {https://ui.adsabs.harvard.edu/abs/2021AJ....162...47B}
}

@ARTICLE{Bouwens2022,
       author = {{Bouwens}, R.~J. and {Illingworth}, G. and {Ellis}, R.~S. and {Oesch}, P. and {Stefanon}, M.},
        title = "{z   2-9 Galaxies Magnified by the Hubble Frontier Field Clusters. II. Luminosity Functions and Constraints on a Faint-end Turnover}",
      journal = {\apj},
         year = 2022,
        month = nov,
       volume = {940},
       number = {1},
          eid = {55},
        pages = {55},
          doi = {10.3847/1538-4357/ac86d1},
archivePrefix = {arXiv},
       eprint = {2205.11526},
 primaryClass = {astro-ph.GA},
       adsurl = {https://ui.adsabs.harvard.edu/abs/2022ApJ...940...55B}
}

@ARTICLE{Bouwens2023a,
       author = {{Bouwens}, Rychard and {Illingworth}, Garth and {Oesch}, Pascal and {Stefanon}, Mauro and {Naidu}, Rohan and {van Leeuwen}, Ivana and {Magee}, Dan},
        title = "{UV luminosity density results at z > 8 from the first JWST/NIRCam fields: limitations of early data sets and the need for spectroscopy}",
      journal = {\mnras},
         year = 2023,
        month = jul,
       volume = {523},
       number = {1},
        pages = {1009-1035},
          doi = {10.1093/mnras/stad1014},
archivePrefix = {arXiv},
       eprint = {2212.06683},
 primaryClass = {astro-ph.CO},
       adsurl = {https://ui.adsabs.harvard.edu/abs/2023MNRAS.523.1009B}
}

@ARTICLE{Bouwens2023b,
       author = {{Bouwens}, Rychard J. and {Stefanon}, Mauro and {Brammer}, Gabriel and {Oesch}, Pascal A. and {Herard-Demanche}, Thomas and {Illingworth}, Garth D. and {Matthee}, Jorryt and {Naidu}, Rohan P. and {van Dokkum}, Pieter G. and {van Leeuwen}, Ivana F.},
        title = "{Evolution of the UV LF from z   15 to z   8 using new JWST NIRCam medium-band observations over the HUDF/XDF}",
      journal = {\mnras},
         year = 2023,
        month = jul,
       volume = {523},
       number = {1},
        pages = {1036-1055},
          doi = {10.1093/mnras/stad1145},
archivePrefix = {arXiv},
       eprint = {2211.02607},
 primaryClass = {astro-ph.GA},
       adsurl = {https://ui.adsabs.harvard.edu/abs/2023MNRAS.523.1036B}
}

@ARTICLE{Bowler2014,
       author = {{Bowler}, R.~A.~A. and {Dunlop}, J.~S. and {McLure}, R.~J. and {Rogers}, A.~B. and {McCracken}, H.~J. and {Milvang-Jensen}, B. and {Furusawa}, H. and {Fynbo}, J.~P.~U. and {Taniguchi}, Y. and {Afonso}, J. and {Bremer}, M.~N. and {Le F{\`e}vre}, O.},
        title = "{The bright end of the galaxy luminosity function at z≃7: before the onset of mass quenching?}",
      journal = {\mnras},
         year = 2014,
        month = may,
       volume = {440},
       number = {3},
        pages = {2810-2842},
          doi = {10.1093/mnras/stu449},
archivePrefix = {arXiv},
       eprint = {1312.5643},
 primaryClass = {astro-ph.CO},
       adsurl = {https://ui.adsabs.harvard.edu/abs/2014MNRAS.440.2810B}
}

@ARTICLE{Bowler2020,
       author = {{Bowler}, R.~A.~A. and {Jarvis}, M.~J. and {Dunlop}, J.~S. and {McLure}, R.~J. and {McLeod}, D.~J. and {Adams}, N.~J. and {Milvang-Jensen}, B. and {McCracken}, H.~J.},
        title = "{A lack of evolution in the very bright end of the galaxy luminosity function from z ≃ 8 to 10}",
      journal = {\mnras},
         year = 2020,
        month = apr,
       volume = {493},
       number = {2},
        pages = {2059-2084},
          doi = {10.1093/mnras/staa313},
archivePrefix = {arXiv},
       eprint = {1911.12832},
 primaryClass = {astro-ph.GA},
       adsurl = {https://ui.adsabs.harvard.edu/abs/2020MNRAS.493.2059B}
}

@ARTICLE{Bromm2004,
       author = {{Bromm}, Volker and {Larson}, Richard B.},
        title = "{The First Stars}",
      journal = {\araa},
         year = 2004,
        month = sep,
       volume = {42},
       number = {1},
        pages = {79-118},
          doi = {10.1146/annurev.astro.42.053102.134034},
archivePrefix = {arXiv},
       eprint = {astro-ph/0311019},
 primaryClass = {astro-ph},
       adsurl = {https://ui.adsabs.harvard.edu/abs/2004ARA&A..42...79B}
}

@ARTICLE{Camps2013,
       author = {{Camps}, P. and {Baes}, M. and {Saftly}, W.},
        title = "{Using 3D Voronoi grids in radiative transfer simulations}",
      journal = {\aap},
         year = 2013,
        month = dec,
       volume = {560},
          eid = {A35},
        pages = {A35},
          doi = {10.1051/0004-6361/201322281},
archivePrefix = {arXiv},
       eprint = {1310.1854},
 primaryClass = {astro-ph.IM},
       adsurl = {https://ui.adsabs.harvard.edu/abs/2013A&A...560A..35C}
}

@ARTICLE{Camps2015,
       author = {{Camps}, P. and {Baes}, M.},
        title = "{SKIRT: An advanced dust radiative transfer code with a user-friendly architecture}",
      journal = {Astronomy and Computing},
         year = 2015,
        month = mar,
       volume = {9},
        pages = {20-33},
          doi = {10.1016/j.ascom.2014.10.004},
archivePrefix = {arXiv},
       eprint = {1410.1629},
 primaryClass = {astro-ph.IM},
       adsurl = {https://ui.adsabs.harvard.edu/abs/2015A&C.....9...20C}
}

@ARTICLE{Carnall2019,
       author = {{Carnall}, Adam C. and {Leja}, Joel and {Johnson}, Benjamin D. and {McLure}, Ross J. and {Dunlop}, James S. and {Conroy}, Charlie},
        title = "{How to Measure Galaxy Star Formation Histories. I. Parametric Models}",
      journal = {\apj},
         year = 2019,
        month = mar,
       volume = {873},
       number = {1},
          eid = {44},
        pages = {44},
          doi = {10.3847/1538-4357/ab04a2},
archivePrefix = {arXiv},
       eprint = {1811.03635},
 primaryClass = {astro-ph.GA},
       adsurl = {https://ui.adsabs.harvard.edu/abs/2019ApJ...873...44C}
}

@ARTICLE{Camps2020,
       author = {{Camps}, P. and {Baes}, M.},
        title = "{SKIRT 9: Redesigning an advanced dust radiative transfer code to allow kinematics, line transfer and polarization by aligned dust grains}",
      journal = {Astronomy and Computing},
         year = 2020,
        month = apr,
       volume = {31},
          eid = {100381},
        pages = {100381},
          doi = {10.1016/j.ascom.2020.100381},
archivePrefix = {arXiv},
       eprint = {2003.00721},
 primaryClass = {astro-ph.GA},
       adsurl = {https://ui.adsabs.harvard.edu/abs/2020A&C....3100381C}
}

@ARTICLE{Casey2024,
       author = {{Casey}, Caitlin M. and {Akins}, Hollis B. and {Shuntov}, Marko and {Ilbert}, Olivier and {Paquereau}, Louise and {Franco}, Maximilien and {Hayward}, Christopher C. and {Finkelstein}, Steven L. and {Boylan-Kolchin}, Michael and {Robertson}, Brant E. and {Allen}, Natalie and {Brinch}, Malte and {Cooper}, Olivia R. and {Ding}, Xuheng and {Drakos}, Nicole E. and {Faisst}, Andreas L. and {Fujimoto}, Seiji and {Gillman}, Steven and {Harish}, Santosh and {Hirschmann}, Michaela and {Jin}, Shuowen and {Kartaltepe}, Jeyhan S. and {Koekemoer}, Anton M. and {Kokorev}, Vasily and {Liu}, Daizhong and {Long}, Arianna S. and {Magdis}, Georgios and {Maraston}, Claudia and {Martin}, Crystal L. and {McCracken}, Henry Joy and {McKinney}, Jed and {Mobasher}, Bahram and {Rhodes}, Jason and {Rich}, R. Michael and {Sanders}, David B. and {Silverman}, John D. and {Toft}, Sune and {Vijayan}, Aswin P. and {Weaver}, John R. and {Wilkins}, Stephen M. and {Yang}, Lilan and {Zavala}, Jorge A.},
        title = "{COSMOS-Web: Intrinsically Luminous z {\ensuremath{\gtrsim}} 10 Galaxy Candidates Test Early Stellar Mass Assembly}",
      journal = {\apj},
         year = 2024,
        month = apr,
       volume = {965},
       number = {1},
          eid = {98},
        pages = {98},
          doi = {10.3847/1538-4357/ad2075},
archivePrefix = {arXiv},
       eprint = {2308.10932},
 primaryClass = {astro-ph.GA},
       adsurl = {https://ui.adsabs.harvard.edu/abs/2024ApJ...965...98C}
}

@ARTICLE{Cameron2024,
       author = {{Cameron}, Alex J. and {Katz}, Harley and {Witten}, Callum and {Saxena}, Aayush and {Laporte}, Nicolas and {Bunker}, Andrew J.},
        title = "{Nebular dominated galaxies: insights into the stellar initial mass function at high redshift}",
      journal = {\mnras},
         year = 2024,
        month = oct,
       volume = {534},
       number = {1},
        pages = {523-543},
          doi = {10.1093/mnras/stae1547},
archivePrefix = {arXiv},
       eprint = {2311.02051},
 primaryClass = {astro-ph.GA},
       adsurl = {https://ui.adsabs.harvard.edu/abs/2024MNRAS.534..523C}
}

@ARTICLE{Carniani2024,
       author = {{Carniani}, Stefano and {Hainline}, Kevin and {D'Eugenio}, Francesco and {Eisenstein}, Daniel J. and {Jakobsen}, Peter and {Witstok}, Joris and {Johnson}, Benjamin D. and {Chevallard}, Jacopo and {Maiolino}, Roberto and {Helton}, Jakob M. and {Willott}, Chris and {Robertson}, Brant and {Alberts}, Stacey and {Arribas}, Santiago and {Baker}, William M. and {Bhatawdekar}, Rachana and {Boyett}, Kristan and {Bunker}, Andrew J. and {Cameron}, Alex J. and {Cargile}, Phillip A. and {Charlot}, St{\'e}phane and {Curti}, Mirko and {Curtis-Lake}, Emma and {Egami}, Eiichi and {Giardino}, Giovanna and {Isaak}, Kate and {Ji}, Zhiyuan and {Jones}, Gareth C. and {Kumari}, Nimisha and {Maseda}, Michael V. and {Parlanti}, Eleonora and {P{\'e}rez-Gonz{\'a}lez}, Pablo G. and {Rawle}, Tim and {Rieke}, George and {Rieke}, Marcia and {Del Pino}, Bruno Rodr{\'\i}guez and {Saxena}, Aayush and {Scholtz}, Jan and {Smit}, Renske and {Sun}, Fengwu and {Tacchella}, Sandro and {{\"U}bler}, Hannah and {Venturi}, Giacomo and {Williams}, Christina C. and {Willmer}, Christopher N.~A.},
        title = "{Spectroscopic confirmation of two luminous galaxies at a redshift of 14}",
      journal = {\nat},
         year = 2024,
        month = sep,
       volume = {633},
       number = {8029},
        pages = {318-322},
          doi = {10.1038/s41586-024-07860-9},
archivePrefix = {arXiv},
       eprint = {2405.18485},
 primaryClass = {astro-ph.GA},
       adsurl = {https://ui.adsabs.harvard.edu/abs/2024Natur.633..318C}
}

@ARTICLE{Castellano2023,
       author = {{Castellano}, Marco and {Fontana}, Adriano and {Treu}, Tommaso and {Merlin}, Emiliano and {Santini}, Paola and {Bergamini}, Pietro and {Grillo}, Claudio and {Rosati}, Piero and {Acebron}, Ana and {Leethochawalit}, Nicha and {Paris}, Diego and {Bonchi}, Andrea and {Belfiori}, Davide and {Calabr{\`o}}, Antonello and {Correnti}, Matteo and {Nonino}, Mario and {Polenta}, Gianluca and {Trenti}, Michele and {Boyett}, Kristan and {Brammer}, G. and {Broadhurst}, Tom and {Caminha}, Gabriel B. and {Chen}, Wenlei and {Filippenko}, Alexei V. and {Fortuni}, Flaminia and {Glazebrook}, Karl and {Mascia}, Sara and {Mason}, Charlotte A. and {Menci}, Nicola and {Meneghetti}, Massimo and {Mercurio}, Amata and {Metha}, Benjamin and {Morishita}, Takahiro and {Nanayakkara}, Themiya and {Pentericci}, Laura and {Roberts-Borsani}, Guido and {Roy}, Namrata and {Vanzella}, Eros and {Vulcani}, Benedetta and {Yang}, Lilan and {Wang}, Xin},
        title = "{Early Results from GLASS-JWST. XIX. A High Density of Bright Galaxies at z {\ensuremath{\approx}} 10 in the A2744 Region}",
      journal = {\apjl},
         year = 2023,
        month = may,
       volume = {948},
       number = {2},
          eid = {L14},
        pages = {L14},
          doi = {10.3847/2041-8213/accea5},
archivePrefix = {arXiv},
       eprint = {2212.06666},
 primaryClass = {astro-ph.GA},
       adsurl = {https://ui.adsabs.harvard.edu/abs/2023ApJ...948L..14C}
}

@ARTICLE{Chabrier2003,
       author = {{Chabrier}, Gilles},
        title = "{Galactic Stellar and Substellar Initial Mass Function}",
      journal = {\pasp},
         year = 2003,
        month = jul,
       volume = {115},
       number = {809},
        pages = {763-795},
          doi = {10.1086/376392},
archivePrefix = {arXiv},
       eprint = {astro-ph/0304382},
 primaryClass = {astro-ph},
       adsurl = {https://ui.adsabs.harvard.edu/abs/2003PASP..115..763C}
}

@ARTICLE{Chemerynska2026,
       author = {{Chemerynska}, Iryna and {Atek}, Hakim and {Furtak}, Lukas J. and {Chisholm}, John and {Endsley}, Ryan and {Kokorev}, Vasily and {Rosdahl}, Joki and {Blaizot}, Jeremy and {Adamo}, Angela and {Bouwens}, Rychard and {Fujimoto}, Seiji and {Korber}, Damien and {Mason}, Charlotte and {McQuinn}, Kristen B.~W. and {Mu{\~n}oz}, Julian B. and {Natarajan}, Priyamvada and {Nelson}, Erica and {Oesch}, Pascal A. and {Pan}, Richard and {Richard}, Johan and {Saldana-Lopez}, Alberto and {Schaerer}, Daniel and {Volonteri}, Marta and {Zitrin}, Adi and {Berg}, Danielle A. and {Claeyssens}, Ad{\'e}la{\"\i}de and {Dessauges-Zavadsky}, Miroslava and {Jecmen}, Michelle and {Labb{\'e}}, Ivo and {Naidu}, Rohan and {Trebitsch}, Maxime},
        title = "{The first GLIMPSE of the faint galaxy population at Cosmic Dawn with JWST: The evolution of the ultraviolet luminosity function across z {\ensuremath{\sim}} 9{\ensuremath{-}}15}",
      journal = {\mnras},
         year = 2026,
        month = feb,
       volume = {546},
       number = {2},
          eid = {staf2267},
        pages = {staf2267},
          doi = {10.1093/mnras/staf2267},
archivePrefix = {arXiv},
       eprint = {2509.24881},
 primaryClass = {astro-ph.GA},
       adsurl = {https://ui.adsabs.harvard.edu/abs/2026MNRAS.546f2267C}
}

@ARTICLE{Cole2000,
       author = {{Cole}, Shaun and {Lacey}, Cedric G. and {Baugh}, Carlton M. and {Frenk}, Carlos S.},
        title = "{Hierarchical galaxy formation}",
      journal = {\mnras},
         year = 2000,
        month = nov,
       volume = {319},
       number = {1},
        pages = {168-204},
          doi = {10.1046/j.1365-8711.2000.03879.x},
archivePrefix = {arXiv},
       eprint = {astro-ph/0007281},
 primaryClass = {astro-ph},
       adsurl = {https://ui.adsabs.harvard.edu/abs/2000MNRAS.319..168C}
}

@ARTICLE{Cowley2018,
       author = {{Cowley}, William I. and {Baugh}, Carlton M. and {Cole}, Shaun and {Frenk}, Carlos S. and {Lacey}, Cedric G.},
        title = "{Predictions for deep galaxy surveys with JWST from {\ensuremath{\Lambda}}CDM}",
      journal = {\mnras},
         year = 2018,
        month = feb,
       volume = {474},
       number = {2},
        pages = {2352-2372},
          doi = {10.1093/mnras/stx2897},
archivePrefix = {arXiv},
       eprint = {1702.02146},
 primaryClass = {astro-ph.GA},
       adsurl = {https://ui.adsabs.harvard.edu/abs/2018MNRAS.474.2352C}
}

@ARTICLE{Crain2015,
       author = {{Crain}, Robert A. and {Schaye}, Joop and {Bower}, Richard G. and {Furlong}, Michelle and {Schaller}, Matthieu and {Theuns}, Tom and {Dalla Vecchia}, Claudio and {Frenk}, Carlos S. and {McCarthy}, Ian G. and {Helly}, John C. and {Jenkins}, Adrian and {Rosas-Guevara}, Yetli M. and {White}, Simon D.~M. and {Trayford}, James W.},
        title = "{The EAGLE simulations of galaxy formation: calibration of subgrid physics and model variations}",
      journal = {\mnras},
         year = 2015,
        month = jun,
       volume = {450},
       number = {2},
        pages = {1937-1961},
          doi = {10.1093/mnras/stv725},
archivePrefix = {arXiv},
       eprint = {1501.01311},
 primaryClass = {astro-ph.GA},
       adsurl = {https://ui.adsabs.harvard.edu/abs/2015MNRAS.450.1937C}
}

@ARTICLE{Cueto2024,
       author = {{Cueto}, Elie R. and {Hutter}, Anne and {Dayal}, Pratika and {Gottl{\"o}ber}, Stefan and {Heintz}, Kasper E. and {Mason}, Charlotte and {Trebitsch}, Maxime and {Yepes}, Gustavo},
        title = "{ASTRAEUS. IX. Impact of an evolving stellar initial mass function on early galaxies and reionisation}",
      journal = {\aap},
         year = 2024,
        month = jun,
       volume = {686},
          eid = {A138},
        pages = {A138},
          doi = {10.1051/0004-6361/202349017},
archivePrefix = {arXiv},
       eprint = {2312.12109},
 primaryClass = {astro-ph.GA},
       adsurl = {https://ui.adsabs.harvard.edu/abs/2024A&A...686A.138C}
}

@ARTICLE{Cullen2023,
       author = {{Cullen}, Fergus and {McLure}, R.~J. and {McLeod}, D.~J. and {Dunlop}, J.~S. and {Donnan}, C.~T. and {Carnall}, A.~C. and {Bowler}, R.~A.~A. and {Begley}, R. and {Hamadouche}, M.~L. and {Stanton}, T.~M.},
        title = "{The ultraviolet continuum slopes ({\ensuremath{\beta}}) of galaxies at z ≃ 8-16 from JWST and ground-based near-infrared imaging}",
      journal = {\mnras},
         year = 2023,
        month = mar,
       volume = {520},
       number = {1},
        pages = {14-23},
          doi = {10.1093/mnras/stad073},
archivePrefix = {arXiv},
       eprint = {2208.04914},
 primaryClass = {astro-ph.GA},
       adsurl = {https://ui.adsabs.harvard.edu/abs/2023MNRAS.520...14C}
}

@ARTICLE{Davis1985,
       author = {{Davis}, M. and {Efstathiou}, G. and {Frenk}, C.~S. and {White}, S.~D.~M.},
        title = "{The evolution of large-scale structure in a universe dominated by cold dark matter}",
      journal = {\apj},
         year = 1985,
        month = may,
       volume = {292},
        pages = {371-394},
          doi = {10.1086/163168},
       adsurl = {https://ui.adsabs.harvard.edu/abs/1985ApJ...292..371D}
}

@ARTICLE{Dave2019,
       author = {{Dav{\'e}}, Romeel and {Angl{\'e}s-Alc{\'a}zar}, Daniel and {Narayanan}, Desika and {Li}, Qi and {Rafieferantsoa}, Mika H. and {Appleby}, Sarah},
        title = "{SIMBA: Cosmological simulations with black hole growth and feedback}",
      journal = {\mnras},
         year = 2019,
        month = jun,
       volume = {486},
       number = {2},
        pages = {2827-2849},
          doi = {10.1093/mnras/stz937},
archivePrefix = {arXiv},
       eprint = {1901.10203},
 primaryClass = {astro-ph.GA},
       adsurl = {https://ui.adsabs.harvard.edu/abs/2019MNRAS.486.2827D}
}

@ARTICLE{DeLucia2024,
       author = {{De Lucia}, Gabriella and {Fontanot}, Fabio and {Xie}, Lizhi and {Hirschmann}, Michaela},
        title = "{Tracing the quenching journey across cosmic time}",
      journal = {\aap},
         year = 2024,
        month = jul,
       volume = {687},
          eid = {A68},
        pages = {A68},
          doi = {10.1051/0004-6361/202349045},
archivePrefix = {arXiv},
       eprint = {2401.06211},
 primaryClass = {astro-ph.GA},
       adsurl = {https://ui.adsabs.harvard.edu/abs/2024A&A...687A..68D}
}

@ARTICLE{Donnan2023,
       author = {{Donnan}, C.~T. and {McLeod}, D.~J. and {Dunlop}, J.~S. and {McLure}, R.~J. and {Carnall}, A.~C. and {Begley}, R. and {Cullen}, F. and {Hamadouche}, M.~L. and {Bowler}, R.~A.~A. and {Magee}, D. and {McCracken}, H.~J. and {Milvang-Jensen}, B. and {Moneti}, A. and {Targett}, T.},
        title = "{The evolution of the galaxy UV luminosity function at redshifts z ≃ 8 - 15 from deep JWST and ground-based near-infrared imaging}",
      journal = {\mnras},
         year = 2023,
        month = feb,
       volume = {518},
       number = {4},
        pages = {6011-6040},
          doi = {10.1093/mnras/stac3472},
archivePrefix = {arXiv},
       eprint = {2207.12356},
 primaryClass = {astro-ph.GA},
       adsurl = {https://ui.adsabs.harvard.edu/abs/2023MNRAS.518.6011D}
}

@ARTICLE{Donnan2024,
       author = {{Donnan}, C.~T. and {McLure}, R.~J. and {Dunlop}, J.~S. and {McLeod}, D.~J. and {Magee}, D. and {Arellano-C{\'o}rdova}, K.~Z. and {Barrufet}, L. and {Begley}, R. and {Bowler}, R.~A.~A. and {Carnall}, A.~C. and {Cullen}, F. and {Ellis}, R.~S. and {Fontana}, A. and {Illingworth}, G.~D. and {Grogin}, N.~A. and {Hamadouche}, M.~L. and {Koekemoer}, A.~M. and {Liu}, F.-Y. and {Mason}, C. and {Santini}, P. and {Stanton}, T.~M.},
        title = "{JWST PRIMER: a new multifield determination of the evolving galaxy UV luminosity function at redshifts z ≃ 9 - 15}",
      journal = {\mnras},
         year = 2024,
        month = sep,
       volume = {533},
       number = {3},
        pages = {3222-3237},
          doi = {10.1093/mnras/stae2037},
archivePrefix = {arXiv},
       eprint = {2403.03171},
 primaryClass = {astro-ph.GA},
       adsurl = {https://ui.adsabs.harvard.edu/abs/2024MNRAS.533.3222D}
}

@ARTICLE{Draine2007,
       author = {{Draine}, B.~T. and {Li}, Aigen},
        title = "{Infrared Emission from Interstellar Dust. IV. The Silicate-Graphite-PAH Model in the Post-Spitzer Era}",
      journal = {\apj},
         year = 2007,
        month = mar,
       volume = {657},
       number = {2},
        pages = {810-837},
          doi = {10.1086/511055},
archivePrefix = {arXiv},
       eprint = {astro-ph/0608003},
 primaryClass = {astro-ph},
       adsurl = {https://ui.adsabs.harvard.edu/abs/2007ApJ...657..810D}
}

@ARTICLE{Dubois2016,
       author = {{Dubois}, Yohan and {Peirani}, S{\'e}bastien and {Pichon}, Christophe and {Devriendt}, Julien and {Gavazzi}, Rapha{\"e}l and {Welker}, Charlotte and {Volonteri}, Marta},
        title = "{The HORIZON-AGN simulation: morphological diversity of galaxies promoted by AGN feedback}",
      journal = {\mnras},
         year = 2016,
        month = dec,
       volume = {463},
       number = {4},
        pages = {3948-3964},
          doi = {10.1093/mnras/stw2265},
archivePrefix = {arXiv},
       eprint = {1606.03086},
 primaryClass = {astro-ph.GA},
       adsurl = {https://ui.adsabs.harvard.edu/abs/2016MNRAS.463.3948D}
}

@ARTICLE{Eddington1913,
       author = {{Eddington}, A.~S.},
        title = "{On a formula for correcting statistics for the effects of a known error of observation}",
      journal = {\mnras},
         year = 1913,
        month = mar,
       volume = {73},
        pages = {359-360},
          doi = {10.1093/mnras/73.5.359},
       adsurl = {https://ui.adsabs.harvard.edu/abs/1913MNRAS..73..359E}
}

@ARTICLE{Eldridge2017,
       author = {{Eldridge}, J.~J. and {Stanway}, E.~R. and {Xiao}, L. and {McClelland}, L.~A.~S. and {Taylor}, G. and {Ng}, M. and {Greis}, S.~M.~L. and {Bray}, J.~C.},
        title = "{Binary Population and Spectral Synthesis Version 2.1: Construction, Observational Verification, and New Results}",
      journal = {\pasa},
         year = 2017,
        month = nov,
       volume = {34},
          eid = {e058},
        pages = {e058},
          doi = {10.1017/pasa.2017.51},
archivePrefix = {arXiv},
       eprint = {1710.02154},
 primaryClass = {astro-ph.SR},
       adsurl = {https://ui.adsabs.harvard.edu/abs/2017PASA...34...58E}
}

@ARTICLE{Ferrara2023,
       author = {{Ferrara}, Andrea and {Pallottini}, Andrea and {Dayal}, Pratika},
        title = "{On the stunning abundance of super-early, luminous galaxies revealed by JWST}",
      journal = {\mnras},
         year = 2023,
        month = jul,
       volume = {522},
       number = {3},
        pages = {3986-3991},
          doi = {10.1093/mnras/stad1095},
archivePrefix = {arXiv},
       eprint = {2208.00720},
 primaryClass = {astro-ph.GA},
       adsurl = {https://ui.adsabs.harvard.edu/abs/2023MNRAS.522.3986F}
}

@ARTICLE{Finkelstein2015,
       author = {{Finkelstein}, Steven L. and {Ryan}, Russell E., Jr. and {Papovich}, Casey and {Dickinson}, Mark and {Song}, Mimi and {Somerville}, Rachel S. and {Ferguson}, Henry C. and {Salmon}, Brett and {Giavalisco}, Mauro and {Koekemoer}, Anton M. and {Ashby}, Matthew L.~N. and {Behroozi}, Peter and {Castellano}, Marco and {Dunlop}, James S. and {Faber}, Sandy M. and {Fazio}, Giovanni G. and {Fontana}, Adriano and {Grogin}, Norman A. and {Hathi}, Nimish and {Jaacks}, Jason and {Kocevski}, Dale D. and {Livermore}, Rachael and {McLure}, Ross J. and {Merlin}, Emiliano and {Mobasher}, Bahram and {Newman}, Jeffrey A. and {Rafelski}, Marc and {Tilvi}, Vithal and {Willner}, S.~P.},
        title = "{The Evolution of the Galaxy Rest-frame Ultraviolet Luminosity Function over the First Two Billion Years}",
      journal = {\apj},
         year = 2015,
        month = sep,
       volume = {810},
       number = {1},
          eid = {71},
        pages = {71},
          doi = {10.1088/0004-637X/810/1/71},
archivePrefix = {arXiv},
       eprint = {1410.5439},
 primaryClass = {astro-ph.GA},
       adsurl = {https://ui.adsabs.harvard.edu/abs/2015ApJ...810...71F}
}

@ARTICLE{Fontanot2026,
       author = {{Fontanot}, Fabio and {De Lucia}, Gabriella and {Xie}, Lizhi and {Zibetti}, Stefano and {La Barbera}, Francesco and {Cantarella}, Sebastiano and {Hirschmann}, Michaela and {Charlot}, Stephane and {Bruzual}, Gustavo},
        title = "{Reinterpreting the puzzling properties of z>6 galaxies within a variable IMF framework}",
      journal = {arXiv e-prints},
         year = 2026,
        month = mar,
          eid = {arXiv:2603.22405},
        pages = {arXiv:2603.22405},
          doi = {10.48550/arXiv.2603.22405},
archivePrefix = {arXiv},
       eprint = {2603.22405},
 primaryClass = {astro-ph.GA},
       adsurl = {https://ui.adsabs.harvard.edu/abs/2026arXiv260322405F}
}

@ARTICLE{Finkelstein2023,
       author = {{Finkelstein}, Steven L. and {Bagley}, Micaela B. and {Ferguson}, Henry C. and {Wilkins}, Stephen M. and {Kartaltepe}, Jeyhan S. and {Papovich}, Casey and {Yung}, L.~Y. Aaron and {Arrabal Haro}, Pablo and {Behroozi}, Peter and {Dickinson}, Mark and {Kocevski}, Dale D. and {Koekemoer}, Anton M. and {Larson}, Rebecca L. and {Le Bail}, Aur{\'e}lien and {Morales}, Alexa M. and {P{\'e}rez-Gonz{\'a}lez}, Pablo G. and {Burgarella}, Denis and {Dav{\'e}}, Romeel and {Hirschmann}, Michaela and {Somerville}, Rachel S. and {Wuyts}, Stijn and {Bromm}, Volker and {Casey}, Caitlin M. and {Fontana}, Adriano and {Fujimoto}, Seiji and {Gardner}, Jonathan P. and {Giavalisco}, Mauro and {Grazian}, Andrea and {Grogin}, Norman A. and {Hathi}, Nimish P. and {Hutchison}, Taylor A. and {Jha}, Saurabh W. and {Jogee}, Shardha and {Kewley}, Lisa J. and {Kirkpatrick}, Allison and {Long}, Arianna S. and {Lotz}, Jennifer M. and {Pentericci}, Laura and {Pierel}, Justin D.~R. and {Pirzkal}, Nor and {Ravindranath}, Swara and {Ryan}, Russell E. and {Trump}, Jonathan R. and {Yang}, Guang and {Bhatawdekar}, Rachana and {Bisigello}, Laura and {Buat}, V{\'e}ronique and {Calabr{\`o}}, Antonello and {Castellano}, Marco and {Cleri}, Nikko J. and {Cooper}, M.~C. and {Croton}, Darren and {Daddi}, Emanuele and {Dekel}, Avishai and {Elbaz}, David and {Franco}, Maximilien and {Gawiser}, Eric and {Holwerda}, Benne W. and {Huertas-Company}, Marc and {Jaskot}, Anne E. and {Leung}, Gene C.~K. and {Lucas}, Ray A. and {Mobasher}, Bahram and {Pandya}, Viraj and {Tacchella}, Sandro and {Weiner}, Benjamin J. and {Zavala}, Jorge A.},
        title = "{CEERS Key Paper. I. An Early Look into the First 500 Myr of Galaxy Formation with JWST}",
      journal = {\apjl},
         year = 2023,
        month = mar,
       volume = {946},
       number = {1},
          eid = {L13},
        pages = {L13},
          doi = {10.3847/2041-8213/acade4},
archivePrefix = {arXiv},
       eprint = {2211.05792},
 primaryClass = {astro-ph.GA},
       adsurl = {https://ui.adsabs.harvard.edu/abs/2023ApJ...946L..13F}
}

@ARTICLE{Finkelstein2024,
       author = {{Finkelstein}, Steven L. and {Leung}, Gene C.~K. and {Bagley}, Micaela B. and {Dickinson}, Mark and {Ferguson}, Henry C. and {Papovich}, Casey and {Akins}, Hollis B. and {Arrabal Haro}, Pablo and {Dav{\'e}}, Romeel and {Dekel}, Avishai and {Kartaltepe}, Jeyhan S. and {Kocevski}, Dale D. and {Koekemoer}, Anton M. and {Pirzkal}, Nor and {Somerville}, Rachel S. and {Yung}, L.~Y. Aaron and {Amor{\'\i}n}, Ricardo O. and {Backhaus}, Bren E. and {Behroozi}, Peter and {Bisigello}, Laura and {Bromm}, Volker and {Casey}, Caitlin M. and {Ch{\'a}vez Ortiz}, {\'O}scar A. and {Cheng}, Yingjie and {Chworowsky}, Katherine and {Cleri}, Nikko J. and {Cooper}, M.~C. and {Davis}, Kelcey and {de la Vega}, Alexander and {Elbaz}, David and {Franco}, Maximilien and {Fontana}, Adriano and {Fujimoto}, Seiji and {Giavalisco}, Mauro and {Grogin}, Norman A. and {Holwerda}, Benne W. and {Huertas-Company}, Marc and {Hirschmann}, Michaela and {Iyer}, Kartheik G. and {Jogee}, Shardha and {Jung}, Intae and {Larson}, Rebecca L. and {Lucas}, Ray A. and {Mobasher}, Bahram and {Morales}, Alexa M. and {Morley}, Caroline V. and {Mukherjee}, Sagnick and {P{\'e}rez-Gonz{\'a}lez}, Pablo G. and {Ravindranath}, Swara and {Rodighiero}, Giulia and {Rowland}, Melanie J. and {Tacchella}, Sandro and {Taylor}, Anthony J. and {Trump}, Jonathan R. and {Wilkins}, Stephen M.},
        title = "{The Complete CEERS Early Universe Galaxy Sample: A Surprisingly Slow Evolution of the Space Density of Bright Galaxies at z {\ensuremath{\sim}} 8.5{\textendash}14.5}",
      journal = {\apjl},
         year = 2024,
        month = jul,
       volume = {969},
       number = {1},
          eid = {L2},
        pages = {L2},
          doi = {10.3847/2041-8213/ad4495},
archivePrefix = {arXiv},
       eprint = {2311.04279},
 primaryClass = {astro-ph.GA},
       adsurl = {https://ui.adsabs.harvard.edu/abs/2024ApJ...969L...2F}
}

@ARTICLE{Franco2025,
       author = {{Franco}, Maximilien and {Casey}, Caitlin M. and {Akins}, Hollis B. and {Ilbert}, Olivier and {Shuntov}, Marko and {Finkelstein}, Steven L. and {Paquereau}, Louise and {Faisst}, Andreas L. and {Koekemoer}, Anton M. and {Hirschmann}, Michaela and {Cantarella}, Sebastiano and {Drakos}, Nicole E. and {Wilkins}, Stephen M. and {McCracken}, Henry Joy and {Kartaltepe}, Jeyhan S. and {Maraston}, Claudia and {Abedini}, Fatemeh and {Achenbach}, Mark J. and {Arango-Toro}, Rafael C. and {Gentile}, Fabrizio and {Gozaliasl}, Ghassem and {Inayoshi}, Kohei and {Kakkad}, Darshan and {Kalantari}, Atousa and {Khostovan}, Ali Ahmad and {Kokorev}, Vasily and {Lambrides}, Erini and {Leroy}, Gavin and {Massey}, Richard and {Mobasher}, Bahram and {Newman}, Sophie L. and {Rhodes}, Jason and {Rich}, R. Michael and {Robertson}, Brant E. and {Sanders}, David B. and {Tanaka}, Takumi S. and {Vijayan}, Aswin P. and {Weaver}, John R. and {Yang}, Lilan and {Yu}, Si-Yue},
        title = "{Physical properties of galaxies and the UV Luminosity Function from $z\sim6$ to $z\sim14$ in COSMOS-Web}",
      journal = {arXiv e-prints},
         year = 2025,
        month = aug,
          eid = {arXiv:2508.04791},
        pages = {arXiv:2508.04791},
          doi = {10.48550/arXiv.2508.04791},
archivePrefix = {arXiv},
       eprint = {2508.04791},
 primaryClass = {astro-ph.GA},
       adsurl = {https://ui.adsabs.harvard.edu/abs/2025arXiv250804791F}
}

@ARTICLE{Faucher-Giguere2020,
       author = {{Faucher-Gigu{\`e}re}, Claude-Andr{\'e}},
        title = "{A cosmic UV/X-ray background model update}",
      journal = {\mnras},
         year = 2020,
        month = apr,
       volume = {493},
       number = {2},
        pages = {1614-1632},
          doi = {10.1093/mnras/staa302},
archivePrefix = {arXiv},
       eprint = {1903.08657},
 primaryClass = {astro-ph.CO},
       adsurl = {https://ui.adsabs.harvard.edu/abs/2020MNRAS.493.1614F}
}

@ARTICLE{Forouhar_Moreno2025,
       author = {{Forouhar Moreno}, Victor J. and {Helly}, John and {McGibbon}, Robert and {Schaye}, Joop and {Schaller}, Matthieu and {Han}, Jiaxin and {Kugel}, Roi and {Bah{\'e}}, Yannick M.},
        title = "{Assessing subhalo finders in cosmological hydrodynamical simulations}",
      journal = {\mnras},
         year = 2025,
        month = oct,
       volume = {543},
       number = {2},
        pages = {1339-1372},
          doi = {10.1093/mnras/staf1478},
archivePrefix = {arXiv},
       eprint = {2502.06932},
 primaryClass = {astro-ph.CO},
       adsurl = {https://ui.adsabs.harvard.edu/abs/2025MNRAS.543.1339F}
}

@ARTICLE{Fujimoto2023,
       author = {{Fujimoto}, Seiji and {Arrabal Haro}, Pablo and {Dickinson}, Mark and {Finkelstein}, Steven L. and {Kartaltepe}, Jeyhan S. and {Larson}, Rebecca L. and {Burgarella}, Denis and {Bagley}, Micaela B. and {Behroozi}, Peter and {Chworowsky}, Katherine and {Hirschmann}, Michaela and {Trump}, Jonathan R. and {Wilkins}, Stephen M. and {Yung}, L.~Y. Aaron and {Koekemoer}, Anton M. and {Papovich}, Casey and {Pirzkal}, Nor and {Ferguson}, Henry C. and {Fontana}, Adriano and {Grogin}, Norman A. and {Grazian}, Andrea and {Kewley}, Lisa J. and {Kocevski}, Dale D. and {Lotz}, Jennifer M. and {Pentericci}, Laura and {Ravindranath}, Swara and {Somerville}, Rachel S. and {Wilkins}, Stephen M. and {Amor{\'\i}n}, Ricardo O. and {Backhaus}, Bren E. and {Calabr{\`o}}, Antonello and {Casey}, Caitlin M. and {Cooper}, M.~C. and {Fern{\'a}ndez}, Vital and {Franco}, Maximilien and {Giavalisco}, Mauro and {Hathi}, Nimish P. and {Harish}, Santosh and {Hutchison}, Taylor A. and {Iyer}, Kartheik G. and {Jung}, Intae and {Lucas}, Ray A. and {Zavala}, Jorge A.},
        title = "{CEERS Spectroscopic Confirmation of NIRCam-selected z {\ensuremath{\gtrsim}} 8 Galaxy Candidates with JWST/NIRSpec: Initial Characterization of Their Properties}",
      journal = {\apjl},
         year = 2023,
        month = jun,
       volume = {949},
       number = {2},
          eid = {L25},
        pages = {L25},
          doi = {10.3847/2041-8213/acd2d9},
archivePrefix = {arXiv},
       eprint = {2301.09482},
 primaryClass = {astro-ph.GA},
       adsurl = {https://ui.adsabs.harvard.edu/abs/2023ApJ...949L..25F}
}

@ARTICLE{Giallongo2015,
       author = {{Giallongo}, E. and {Grazian}, A. and {Fiore}, F. and {Fontana}, A. and {Pentericci}, L. and {Vanzella}, E. and {Dickinson}, M. and {Kocevski}, D. and {Castellano}, M. and {Cristiani}, S. and {Ferguson}, H. and {Finkelstein}, S. and {Grogin}, N. and {Hathi}, N. and {Koekemoer}, A.~M. and {Newman}, J.~A. and {Salvato}, M.},
        title = "{Faint AGNs at z > 4 in the CANDELS GOODS-S field: looking for contributors to the reionization of the Universe}",
      journal = {\aap},
         year = 2015,
        month = jun,
       volume = {578},
          eid = {A83},
        pages = {A83},
          doi = {10.1051/0004-6361/201425334},
archivePrefix = {arXiv},
       eprint = {1502.02562},
 primaryClass = {astro-ph.CO},
       adsurl = {https://ui.adsabs.harvard.edu/abs/2015A&A...578A..83G}
}

@article{Hahn2020,
  title={MUSIC2-monofonIC: 3LPT initial condition generator},
  author={{Hahn}, Oliver and {Michaux}, Micha{\"e}l and {Rampf}, Cornelius and {Uhlemann}, Cora and {Angulo}, Raul E},
  journal={Astrophysics Source Code Library},
  pages={ascl--2008},
  year={2020}
}

@ARTICLE{Han2018,
       author = {{Han}, Jiaxin and {Cole}, Shaun and {Frenk}, Carlos S. and {Benitez-Llambay}, Alejandro and {Helly}, John},
        title = "{HBT+: an improved code for finding subhaloes and building merger trees in cosmological simulations}",
      journal = {\mnras},
         year = 2018,
        month = feb,
       volume = {474},
       number = {1},
        pages = {604-617},
          doi = {10.1093/mnras/stx2792},
archivePrefix = {arXiv},
       eprint = {1708.03646},
 primaryClass = {astro-ph.CO},
       adsurl = {https://ui.adsabs.harvard.edu/abs/2018MNRAS.474..604H}
}

@ARTICLE{Harikane2022,
       author = {{Harikane}, Yuichi and {Ono}, Yoshiaki and {Ouchi}, Masami and {Liu}, Chengze and {Sawicki}, Marcin and {Shibuya}, Takatoshi and {Behroozi}, Peter S. and {He}, Wanqiu and {Shimasaku}, Kazuhiro and {Arnouts}, Stephane and {Coupon}, Jean and {Fujimoto}, Seiji and {Gwyn}, Stephen and {Huang}, Jiasheng and {Inoue}, Akio K. and {Kashikawa}, Nobunari and {Komiyama}, Yutaka and {Matsuoka}, Yoshiki and {Willott}, Chris J.},
        title = "{GOLDRUSH. IV. Luminosity Functions and Clustering Revealed with  4,000,000 Galaxies at z   2-7: Galaxy-AGN Transition, Star Formation Efficiency, and Implication for Evolution at z > 10}",
      journal = {\apjs},
         year = 2022,
        month = mar,
       volume = {259},
       number = {1},
          eid = {20},
        pages = {20},
          doi = {10.3847/1538-4365/ac3dfc},
archivePrefix = {arXiv},
       eprint = {2108.01090},
 primaryClass = {astro-ph.GA},
       adsurl = {https://ui.adsabs.harvard.edu/abs/2022ApJS..259...20H}
}

@ARTICLE{Harikane2025,
       author = {{Harikane}, Yuichi and {Inoue}, Akio K. and {Ellis}, Richard S. and {Ouchi}, Masami and {Nakazato}, Yurina and {Yoshida}, Naoki and {Ono}, Yoshiaki and {Sun}, Fengwu and {Sato}, Riku A. and {Ferrami}, Giovanni and {Fujimoto}, Seiji and {Kashikawa}, Nobunari and {McLeod}, Derek J. and {P{\'e}rez-Gonz{\'a}lez}, Pablo G. and {Sawicki}, Marcin and {Sugahara}, Yuma and {Xu}, Yi and {Yamanaka}, Satoshi and {Carnall}, Adam C. and {Cullen}, Fergus and {Dunlop}, James S. and {Egami}, Eiichi and {Grogin}, Norman and {Isobe}, Yuki and {Koekemoer}, Anton M. and {Laporte}, Nicolas and {Lee}, Chien-Hsiu and {Magee}, Dan and {Matsuo}, Hiroshi and {Matsuoka}, Yoshiki and {Mawatari}, Ken and {Nakajima}, Kimihiko and {Nakane}, Minami and {Tamura}, Yoichi and {Umeda}, Hiroya and {Yanagisawa}, Hiroto},
        title = "{JWST, ALMA, and Keck Spectroscopic Constraints on the UV Luminosity Functions at z {\ensuremath{\sim}} 7─14: Clumpiness and Compactness of the Brightest Galaxies in the Early Universe}",
      journal = {\apj},
         year = 2025,
        month = feb,
       volume = {980},
       number = {1},
          eid = {138},
        pages = {138},
          doi = {10.3847/1538-4357/ad9b2c},
archivePrefix = {arXiv},
       eprint = {2406.18352},
 primaryClass = {astro-ph.GA},
       adsurl = {https://ui.adsabs.harvard.edu/abs/2025ApJ...980..138H}
}

@ARTICLE{Hutter2025,
       author = {{Hutter}, Anne and {Cueto}, Elie R. and {Dayal}, Pratika and {Gottl{\"o}ber}, Stefan and {Trebitsch}, Maxime and {Yepes}, Gustavo},
        title = "{ASTRAEUS: X. Indications of a top-heavy initial mass function in highly star-forming galaxies from JWST observations at z > 10}",
      journal = {\aap},
         year = 2025,
        month = feb,
       volume = {694},
          eid = {A254},
        pages = {A254},
          doi = {10.1051/0004-6361/202452460},
archivePrefix = {arXiv},
       eprint = {2410.00730},
 primaryClass = {astro-ph.GA},
       adsurl = {https://ui.adsabs.harvard.edu/abs/2025A&A...694A.254H}
}

@ARTICLE{Harikane2023,
       author = {{Harikane}, Yuichi and {Ouchi}, Masami and {Oguri}, Masamune and {Ono}, Yoshiaki and {Nakajima}, Kimihiko and {Isobe}, Yuki and {Umeda}, Hiroya and {Mawatari}, Ken and {Zhang}, Yechi},
        title = "{A Comprehensive Study of Galaxies at z   9-16 Found in the Early JWST Data: Ultraviolet Luminosity Functions and Cosmic Star Formation History at the Pre-reionization Epoch}",
      journal = {\apjs},
         year = 2023,
        month = mar,
       volume = {265},
       number = {1},
          eid = {5},
        pages = {5},
          doi = {10.3847/1538-4365/acaaa9},
archivePrefix = {arXiv},
       eprint = {2208.01612},
 primaryClass = {astro-ph.GA},
       adsurl = {https://ui.adsabs.harvard.edu/abs/2023ApJS..265....5H}
}

@ARTICLE{Henden2018,
       author = {{Henden}, Nicholas A. and {Puchwein}, Ewald and {Shen}, Sijing and {Sijacki}, Debora},
        title = "{The FABLE simulations: a feedback model for galaxies, groups, and clusters}",
      journal = {\mnras},
         year = 2018,
        month = oct,
       volume = {479},
       number = {4},
        pages = {5385-5412},
          doi = {10.1093/mnras/sty1780},
archivePrefix = {arXiv},
       eprint = {1804.05064},
 primaryClass = {astro-ph.GA},
       adsurl = {https://ui.adsabs.harvard.edu/abs/2018MNRAS.479.5385H}
}

@ARTICLE{Harikane2024,
       author = {{Harikane}, Yuichi and {Nakajima}, Kimihiko and {Ouchi}, Masami and {Umeda}, Hiroya and {Isobe}, Yuki and {Ono}, Yoshiaki and {Xu}, Yi and {Zhang}, Yechi},
        title = "{Pure Spectroscopic Constraints on UV Luminosity Functions and Cosmic Star Formation History from 25 Galaxies at z $_{spec}$ = 8.61-13.20 Confirmed with JWST/NIRSpec}",
      journal = {\apj},
         year = 2024,
        month = jan,
       volume = {960},
       number = {1},
          eid = {56},
        pages = {56},
          doi = {10.3847/1538-4357/ad0b7e},
archivePrefix = {arXiv},
       eprint = {2304.06658},
 primaryClass = {astro-ph.GA},
       adsurl = {https://ui.adsabs.harvard.edu/abs/2024ApJ...960...56H}
}

@ARTICLE{Iocco2024,
       author = {{Iocco}, Fabio and {Visinelli}, Luca},
        title = "{Compatibility of JWST results with exotic halos}",
      journal = {Physics of the Dark Universe},
         year = 2024,
        month = may,
       volume = {44},
          eid = {101496},
        pages = {101496},
          doi = {10.1016/j.dark.2024.101496},
archivePrefix = {arXiv},
       eprint = {2403.13068},
 primaryClass = {astro-ph.CO},
       adsurl = {https://ui.adsabs.harvard.edu/abs/2024PDU....4401496I}
}

@ARTICLE{Kannan2023,
       author = {{Kannan}, Rahul and {Springel}, Volker and {Hernquist}, Lars and {Pakmor}, R{\"u}diger and {Delgado}, Ana Maria and {Hadzhiyska}, Boryana and {Hern{\'a}ndez-Aguayo}, C{\'e}sar and {Barrera}, Monica and {Ferlito}, Fulvio and {Bose}, Sownak and {White}, Simon D.~M. and {Frenk}, Carlos and {Smith}, Aaron and {Garaldi}, Enrico},
        title = "{The MillenniumTNG project: the galaxy population at z {\ensuremath{\geq}} 8}",
      journal = {\mnras},
         year = 2023,
        month = sep,
       volume = {524},
       number = {2},
        pages = {2594-2605},
          doi = {10.1093/mnras/stac3743},
archivePrefix = {arXiv},
       eprint = {2210.10066},
 primaryClass = {astro-ph.GA},
       adsurl = {https://ui.adsabs.harvard.edu/abs/2023MNRAS.524.2594K}
}

@ARTICLE{Kapoor2023,
       author = {{Kapoor}, Anand Utsav and {Baes}, Maarten and {van der Wel}, Arjen and {Gebek}, Andrea and {Camps}, Peter and {Nersesian}, Angelos and {Meidt}, Sharon E. and {Smith}, Aaron and {Vicens}, Sebastien and {D'Eugenio}, Francesco and {Martorano}, Marco and {Barrientos}, Daniela and {Sartorio}, Nina Sanches},
        title = "{TODDLERS: a new UV-mm emission library for star-forming regions - I. Integration with SKIRT and public release}",
      journal = {\mnras},
         year = 2023,
        month = dec,
       volume = {526},
       number = {3},
        pages = {3871-3901},
          doi = {10.1093/mnras/stad2977},
archivePrefix = {arXiv},
       eprint = {2310.00388},
 primaryClass = {astro-ph.GA},
       adsurl = {https://ui.adsabs.harvard.edu/abs/2023MNRAS.526.3871K}
}

@ARTICLE{Kapoor2024,
       author = {{Kapoor}, Anand Utsav and {Baes}, Maarten and {van der Wel}, Arjen and {Gebek}, Andrea and {Camps}, Peter and {Smith}, Aaron and {Boquien}, M{\'e}d{\'e}ric and {Andreadis}, Nick and {Vicens}, Sebastien},
        title = "{TODDLERS: A new UV-millimeter emission library for star-forming regions: II. Star-formation rate indicators using Auriga zoom simulations}",
      journal = {\aap},
         year = 2024,
        month = dec,
       volume = {692},
          eid = {A79},
        pages = {A79},
          doi = {10.1051/0004-6361/202451207},
archivePrefix = {arXiv},
       eprint = {2410.01067},
 primaryClass = {astro-ph.GA},
       adsurl = {https://ui.adsabs.harvard.edu/abs/2024A&A...692A..79K}
}

@ARTICLE{Kaviraj2017,
       author = {{Kaviraj}, S. and {Laigle}, C. and {Kimm}, T. and {Devriendt}, J.~E.~G. and {Dubois}, Y. and {Pichon}, C. and {Slyz}, A. and {Chisari}, E. and {Peirani}, S.},
        title = "{The Horizon-AGN simulation: evolution of galaxy properties over cosmic time}",
      journal = {\mnras},
         year = 2017,
        month = jun,
       volume = {467},
       number = {4},
        pages = {4739-4752},
          doi = {10.1093/mnras/stx126},
archivePrefix = {arXiv},
       eprint = {1605.09379},
 primaryClass = {astro-ph.GA},
       adsurl = {https://ui.adsabs.harvard.edu/abs/2017MNRAS.467.4739K}
}

@ARTICLE{Kreilgaard2026,
       author = {{Kreilgaard}, Kimi C. and {Mason}, Charlotte A. and {Morishita}, Takahiro and {Zhang}, Yechi and {Gelli}, Viola and {Leethochawalit}, Nicha and {Treu}, Tommaso and {Trenti}, Michele and {Abdurro'uf} and {Atek}, Hakim and {Brada{\v{c}}}, Maru{\v{s}}a and {Bradley}, Larry D. and {Bunker}, Andrew J. and {Saputra Haryana}, Novan and {Hayes}, Matthew J. and {Liu}, Zhaoran and {Mehta}, Vihang and {Rafelski}, Marc and {Roberts-Borsani}, Guido and {Scarlata}, Claudia and {Stiavelli}, Massimo and {Sutanto}, Ryo A. and {Takahashi}, Kosuke and {Vulcani}, Benedetta},
        title = "{BEACON: JWST NIRCam Pure-parallel Imaging Survey. III. Constraints on the UV LF and the Clustering of z\raisebox{-0.5ex}\textasciitilde7-14 Galaxies}",
      journal = {arXiv e-prints},
         year = 2026,
        month = apr,
          eid = {arXiv:2604.17963},
        pages = {arXiv:2604.17963},
archivePrefix = {arXiv},
       eprint = {2604.17963},
 primaryClass = {astro-ph.GA},
       adsurl = {https://ui.adsabs.harvard.edu/abs/2026arXiv260417963K}
}

@ARTICLE{Leja2019,
       author = {{Leja}, Joel and {Carnall}, Adam C. and {Johnson}, Benjamin D. and {Conroy}, Charlie and {Speagle}, Joshua S.},
        title = "{How to Measure Galaxy Star Formation Histories. II. Nonparametric Models}",
      journal = {\apj},
         year = 2019,
        month = may,
       volume = {876},
       number = {1},
          eid = {3},
        pages = {3},
          doi = {10.3847/1538-4357/ab133c},
archivePrefix = {arXiv},
       eprint = {1811.03637},
 primaryClass = {astro-ph.GA},
       adsurl = {https://ui.adsabs.harvard.edu/abs/2019ApJ...876....3L}
}

@ARTICLE{Leung2023,
       author = {{Leung}, Gene C.~K. and {Bagley}, Micaela B. and {Finkelstein}, Steven L. and {Ferguson}, Henry C. and {Koekemoer}, Anton M. and {P{\'e}rez-Gonz{\'a}lez}, Pablo G. and {Morales}, Alexa and {Kocevski}, Dale D. and {Yang}, Guang and {Somerville}, Rachel S. and {Wilkins}, Stephen M. and {Yung}, L.~Y. Aaron and {Fujimoto}, Seiji and {Larson}, Rebecca L. and {Papovich}, Casey and {Pirzkal}, Nor and {Berg}, Danielle A. and {Lotz}, Jennifer M. and {Castellano}, Marco and {Ch{\'a}vez Ortiz}, {\'O}scar A. and {Cheng}, Yingjie and {Dickinson}, Mark and {Giavalisco}, Mauro and {Hathi}, Nimish P. and {Hutchison}, Taylor A. and {Jung}, Intae and {Kartaltepe}, Jeyhan S. and {Natarajan}, Priyamvada and {Rothberg}, Barry},
        title = "{NGDEEP Epoch 1: The Faint End of the Luminosity Function at z   9-12 from Ultradeep JWST Imaging}",
      journal = {\apjl},
         year = 2023,
        month = sep,
       volume = {954},
       number = {2},
          eid = {L46},
        pages = {L46},
          doi = {10.3847/2041-8213/acf365},
archivePrefix = {arXiv},
       eprint = {2306.06244},
 primaryClass = {astro-ph.GA},
       adsurl = {https://ui.adsabs.harvard.edu/abs/2023ApJ...954L..46L}
}

@ARTICLE{Lacey2016,
       author = {{Lacey}, Cedric G. and {Baugh}, Carlton M. and {Frenk}, Carlos S. and {Benson}, Andrew J. and {Bower}, Richard G. and {Cole}, Shaun and {Gonzalez-Perez}, Violeta and {Helly}, John C. and {Lagos}, Claudia D.~P. and {Mitchell}, Peter D.},
        title = "{A unified multiwavelength model of galaxy formation}",
      journal = {\mnras},
         year = 2016,
        month = nov,
       volume = {462},
       number = {4},
        pages = {3854-3911},
          doi = {10.1093/mnras/stw1888},
archivePrefix = {arXiv},
       eprint = {1509.08473},
 primaryClass = {astro-ph.GA},
       adsurl = {https://ui.adsabs.harvard.edu/abs/2016MNRAS.462.3854L}
}

@ARTICLE{Lagos2025,
       author = {{Lagos}, Claudia del P. and {Schaye}, Joop and {Schaller}, Matthieu and {Obreschkow}, Danail and {Bahe}, Yannick M. and {Benitez-Llambay}, Alejandro and {Chaikin}, Evgenii and {Correa}, Camila and {Davis}, Timothy A. and {Frenk}, Carlos S. and {Husko}, Filip and {Kaasinen}, Melanie and {McGibbon}, Robert J. and {Oman}, Kyle and {Ploeckinger}, Sylvia and {Richings}, Alexander J. and {Trayford}, James W. and {Wang}, Jing and {Wright}, Ruby J.},
        title = "{Kennicutt-Schmidt relation of galaxies over 13 billion years in the COLIBRE hydrodynamical simulations}",
      journal = {arXiv e-prints},
         year = 2025,
        month = dec,
          eid = {arXiv:2512.11309},
        pages = {arXiv:2512.11309},
          doi = {10.48550/arXiv.2512.11309},
archivePrefix = {arXiv},
       eprint = {2512.11309},
 primaryClass = {astro-ph.GA},
       adsurl = {https://ui.adsabs.harvard.edu/abs/2025arXiv251211309L}
}

@ARTICLE{Liang2021,
       author = {{Liang}, Fu-Heng and {Li}, Cheng and {Li}, Niu and {Zhou}, Shuang and {Yan}, Renbin and {Mo}, Houjun and {Zhang}, Wei},
        title = "{Wolf-Rayet Galaxies in SDSS-IV MaNGA. II. Metallicity Dependence of the High-mass Slope of the Stellar Initial Mass Function}",
      journal = {\apj},
         year = 2021,
        month = dec,
       volume = {923},
       number = {1},
          eid = {120},
        pages = {120},
          doi = {10.3847/1538-4357/ac2bff},
archivePrefix = {arXiv},
       eprint = {2101.03217},
 primaryClass = {astro-ph.GA},
       adsurl = {https://ui.adsabs.harvard.edu/abs/2021ApJ...923..120L}
}

@ARTICLE{Laigle2016,
       author = {{Laigle}, C. and {McCracken}, H.~J. and {Ilbert}, O. and {Hsieh}, B.~C. and {Davidzon}, I. and {Capak}, P. and {Hasinger}, G. and {Silverman}, J.~D. and {Pichon}, C. and {Coupon}, J. and {Aussel}, H. and {Le Borgne}, D. and {Caputi}, K. and {Cassata}, P. and {Chang}, Y. -Y. and {Civano}, F. and {Dunlop}, J. and {Fynbo}, J. and {Kartaltepe}, J.~S. and {Koekemoer}, A. and {Le F{\`e}vre}, O. and {Le Floc'h}, E. and {Leauthaud}, A. and {Lilly}, S. and {Lin}, L. and {Marchesi}, S. and {Milvang-Jensen}, B. and {Salvato}, M. and {Sanders}, D.~B. and {Scoville}, N. and {Smolcic}, V. and {Stockmann}, M. and {Taniguchi}, Y. and {Tasca}, L. and {Toft}, S. and {Vaccari}, Mattia and {Zabl}, J.},
        title = "{The COSMOS2015 Catalog: Exploring the 1 < z < 6 Universe with Half a Million Galaxies}",
      journal = {\apjs},
         year = 2016,
        month = jun,
       volume = {224},
       number = {2},
          eid = {24},
        pages = {24},
          doi = {10.3847/0067-0049/224/2/24},
archivePrefix = {arXiv},
       eprint = {1604.02350},
 primaryClass = {astro-ph.GA},
       adsurl = {https://ui.adsabs.harvard.edu/abs/2016ApJS..224...24L}
}

@ARTICLE{Ludlow2019,
       author = {{Ludlow}, Aaron D. and {Schaye}, Joop and {Schaller}, Matthieu and {Richings}, Jack},
        title = "{Energy equipartition between stellar and dark matter particles in cosmological simulations results in spurious growth of galaxy sizes}",
      journal = {\mnras},
         year = 2019,
        month = sep,
       volume = {488},
       number = {1},
        pages = {L123-L128},
          doi = {10.1093/mnrasl/slz110},
archivePrefix = {arXiv},
       eprint = {1903.10110},
 primaryClass = {astro-ph.GA},
       adsurl = {https://ui.adsabs.harvard.edu/abs/2019MNRAS.488L.123L}
}

@ARTICLE{Ludlow2023,
       author = {{Ludlow}, Aaron D. and {Fall}, S. Michael and {Wilkinson}, Matthew J. and {Schaye}, Joop and {Obreschkow}, Danail},
        title = "{Spurious heating of stellar motions by dark matter particles in cosmological simulations of galaxy formation}",
      journal = {\mnras},
         year = 2023,
        month = nov,
       volume = {525},
       number = {4},
        pages = {5614-5630},
          doi = {10.1093/mnras/stad2615},
archivePrefix = {arXiv},
       eprint = {2306.05753},
 primaryClass = {astro-ph.GA},
       adsurl = {https://ui.adsabs.harvard.edu/abs/2023MNRAS.525.5614L}
}

@ARTICLE{Ludlow2026,
       author = {{Ludlow}, Aaron D. and {Proctor}, Katy L. and {Schaye}, Joop and {Hu{\v{s}}ko}, Filip and {Forouhar Moreno}, Victor J. and {Obreschkow}, Danail and {Chaikin}, Evgenii and {Schaller}, Matthieu and {Ploeckinger}, Sylvia and {Ben{\'\i}tez-Llambay}, Alejandro and {Oman}, Kyle A. and {McGibbon}, Robert J. and {Trayford}, James W. and {Frenk}, Carlos S. and {Richings}, Alexander J.},
        title = "{The evolution of the sizes and angular momentum content of galaxies in the COLIBRE simulations}",
      journal = {arXiv e-prints},
         year = 2026,
        month = mar,
          eid = {arXiv:2603.26200},
        pages = {arXiv:2603.26200},
          doi = {10.48550/arXiv.2603.26200},
archivePrefix = {arXiv},
       eprint = {2603.26200},
 primaryClass = {astro-ph.GA},
       adsurl = {https://ui.adsabs.harvard.edu/abs/2026arXiv260326200L}
}

@ARTICLE{Lo_Faro2017,
       author = {{Lo Faro}, B. and {Buat}, V. and {Roehlly}, Y. and {Alvarez-Marquez}, J. and {Burgarella}, D. and {Silva}, L. and {Efstathiou}, A.},
        title = "{Characterizing the UV-to-NIR shape of the dust attenuation curve of IR luminous galaxies up to z {\ensuremath{\sim}} 2}",
      journal = {\mnras},
         year = 2017,
        month = dec,
       volume = {472},
       number = {2},
        pages = {1372-1391},
          doi = {10.1093/mnras/stx1901},
archivePrefix = {arXiv},
       eprint = {1707.09805},
 primaryClass = {astro-ph.GA},
       adsurl = {https://ui.adsabs.harvard.edu/abs/2017MNRAS.472.1372L}
}

@ARTICLE{Lu2025,
       author = {{Lu}, Shengdong and {Frenk}, Carlos S. and {Bose}, Sownak and {Lacey}, Cedric G. and {Cole}, Shaun and {Baugh}, Carlton M. and {Helly}, John C.},
        title = "{A comparison of pre-existing {\ensuremath{\Lambda}}CDM predictions with the abundance of JWST galaxies at high redshift}",
      journal = {\mnras},
         year = 2025,
        month = jan,
       volume = {536},
       number = {1},
        pages = {1018-1034},
          doi = {10.1093/mnras/stae2646},
archivePrefix = {arXiv},
       eprint = {2406.02672},
 primaryClass = {astro-ph.GA},
       adsurl = {https://ui.adsabs.harvard.edu/abs/2025MNRAS.536.1018L}
}

@ARTICLE{Matsuoka2018,
       author = {{Matsuoka}, Yoshiki and {Strauss}, Michael A. and {Kashikawa}, Nobunari and {Onoue}, Masafusa and {Iwasawa}, Kazushi and {Tang}, Ji-Jia and {Lee}, Chien-Hsiu and {Imanishi}, Masatoshi and {Nagao}, Tohru and {Akiyama}, Masayuki and {Asami}, Naoko and {Bosch}, James and {Furusawa}, Hisanori and {Goto}, Tomotsugu and {Gunn}, James E. and {Harikane}, Yuichi and {Ikeda}, Hiroyuki and {Izumi}, Takuma and {Kawaguchi}, Toshihiro and {Kato}, Nanako and {Kikuta}, Satoshi and {Kohno}, Kotaro and {Komiyama}, Yutaka and {Lupton}, Robert H. and {Minezaki}, Takeo and {Miyazaki}, Satoshi and {Murayama}, Hitoshi and {Niida}, Mana and {Nishizawa}, Atsushi J. and {Noboriguchi}, Akatoki and {Oguri}, Masamune and {Ono}, Yoshiaki and {Ouchi}, Masami and {Price}, Paul A. and {Sameshima}, Hiroaki and {Schulze}, Andreas and {Shirakata}, Hikari and {Silverman}, John D. and {Sugiyama}, Naoshi and {Tait}, Philip J. and {Takada}, Masahiro and {Takata}, Tadafumi and {Tanaka}, Masayuki and {Toba}, Yoshiki and {Utsumi}, Yousuke and {Wang}, Shiang-Yu and {Yamashita}, Takuji},
        title = "{Subaru High-z  Exploration of Low-luminosity Quasars (SHELLQs). V. Quasar Luminosity Function and Contribution to Cosmic Reionization at z = 6}",
      journal = {\apj},
         year = 2018,
        month = dec,
       volume = {869},
       number = {2},
          eid = {150},
        pages = {150},
          doi = {10.3847/1538-4357/aaee7a},
archivePrefix = {arXiv},
       eprint = {1811.01963},
 primaryClass = {astro-ph.GA},
       adsurl = {https://ui.adsabs.harvard.edu/abs/2018ApJ...869..150M}
}

@ARTICLE{Ma2018a,
       author = {{Ma}, Xiangcheng and {Hopkins}, Philip F. and {Boylan-Kolchin}, Michael and {Faucher-Gigu{\`e}re}, Claude-Andr{\'e} and {Quataert}, Eliot and {Feldmann}, Robert and {Garrison-Kimmel}, Shea and {Hayward}, Christopher C. and {Kere{\v{s}}}, Du{\v{s}}an and {Wetzel}, Andrew},
        title = "{Simulating galaxies in the reionization era with FIRE-2: morphologies and sizes}",
      journal = {\mnras},
         year = 2018,
        month = jun,
       volume = {477},
       number = {1},
        pages = {219-229},
          doi = {10.1093/mnras/sty684},
archivePrefix = {arXiv},
       eprint = {1710.00008},
 primaryClass = {astro-ph.GA},
       adsurl = {https://ui.adsabs.harvard.edu/abs/2018MNRAS.477..219M}
}

@ARTICLE{Ma2018b,
       author = {{Ma}, Xiangcheng and {Hopkins}, Philip F. and {Garrison-Kimmel}, Shea and {Faucher-Gigu{\`e}re}, Claude-Andr{\'e} and {Quataert}, Eliot and {Boylan-Kolchin}, Michael and {Hayward}, Christopher C. and {Feldmann}, Robert and {Kere{\v{s}}}, Du{\v{s}}an},
        title = "{Simulating galaxies in the reionization era with FIRE-2: galaxy scaling relations, stellar mass functions, and luminosity functions}",
      journal = {\mnras},
         year = 2018,
        month = aug,
       volume = {478},
       number = {2},
        pages = {1694-1715},
          doi = {10.1093/mnras/sty1024},
archivePrefix = {arXiv},
       eprint = {1706.06605},
 primaryClass = {astro-ph.GA},
       adsurl = {https://ui.adsabs.harvard.edu/abs/2018MNRAS.478.1694M}
}

@ARTICLE{Mason2023,
       author = {{Mason}, Charlotte A. and {Trenti}, Michele and {Treu}, Tommaso},
        title = "{The brightest galaxies at cosmic dawn}",
      journal = {\mnras},
         year = 2023,
        month = may,
       volume = {521},
       number = {1},
        pages = {497-503},
          doi = {10.1093/mnras/stad035},
archivePrefix = {arXiv},
       eprint = {2207.14808},
 primaryClass = {astro-ph.GA},
       adsurl = {https://ui.adsabs.harvard.edu/abs/2023MNRAS.521..497M}
}

@ARTICLE{Maiolino2024a,
       author = {{Maiolino}, Roberto and {Scholtz}, Jan and {Witstok}, Joris and {Carniani}, Stefano and {D'Eugenio}, Francesco and {de Graaff}, Anna and {{\"U}bler}, Hannah and {Tacchella}, Sandro and {Curtis-Lake}, Emma and {Arribas}, Santiago and {Bunker}, Andrew and {Charlot}, St{\'e}phane and {Chevallard}, Jacopo and {Curti}, Mirko and {Looser}, Tobias J. and {Maseda}, Michael V. and {Rawle}, Timothy D. and {Rodr{\'\i}guez del Pino}, Bruno and {Willott}, Chris J. and {Egami}, Eiichi and {Eisenstein}, Daniel J. and {Hainline}, Kevin N. and {Robertson}, Brant and {Williams}, Christina C. and {Willmer}, Christopher N.~A. and {Baker}, William M. and {Boyett}, Kristan and {DeCoursey}, Christa and {Fabian}, Andrew C. and {Helton}, Jakob M. and {Ji}, Zhiyuan and {Jones}, Gareth C. and {Kumari}, Nimisha and {Laporte}, Nicolas and {Nelson}, Erica J. and {Perna}, Michele and {Sandles}, Lester and {Shivaei}, Irene and {Sun}, Fengwu},
        title = "{A small and vigorous black hole in the early Universe}",
      journal = {\nat},
         year = 2024,
        month = mar,
       volume = {627},
       number = {8002},
        pages = {59-63},
          doi = {10.1038/s41586-024-07052-5},
archivePrefix = {arXiv},
       eprint = {2305.12492},
 primaryClass = {astro-ph.GA},
       adsurl = {https://ui.adsabs.harvard.edu/abs/2024Natur.627...59M}
}

@ARTICLE{Meyer24,
       author = {{Meyer}, R.~A. and {Oesch}, P.~A. and {Giovinazzo}, E. and {Weibel}, A. and {Brammer}, G. and {Matthee}, J. and {Naidu}, R.~P. and {Bouwens}, R.~J. and {Chisholm}, J. and {Covelo-Paz}, A. and {Fudamoto}, Y. and {Maseda}, M. and {Nelson}, E. and {Shivaei}, I. and {Xiao}, M. and {Herard-Demanche}, T. and {Illingworth}, G.~D. and {Kerutt}, J. and {Kramarenko}, I. and {Labbe}, I. and {Leonova}, E. and {Magee}, D. and {Matharu}, J. and {Prieto Lyon}, G. and {Reddy}, N. and {Schaerer}, D. and {Shapley}, A. and {Stefanon}, M. and {Wozniak}, M.~A. and {Wuyts}, S.},
        title = "{JWST FRESCO: a comprehensive census of H {\ensuremath{\beta}} + [O III] emitters at 6.8 < z < 9.0 in the GOODS fields}",
      journal = {\mnras},
         year = 2024,
        month = nov,
       volume = {535},
       number = {1},
        pages = {1067-1094},
          doi = {10.1093/mnras/stae2353},
archivePrefix = {arXiv},
       eprint = {2405.05111},
 primaryClass = {astro-ph.GA},
       adsurl = {https://ui.adsabs.harvard.edu/abs/2024MNRAS.535.1067M}
}

@ARTICLE{McLure2013,
       author = {{McLure}, R.~J. and {Dunlop}, J.~S. and {Bowler}, R.~A.~A. and {Curtis-Lake}, E. and {Schenker}, M. and {Ellis}, R.~S. and {Robertson}, B.~E. and {Koekemoer}, A.~M. and {Rogers}, A.~B. and {Ono}, Y. and {Ouchi}, M. and {Charlot}, S. and {Wild}, V. and {Stark}, D.~P. and {Furlanetto}, S.~R. and {Cirasuolo}, M. and {Targett}, T.~A.},
        title = "{A new multifield determination of the galaxy luminosity function at z = 7-9 incorporating the 2012 Hubble Ultra-Deep Field imaging}",
      journal = {\mnras},
         year = 2013,
        month = jul,
       volume = {432},
       number = {4},
        pages = {2696-2716},
          doi = {10.1093/mnras/stt627},
archivePrefix = {arXiv},
       eprint = {1212.5222},
 primaryClass = {astro-ph.CO},
       adsurl = {https://ui.adsabs.harvard.edu/abs/2013MNRAS.432.2696M}
}

@ARTICLE{McLeod2024,
       author = {{McLeod}, D.~J. and {Donnan}, C.~T. and {McLure}, R.~J. and {Dunlop}, J.~S. and {Magee}, D. and {Begley}, R. and {Carnall}, A.~C. and {Cullen}, F. and {Ellis}, R.~S. and {Hamadouche}, M.~L. and {Stanton}, T.~M.},
        title = "{The galaxy UV luminosity function at z ≃ 11 from a suite of public JWST ERS, ERO, and Cycle-1 programs}",
      journal = {\mnras},
         year = 2024,
        month = jan,
       volume = {527},
       number = {3},
        pages = {5004-5022},
          doi = {10.1093/mnras/stad3471},
archivePrefix = {arXiv},
       eprint = {2304.14469},
 primaryClass = {astro-ph.GA},
       adsurl = {https://ui.adsabs.harvard.edu/abs/2024MNRAS.527.5004M}
}

@ARTICLE{McLeod2026,
       author = {{McLeod}, D.~J. and {Dunlop}, J.~S. and {McLure}, R.~J. and {Donnan}, C.~T. and {Begley}, R. and {Antonogiannaki}, S. and {Magee}, D. and {Illingworth}, G.~D. and {Arrabal Haro}, P. and {Bondestam}, C. and {Carnall}, A.~C. and {Cullen}, F. and {Dickinson}, M. and {Ellis}, R.~S. and {Frye}, B.~L. and {Golawska}, H. and {Grogin}, N.~A. and {Holst}, I.~J.~B. and {Kamieneski}, P.~S. and {Leung}, H.-H. and {Liu}, F.-Y. and {Stanton}, T.~M. and {Tittley}, E.~R.},
        title = "{A search for the first galaxies across $>0.6$ deg$^2$ of JWST imaging: new evidence for a rapid decline in star-formation activity at $z>12$}",
      journal = {arXiv e-prints},
         year = 2026,
        month = apr,
          eid = {arXiv:2604.16666},
        pages = {arXiv:2604.16666},
archivePrefix = {arXiv},
       eprint = {2604.16666},
 primaryClass = {astro-ph.GA},
       adsurl = {https://ui.adsabs.harvard.edu/abs/2026arXiv260416666M}
}

@ARTICLE{Michaux2021,
       author = {{Michaux}, Micha{\"e}l and {Hahn}, Oliver and {Rampf}, Cornelius and {Angulo}, Raul E.},
        title = "{Accurate initial conditions for cosmological N-body simulations: minimizing truncation and discreteness errors}",
      journal = {\mnras},
         year = 2021,
        month = jan,
       volume = {500},
       number = {1},
        pages = {663-683},
          doi = {10.1093/mnras/staa3149},
archivePrefix = {arXiv},
       eprint = {2008.09588},
 primaryClass = {astro-ph.CO},
       adsurl = {https://ui.adsabs.harvard.edu/abs/2021MNRAS.500..663M}
}

@ARTICLE{Maiolino2024b,
       author = {{Maiolino}, Roberto and {Scholtz}, Jan and {Curtis-Lake}, Emma and {Carniani}, Stefano and {Baker}, William and {de Graaff}, Anna and {Tacchella}, Sandro and {{\"U}bler}, Hannah and {D'Eugenio}, Francesco and {Witstok}, Joris and {Curti}, Mirko and {Arribas}, Santiago and {Bunker}, Andrew J. and {Charlot}, St{\'e}phane and {Chevallard}, Jacopo and {Eisenstein}, Daniel J. and {Egami}, Eiichi and {Ji}, Zhiyuan and {Jones}, Gareth C. and {Lyu}, Jianwei and {Rawle}, Tim and {Robertson}, Brant and {Rujopakarn}, Wiphu and {Perna}, Michele and {Sun}, Fengwu and {Venturi}, Giacomo and {Williams}, Christina C. and {Willott}, Chris},
        title = "{JADES: The diverse population of infant black holes at 4 < z < 11: Merging, tiny, poor, but mighty}",
      journal = {\aap},
         year = 2024,
        month = nov,
       volume = {691},
          eid = {A145},
        pages = {A145},
          doi = {10.1051/0004-6361/202347640},
archivePrefix = {arXiv},
       eprint = {2308.01230},
 primaryClass = {astro-ph.GA},
       adsurl = {https://ui.adsabs.harvard.edu/abs/2024A&A...691A.145M}
}

@ARTICLE{McGibbon2025,
       author = {{McGibbon}, Robert and {Helly}, John and {Schaye}, Joop and {Schaller}, Matthieu and {Vandenbroucke}, Bert},
        title = "{SOAP: A Python Package for Calculating the Properties of Galaxies and Halos Formed in Cosmological Simulations}",
      journal = {The Journal of Open Source Software},
         year = 2025,
        month = jul,
       volume = {10},
       number = {111},
          eid = {8252},
        pages = {8252},
          doi = {10.21105/joss.08252},
archivePrefix = {arXiv},
       eprint = {2507.22669},
 primaryClass = {astro-ph.IM},
       adsurl = {https://ui.adsabs.harvard.edu/abs/2025JOSS...10.8252M}
}

@ARTICLE{Morishita2018,
       author = {{Morishita}, T. and {Trenti}, M. and {Stiavelli}, M. and {Bradley}, L.~D. and {Coe}, D. and {Oesch}, P.~A. and {Mason}, C.~A. and {Bridge}, J.~S. and {Holwerda}, B.~W. and {Livermore}, R.~C. and {Salmon}, B. and {Schmidt}, K.~B. and {Shull}, J.~M. and {Treu}, T.},
        title = "{The Bright-end Galaxy Candidates at z {\ensuremath{\sim}} 9 from 79 Independent HST Fields}",
      journal = {\apj},
         year = 2018,
        month = nov,
       volume = {867},
       number = {2},
          eid = {150},
        pages = {150},
          doi = {10.3847/1538-4357/aae68c},
archivePrefix = {arXiv},
       eprint = {1809.07604},
 primaryClass = {astro-ph.GA},
       adsurl = {https://ui.adsabs.harvard.edu/abs/2018ApJ...867..150M}
}

@ARTICLE{Mitchell2013,
       author = {{Mitchell}, Peter D. and {Lacey}, Cedric G. and {Baugh}, Carlton M. and {Cole}, Shaun},
        title = "{How well can we really estimate the stellar masses of galaxies from broad-band photometry?}",
      journal = {\mnras},
         year = 2013,
        month = oct,
       volume = {435},
       number = {1},
        pages = {87-114},
          doi = {10.1093/mnras/stt1280},
archivePrefix = {arXiv},
       eprint = {1303.7228},
 primaryClass = {astro-ph.CO},
       adsurl = {https://ui.adsabs.harvard.edu/abs/2013MNRAS.435...87M}
}

@ARTICLE{Madau2014,
       author = {{Madau}, Piero and {Dickinson}, Mark},
        title = "{Cosmic Star-Formation History}",
      journal = {\araa},
         year = 2014,
        month = aug,
       volume = {52},
        pages = {415-486},
          doi = {10.1146/annurev-astro-081811-125615},
archivePrefix = {arXiv},
       eprint = {1403.0007},
 primaryClass = {astro-ph.CO},
       adsurl = {https://ui.adsabs.harvard.edu/abs/2014ARA&A..52..415M}
}

@ARTICLE{Mitsuhashi2026,
       author = {{Mitsuhashi}, Ikki and {Zavala}, Jorge A. and {Bakx}, Tom J.~L.~C. and {Inoue}, Akio K. and {Castellano}, Marco and {Calabr{\`o}}, Antonello and {Casey}, Caitlin M. and {Franco}, Maximilien and {Hatsukade}, Bunyo and {Hathi}, Nimish P. and {Ikeda}, Ryota and {Koekemoer}, Anton M. and {Kartaltepe}, Jeyhan and {Knudsen}, Kirsten K. and {Santini}, Paola and {Saito}, Toshiki and {Terlevich}, Elena and {Terlevich}, Roberto and {Yung}, L.~Y. Aaron},
        title = "{Low Dust Mass and High Star Formation Efficiency at z > 12 from Deep ALMA Observations}",
      journal = {\apj},
         year = 2026,
        month = apr,
       volume = {1000},
       number = {2},
          eid = {159},
        pages = {159},
          doi = {10.3847/1538-4357/ae4511},
archivePrefix = {arXiv},
       eprint = {2501.19384},
 primaryClass = {astro-ph.GA},
       adsurl = {https://ui.adsabs.harvard.edu/abs/2026ApJ..1000..159M}
}

@ARTICLE{Naidu2026,
       author = {{Naidu}, Rohan P. and {Oesch}, Pascal A. and {Brammer}, Gabriel and {Weibel}, Andrea and {Li}, Yijia and {Matthee}, Jorryt and {Chisolm}, John and {Pollock}, Clara L. and {Heintz}, Kasper E. and {Johnson}, Benjamin D. and {Shen}, Xuejian and {Hviding}, Raphael E. and {Leja}, Joel and {Tacchella}, Sandro and {Ganguly}, Arpita and {Witten}, Callum and {Atek}, Hakim and {Belli}, Siro and {Bose}, Sownak and {Bouwens}, Rychard and {Dayal}, Pratika and {Decarli}, Roberto and {de Graaff}, Anna and {Fudamoto}, Yoshinobu and {Giovinazzo}, Emma and {Greene}, Jenny E. and {Illingworth}, Garth and {Inoue}, Akio K. and {Kane}, Sarah G. and {Labbe}, Ivo and {Leonova}, Ecaterina and {Marques-Chaves}, Rui and {Meyer}, Roman A. and {Nelson}, Erica J. and {Roberts-Borsani}, Guido and {Schaerer}, Daniel and {Simcoe}, Robert A. and {Stefanon}, Mauro and {Sugahara}, Yuma and {Toft}, Sune and {van der Wel}, Arjen and {van Dokkum}, Pieter and {Walter}, Fabian and {Watson}, Darrach and {Weaver}, John R. and {Whitaker}, Katherine E.},
        title = "{A Cosmic Miracle: A Remarkably Luminous Galaxy at zspec = 14.44 Confirmed with JWST}",
      journal = {The Open Journal of Astrophysics},
         year = 2026,
        month = jan,
       volume = {9},
        pages = {56033},
          doi = {10.33232/001c.156033},
archivePrefix = {arXiv},
       eprint = {2505.11263},
 primaryClass = {astro-ph.GA},
       adsurl = {https://ui.adsabs.harvard.edu/abs/2026OJAp....956033N}
}

@ARTICLE{Nelson2019,
       author = {{Nelson}, Dylan and {Springel}, Volker and {Pillepich}, Annalisa and {Rodriguez-Gomez}, Vicente and {Torrey}, Paul and {Genel}, Shy and {Vogelsberger}, Mark and {Pakmor}, Ruediger and {Marinacci}, Federico and {Weinberger}, Rainer and {Kelley}, Luke and {Lovell}, Mark and {Diemer}, Benedikt and {Hernquist}, Lars},
        title = "{The IllustrisTNG simulations: public data release}",
      journal = {Computational Astrophysics and Cosmology},
         year = 2019,
        month = may,
       volume = {6},
       number = {1},
          eid = {2},
        pages = {2},
          doi = {10.1186/s40668-019-0028-x},
archivePrefix = {arXiv},
       eprint = {1812.05609},
 primaryClass = {astro-ph.GA},
       adsurl = {https://ui.adsabs.harvard.edu/abs/2019ComAC...6....2N}
}

@ARTICLE{Nobels2024,
       author = {{Nobels}, Folkert S.~J. and {Schaye}, Joop and {Schaller}, Matthieu and {Ploeckinger}, Sylvia and {Chaikin}, Evgenii and {Richings}, Alexander J.},
        title = "{Tests of subgrid models for star formation using simulations of isolated disc galaxies}",
      journal = {\mnras},
         year = 2024,
        month = aug,
       volume = {532},
       number = {3},
        pages = {3299-3321},
          doi = {10.1093/mnras/stae1390},
archivePrefix = {arXiv},
       eprint = {2309.13750},
 primaryClass = {astro-ph.GA},
       adsurl = {https://ui.adsabs.harvard.edu/abs/2024MNRAS.532.3299N}
}

@ARTICLE{Oesch2014,
       author = {{Oesch}, P.~A. and {Bouwens}, R.~J. and {Illingworth}, G.~D. and {Labb{\'e}}, I. and {Smit}, R. and {Franx}, M. and {van Dokkum}, P.~G. and {Momcheva}, I. and {Ashby}, M.~L.~N. and {Fazio}, G.~G. and {Huang}, J. -S. and {Willner}, S.~P. and {Gonzalez}, V. and {Magee}, D. and {Trenti}, M. and {Brammer}, G.~B. and {Skelton}, R.~E. and {Spitler}, L.~R.},
        title = "{The Most Luminous z \raisebox{-0.5ex}\textasciitilde 9-10 Galaxy Candidates Yet Found: The Luminosity Function, Cosmic Star-formation Rate, and the First Mass Density Estimate at 500 Myr}",
      journal = {\apj},
         year = 2014,
        month = may,
       volume = {786},
       number = {2},
          eid = {108},
        pages = {108},
          doi = {10.1088/0004-637X/786/2/108},
archivePrefix = {arXiv},
       eprint = {1309.2280},
 primaryClass = {astro-ph.CO},
       adsurl = {https://ui.adsabs.harvard.edu/abs/2014ApJ...786..108O}
}

@ARTICLE{Oesch2016,
       author = {{Oesch}, P.~A. and {Brammer}, G. and {van Dokkum}, P.~G. and {Illingworth}, G.~D. and {Bouwens}, R.~J. and {Labb{\'e}}, I. and {Franx}, M. and {Momcheva}, I. and {Ashby}, M.~L.~N. and {Fazio}, G.~G. and {Gonzalez}, V. and {Holden}, B. and {Magee}, D. and {Skelton}, R.~E. and {Smit}, R. and {Spitler}, L.~R. and {Trenti}, M. and {Willner}, S.~P.},
        title = "{A Remarkably Luminous Galaxy at z=11.1 Measured with Hubble Space Telescope Grism Spectroscopy}",
      journal = {\apj},
         year = 2016,
        month = mar,
       volume = {819},
       number = {2},
          eid = {129},
        pages = {129},
          doi = {10.3847/0004-637X/819/2/129},
archivePrefix = {arXiv},
       eprint = {1603.00461},
 primaryClass = {astro-ph.GA},
       adsurl = {https://ui.adsabs.harvard.edu/abs/2016ApJ...819..129O}
}

@ARTICLE{Oesch2018,
       author = {{Oesch}, P.~A. and {Bouwens}, R.~J. and {Illingworth}, G.~D. and {Labb{\'e}}, I. and {Stefanon}, M.},
        title = "{The Dearth of z {\ensuremath{\sim}} 10 Galaxies in All HST Legacy Fields{\textemdash}The Rapid Evolution of the Galaxy Population in the First 500 Myr}",
      journal = {\apj},
         year = 2018,
        month = mar,
       volume = {855},
       number = {2},
          eid = {105},
        pages = {105},
          doi = {10.3847/1538-4357/aab03f},
archivePrefix = {arXiv},
       eprint = {1710.11131},
 primaryClass = {astro-ph.GA},
       adsurl = {https://ui.adsabs.harvard.edu/abs/2018ApJ...855..105O}
}

@ARTICLE{Page2000,
       author = {{Page}, M.~J. and {Carrera}, F.~J.},
        title = "{An improved method of constructing binned luminosity functions}",
      journal = {\mnras},
         year = 2000,
        month = jan,
       volume = {311},
       number = {2},
        pages = {433-440},
          doi = {10.1046/j.1365-8711.2000.03105.x},
archivePrefix = {arXiv},
       eprint = {astro-ph/9909434},
 primaryClass = {astro-ph},
       adsurl = {https://ui.adsabs.harvard.edu/abs/2000MNRAS.311..433P}
}

@ARTICLE{Perez-Gonzalez2023,
       author = {{P{\'e}rez-Gonz{\'a}lez}, Pablo G. and {Costantin}, Luca and {Langeroodi}, Danial and {Rinaldi}, Pierluigi and {Annunziatella}, Marianna and {Ilbert}, Olivier and {Colina}, Luis and {N{\o}rgaard-Nielsen}, Hans Ulrik and {Greve}, Thomas R. and {{\"O}stlin}, G{\"o}ran and {Wright}, Gillian and {Alonso-Herrero}, Almudena and {{\'A}lvarez-M{\'a}rquez}, Javier and {Caputi}, Karina I. and {Eckart}, Andreas and {Le F{\`e}vre}, Olivier and {Labiano}, {\'A}lvaro and {Garc{\'\i}a-Mar{\'\i}n}, Macarena and {Hjorth}, Jens and {Kendrew}, Sarah and {Pye}, John P. and {Tikkanen}, Tuomo and {van der Werf}, Paul and {Walter}, Fabian and {Ward}, Martin and {Bik}, Arjan and {Boogaard}, Leindert and {Bosman}, Sarah E.~I. and {G{\'o}mez}, Alejandro Crespo and {Gillman}, Steven and {Iani}, Edoardo and {Jermann}, Iris and {Melinder}, Jens and {Meyer}, Romain A. and {Moutard}, Thibaud and {van Dishoek}, Ewine and {Henning}, Thomas and {Lagage}, Pierre-Olivier and {Guedel}, Manuel and {Peissker}, Florian and {Ray}, Tom and {Vandenbussche}, Bart and {Garc{\'\i}a-Argum{\'a}nez}, {\'A}ngela and {Mar{\'\i}a M{\'e}rida}, Rosa},
        title = "{Life beyond 30: Probing the -20 < M $_{UV}$ < -17 Luminosity Function at 8 < z < 13 with the NIRCam Parallel Field of the MIRI Deep Survey}",
      journal = {\apjl},
         year = 2023,
        month = jul,
       volume = {951},
       number = {1},
          eid = {L1},
        pages = {L1},
          doi = {10.3847/2041-8213/acd9d0},
archivePrefix = {arXiv},
       eprint = {2302.02429},
 primaryClass = {astro-ph.GA},
       adsurl = {https://ui.adsabs.harvard.edu/abs/2023ApJ...951L...1P}
}

@ARTICLE{Prada2026,
       author = {{Prada}, Francisco and {Behroozi}, Peter and {Ishiyama}, Tomoaki and {P{\'e}rez}, Enrique and {Klypin}, Anatoly and {Wang}, Xin and {Ruedas}, Jos{\'e}},
        title = "{Galaxies at z > 10: {\ensuremath{\Lambda}}CDM predicts increased Star Formation Efficiency}",
      journal = {arXiv e-prints},
         year = 2026,
        month = apr,
          eid = {arXiv:2604.18683},
        pages = {arXiv:2604.18683},
          doi = {10.48550/arXiv.2604.18683},
archivePrefix = {arXiv},
       eprint = {2604.18683},
 primaryClass = {astro-ph.GA},
       adsurl = {https://ui.adsabs.harvard.edu/abs/2026arXiv260418683P}
}

@ARTICLE{Pacifici2023,
       author = {{Pacifici}, Camilla and {Iyer}, Kartheik G. and {Mobasher}, Bahram and {da Cunha}, Elisabete and {Acquaviva}, Viviana and {Burgarella}, Denis and {Calistro Rivera}, Gabriela and {Carnall}, Adam C. and {Chang}, Yu-Yen and {Chartab}, Nima and {Cooke}, Kevin C. and {Fairhurst}, Ciaran and {Kartaltepe}, Jeyhan and {Leja}, Joel and {Ma{\l}ek}, Katarzyna and {Salmon}, Brett and {Torelli}, Marianna and {Vidal-Garc{\'\i}a}, Alba and {Boquien}, M{\'e}d{\'e}ric and {Brammer}, Gabriel G. and {Brown}, Michael J.~I. and {Capak}, Peter L. and {Chevallard}, Jacopo and {Circosta}, Chiara and {Croton}, Darren and {Davidzon}, Iary and {Dickinson}, Mark and {Duncan}, Kenneth J. and {Faber}, Sandra M. and {Ferguson}, Harry C. and {Fontana}, Adriano and {Guo}, Yicheng and {Haeussler}, Boris and {Hemmati}, Shoubaneh and {Jafariyazani}, Marziye and {Kassin}, Susan A. and {Larson}, Rebecca L. and {Lee}, Bomee and {Mantha}, Kameswara Bharadwaj and {Marchi}, Francesca and {Nayyeri}, Hooshang and {Newman}, Jeffrey A. and {Pandya}, Viraj and {Pforr}, Janine and {Reddy}, Naveen and {Sanders}, Ryan and {Shah}, Ekta and {Shahidi}, Abtin and {Stevans}, Matthew L. and {Triani}, Dian Puspita and {Tyler}, Krystal D. and {Vanderhoof}, Brittany N. and {de la Vega}, Alexander and {Wang}, Weichen and {Weston}, Madalyn E.},
        title = "{The Art of Measuring Physical Parameters in Galaxies: A Critical Assessment of Spectral Energy Distribution Fitting Techniques}",
      journal = {\apj},
         year = 2023,
        month = feb,
       volume = {944},
       number = {2},
          eid = {141},
        pages = {141},
          doi = {10.3847/1538-4357/acacff},
archivePrefix = {arXiv},
       eprint = {2212.01915},
 primaryClass = {astro-ph.GA},
       adsurl = {https://ui.adsabs.harvard.edu/abs/2023ApJ...944..141P}
}

@ARTICLE{Pakmor2023,
       author = {{Pakmor}, R{\"u}diger and {Springel}, Volker and {Coles}, Jonathan P. and {Guillet}, Thomas and {Pfrommer}, Christoph and {Bose}, Sownak and {Barrera}, Monica and {Delgado}, Ana Maria and {Ferlito}, Fulvio and {Frenk}, Carlos and {Hadzhiyska}, Boryana and {Hern{\'a}ndez-Aguayo}, C{\'e}sar and {Hernquist}, Lars and {Kannan}, Rahul and {White}, Simon D.~M.},
        title = "{The MillenniumTNG Project: the hydrodynamical full physics simulation and a first look at its galaxy clusters}",
      journal = {\mnras},
         year = 2023,
        month = sep,
       volume = {524},
       number = {2},
        pages = {2539-2555},
          doi = {10.1093/mnras/stac3620},
archivePrefix = {arXiv},
       eprint = {2210.10060},
 primaryClass = {astro-ph.CO},
       adsurl = {https://ui.adsabs.harvard.edu/abs/2023MNRAS.524.2539P}
}

@ARTICLE{Raiter2010,
       author = {{Raiter}, A. and {Fosbury}, R.~A.~E. and {Teimoorinia}, H.},
        title = "{Ly{\ensuremath{\alpha}} emitters in the GOODS-S field. A powerful pure nebular SED with N IV] emission at z = 5.563}",
      journal = {\aap},
         year = 2010,
        month = feb,
       volume = {510},
          eid = {A109},
        pages = {A109},
          doi = {10.1051/0004-6361/200912429},
archivePrefix = {arXiv},
       eprint = {0912.4305},
 primaryClass = {astro-ph.CO},
       adsurl = {https://ui.adsabs.harvard.edu/abs/2010A&A...510A.109R}
}

@ARTICLE{Robertson2015,
       author = {{Robertson}, Brant E. and {Ellis}, Richard S. and {Furlanetto}, Steven R. and {Dunlop}, James S.},
        title = "{Cosmic Reionization and Early Star-forming Galaxies: A Joint Analysis of New Constraints from Planck and the Hubble Space Telescope}",
      journal = {\apjl},
         year = 2015,
        month = apr,
       volume = {802},
       number = {2},
          eid = {L19},
        pages = {L19},
          doi = {10.1088/2041-8205/802/2/L19},
archivePrefix = {arXiv},
       eprint = {1502.02024},
 primaryClass = {astro-ph.CO},
       adsurl = {https://ui.adsabs.harvard.edu/abs/2015ApJ...802L..19R}
}

@ARTICLE{Robertson2024,
       author = {{Robertson}, Brant and {Johnson}, Benjamin D. and {Tacchella}, Sandro and {Eisenstein}, Daniel J. and {Hainline}, Kevin and {Arribas}, Santiago and {Baker}, William M. and {Bunker}, Andrew J. and {Carniani}, Stefano and {Cargile}, Phillip A. and {Carreira}, Courtney and {Charlot}, Stephane and {Chevallard}, Jacopo and {Curti}, Mirko and {Curtis-Lake}, Emma and {D'Eugenio}, Francesco and {Egami}, Eiichi and {Hausen}, Ryan and {Helton}, Jakob M. and {Jakobsen}, Peter and {Ji}, Zhiyuan and {Jones}, Gareth C. and {Maiolino}, Roberto and {Maseda}, Michael V. and {Nelson}, Erica and {P{\'e}rez-Gonz{\'a}lez}, Pablo G. and {Pusk{\'a}s}, D{\'a}vid and {Rieke}, Marcia and {Smit}, Renske and {Sun}, Fengwu and {{\"U}bler}, Hannah and {Whitler}, Lily and {Williams}, Christina C. and {Willmer}, Christopher N.~A. and {Willott}, Chris and {Witstok}, Joris},
        title = "{Earliest Galaxies in the JADES Origins Field: Luminosity Function and Cosmic Star Formation Rate Density 300 Myr after the Big Bang}",
      journal = {\apj},
         year = 2024,
        month = jul,
       volume = {970},
       number = {1},
          eid = {31},
        pages = {31},
          doi = {10.3847/1538-4357/ad463d},
archivePrefix = {arXiv},
       eprint = {2312.10033},
 primaryClass = {astro-ph.GA},
       adsurl = {https://ui.adsabs.harvard.edu/abs/2024ApJ...970...31R}
}

@ARTICLE{Roberts-Borsani2016,
       author = {{Roberts-Borsani}, G.~W. and {Bouwens}, R.~J. and {Oesch}, P.~A. and {Labbe}, I. and {Smit}, R. and {Illingworth}, G.~D. and {van Dokkum}, P. and {Holden}, B. and {Gonzalez}, V. and {Stefanon}, M. and {Holwerda}, B. and {Wilkins}, S.},
        title = "{z {\ensuremath{\gtrsim}} 7 Galaxies with Red Spitzer/IRAC [3.6]-[4.5] Colors in the Full CANDELS Data Set: The Brightest-Known Galaxies at z \raisebox{-0.5ex}\textasciitilde 7-9 and a Probable Spectroscopic Confirmation at z = 7.48}",
      journal = {\apj},
         year = 2016,
        month = jun,
       volume = {823},
       number = {2},
          eid = {143},
        pages = {143},
          doi = {10.3847/0004-637X/823/2/143},
archivePrefix = {arXiv},
       eprint = {1506.00854},
 primaryClass = {astro-ph.GA},
       adsurl = {https://ui.adsabs.harvard.edu/abs/2016ApJ...823..143R}
}

@ARTICLE{Roberts-Borsani2020,
       author = {{Roberts-Borsani}, G.~W. and {Ellis}, R.~S. and {Laporte}, N.},
        title = "{Interpreting the Spitzer/IRAC colours of 7 {\ensuremath{\leq}} z {\ensuremath{\leq}} 9 galaxies: distinguishing between line emission and starlight using ALMA}",
      journal = {\mnras},
         year = 2020,
        month = sep,
       volume = {497},
       number = {3},
        pages = {3440-3450},
          doi = {10.1093/mnras/staa2085},
archivePrefix = {arXiv},
       eprint = {2002.02968},
 primaryClass = {astro-ph.GA},
       adsurl = {https://ui.adsabs.harvard.edu/abs/2020MNRAS.497.3440R}
}

@ARTICLE{Romano2017,
       author = {{Romano}, D. and {Matteucci}, F. and {Zhang}, Z.-Y. and {Papadopoulos}, P.~P. and {Ivison}, R.~J.},
        title = "{The evolution of CNO isotopes: a new window on cosmic star formation history and the stellar IMF in the age of ALMA}",
      journal = {\mnras},
         year = 2017,
        month = sep,
       volume = {470},
       number = {1},
        pages = {401-415},
          doi = {10.1093/mnras/stx1197},
archivePrefix = {arXiv},
       eprint = {1704.06701},
 primaryClass = {astro-ph.GA},
       adsurl = {https://ui.adsabs.harvard.edu/abs/2017MNRAS.470..401R}
}

@ARTICLE{RojasRuiz2025,
       author = {{Rojas-Ruiz}, Sof{\'\i}a and {Bagley}, Micaela and {Roberts-Borsani}, Guido and {Treu}, Tommaso and {Finkelstein}, Steven L. and {Morishita}, Takahiro and {Leethochawalit}, Nicha and {Mason}, Charlotte and {Ba{\~n}ados}, Eduardo and {Trenti}, Michele and {Stiavelli}, Massimo and {Yung}, L.~Y. Aaron and {Arrabal Haro}, Pablo and {Somerville}, Rachel S. and {Soto}, Christian},
        title = "{The BoRG-JWST Survey: Abundance and Mass-to-light Ratio of Luminous z = 7─9 Galaxies from Independent Sight Lines with NIRSpec}",
      journal = {\apj},
         year = 2025,
        month = may,
       volume = {985},
       number = {1},
          eid = {80},
        pages = {80},
          doi = {10.3847/1538-4357/adc67d},
archivePrefix = {arXiv},
       eprint = {2408.00843},
 primaryClass = {astro-ph.GA},
       adsurl = {https://ui.adsabs.harvard.edu/abs/2025ApJ...985...80R}
}

@ARTICLE{Saftly2013,
       author = {{Saftly}, W. and {Camps}, P. and {Baes}, M. and {Gordon}, K.~D. and {Vandewoude}, S. and {Rahimi}, A. and {Stalevski}, M.},
        title = "{Using hierarchical octrees in Monte Carlo radiative transfer simulations}",
      journal = {\aap},
         year = 2013,
        month = jun,
       volume = {554},
          eid = {A10},
        pages = {A10},
          doi = {10.1051/0004-6361/201220854},
archivePrefix = {arXiv},
       eprint = {1304.2896},
 primaryClass = {astro-ph.IM},
       adsurl = {https://ui.adsabs.harvard.edu/abs/2013A&A...554A..10S}
}

@ARTICLE{Saftly2014,
       author = {{Saftly}, W. and {Baes}, M. and {Camps}, P.},
        title = "{Hierarchical octree and k-d tree grids for 3D radiative transfer simulations}",
      journal = {\aap},
         year = 2014,
        month = jan,
       volume = {561},
          eid = {A77},
        pages = {A77},
          doi = {10.1051/0004-6361/201322593},
archivePrefix = {arXiv},
       eprint = {1311.0705},
 primaryClass = {astro-ph.IM},
       adsurl = {https://ui.adsabs.harvard.edu/abs/2014A&A...561A..77S}
}

@ARTICLE{Salpeter1955,
       author = {{Salpeter}, Edwin E.},
        title = "{The Luminosity Function and Stellar Evolution.}",
      journal = {\apj},
         year = 1955,
        month = jan,
       volume = {121},
        pages = {161},
          doi = {10.1086/145971},
       adsurl = {https://ui.adsabs.harvard.edu/abs/1955ApJ...121..161S}
}

@ARTICLE{Schaerer2003,
       author = {{Schaerer}, D.},
        title = "{The transition from Population III to normal galaxies: Lyalpha and He II emission and the ionising properties of high redshift starburst galaxies}",
      journal = {\aap},
         year = 2003,
        month = jan,
       volume = {397},
        pages = {527-538},
          doi = {10.1051/0004-6361:20021525},
archivePrefix = {arXiv},
       eprint = {astro-ph/0210462},
 primaryClass = {astro-ph},
       adsurl = {https://ui.adsabs.harvard.edu/abs/2003A&A...397..527S}
}

@ARTICLE{Schaller2024,
       author = {{Schaller}, Matthieu and {Borrow}, Josh and {Draper}, Peter W. and {Ivkovic}, Mladen and {McAlpine}, Stuart and {Vandenbroucke}, Bert and {Bah{\'e}}, Yannick and {Chaikin}, Evgenii and {Chalk}, Aidan B.~G. and {Chan}, Tsang Keung and {Correa}, Camila and {van Daalen}, Marcel and {Elbers}, Willem and {Gonnet}, Pedro and {Hausammann}, Lo{\"\i}c and {Helly}, John and {Hu{\v{s}}ko}, Filip and {Kegerreis}, Jacob A. and {Nobels}, Folkert S.~J. and {Ploeckinger}, Sylvia and {Revaz}, Yves and {Roper}, William J. and {Ruiz-Bonilla}, Sergio and {Sandnes}, Thomas D. and {Uyttenhove}, Yolan and {Willis}, James S. and {Xiang}, Zhen},
        title = "{SWIFT: A modern highly-parallel gravity and smoothed particle hydrodynamics solver for astrophysical and cosmological applications}",
      journal = {\mnras},
         year = 2024,
        month = may,
       volume = {530},
       number = {2},
        pages = {2378-2419},
          doi = {10.1093/mnras/stae922},
archivePrefix = {arXiv},
       eprint = {2305.13380},
 primaryClass = {astro-ph.IM},
       adsurl = {https://ui.adsabs.harvard.edu/abs/2024MNRAS.530.2378S}
}

@ARTICLE{Schaye2015,
       author = {{Schaye}, Joop and {Crain}, Robert A. and {Bower}, Richard G. and {Furlong}, Michelle and {Schaller}, Matthieu and {Theuns}, Tom and {Dalla Vecchia}, Claudio and {Frenk}, Carlos S. and {McCarthy}, I.~G. and {Helly}, John C. and {Jenkins}, Adrian and {Rosas-Guevara}, Y.~M. and {White}, Simon D.~M. and {Baes}, Maarten and {Booth}, C.~M. and {Camps}, Peter and {Navarro}, Julio F. and {Qu}, Yan and {Rahmati}, Alireza and {Sawala}, Till and {Thomas}, Peter A. and {Trayford}, James},
        title = "{The EAGLE project: simulating the evolution and assembly of galaxies and their environments}",
      journal = {\mnras},
         year = 2015,
        month = jan,
       volume = {446},
       number = {1},
        pages = {521-554},
          doi = {10.1093/mnras/stu2058},
archivePrefix = {arXiv},
       eprint = {1407.7040},
 primaryClass = {astro-ph.GA},
       adsurl = {https://ui.adsabs.harvard.edu/abs/2015MNRAS.446..521S}
}

@ARTICLE{Schaye2023,
       author = {{Schaye}, Joop and {Kugel}, Roi and {Schaller}, Matthieu and {Helly}, John C. and {Braspenning}, Joey and {Elbers}, Willem and {McCarthy}, Ian G. and {van Daalen}, Marcel P. and {Vandenbroucke}, Bert and {Frenk}, Carlos S. and {Kwan}, Juliana and {Salcido}, Jaime and {Bah{\'e}}, Yannick M. and {Borrow}, Josh and {Chaikin}, Evgenii and {Hahn}, Oliver and {Hu{\v{s}}ko}, Filip and {Jenkins}, Adrian and {Lacey}, Cedric G. and {Nobels}, Folkert S.~J.},
        title = "{The FLAMINGO project: cosmological hydrodynamical simulations for large-scale structure and galaxy cluster surveys}",
      journal = {\mnras},
         year = 2023,
        month = dec,
       volume = {526},
       number = {4},
        pages = {4978-5020},
          doi = {10.1093/mnras/stad2419},
archivePrefix = {arXiv},
       eprint = {2306.04024},
 primaryClass = {astro-ph.CO},
       adsurl = {https://ui.adsabs.harvard.edu/abs/2023MNRAS.526.4978S}
}

@ARTICLE{Schenker2013,
       author = {{Schenker}, Matthew A. and {Robertson}, Brant E. and {Ellis}, Richard S. and {Ono}, Yoshiaki and {McLure}, Ross J. and {Dunlop}, James S. and {Koekemoer}, Anton and {Bowler}, Rebecca A.~A. and {Ouchi}, Masami and {Curtis-Lake}, Emma and {Rogers}, Alexander B. and {Schneider}, Evan and {Charlot}, Stephane and {Stark}, Daniel P. and {Furlanetto}, Steven R. and {Cirasuolo}, Michele},
        title = "{The UV Luminosity Function of Star-forming Galaxies via Dropout Selection at Redshifts z \raisebox{-0.5ex}\textasciitilde 7 and 8 from the 2012 Ultra Deep Field Campaign}",
      journal = {\apj},
         year = 2013,
        month = may,
       volume = {768},
       number = {2},
          eid = {196},
        pages = {196},
          doi = {10.1088/0004-637X/768/2/196},
archivePrefix = {arXiv},
       eprint = {1212.4819},
 primaryClass = {astro-ph.CO},
       adsurl = {https://ui.adsabs.harvard.edu/abs/2013ApJ...768..196S}
}

@ARTICLE{Schechter1976,
       author = {{Schechter}, P.},
        title = "{An analytic expression for the luminosity function for galaxies.}",
      journal = {\apj},
         year = 1976,
        month = jan,
       volume = {203},
        pages = {297-306},
          doi = {10.1086/154079},
       adsurl = {https://ui.adsabs.harvard.edu/abs/1976ApJ...203..297S}
}

@ARTICLE{Serjeant2023,
       author = {{Serjeant}, Stephen and {Bakx}, Tom J.~L.~C.},
        title = "{Objects in JWST's mirrors are closer than they appear}",
      journal = {Nature Astronomy},
         year = 2023,
        month = oct,
       volume = {7},
        pages = {1143-1144},
          doi = {10.1038/s41550-023-02093-8},
archivePrefix = {arXiv},
       eprint = {2308.13347},
 primaryClass = {astro-ph.GA},
       adsurl = {https://ui.adsabs.harvard.edu/abs/2023NatAs...7.1143S}
}

@ARTICLE{Shen2023,
       author = {{Shen}, Xuejian and {Vogelsberger}, Mark and {Boylan-Kolchin}, Michael and {Tacchella}, Sandro and {Kannan}, Rahul},
        title = "{The impact of UV variability on the abundance of bright galaxies at z {\ensuremath{\geq}} 9}",
      journal = {\mnras},
         year = 2023,
        month = nov,
       volume = {525},
       number = {3},
        pages = {3254-3261},
          doi = {10.1093/mnras/stad2508},
archivePrefix = {arXiv},
       eprint = {2305.05679},
 primaryClass = {astro-ph.GA},
       adsurl = {https://ui.adsabs.harvard.edu/abs/2023MNRAS.525.3254S}
}

@ARTICLE{Sneppen2022,
       author = {{Sneppen}, Albert and {Steinhardt}, Charles L. and {Hensley}, Hagan and {Jermyn}, Adam S. and {Mostafa}, Basel and {Weaver}, John R.},
        title = "{Implications of a Temperature-dependent Initial Mass Function. I. Photometric Template Fitting}",
      journal = {\apj},
         year = 2022,
        month = may,
       volume = {931},
       number = {1},
          eid = {57},
        pages = {57},
          doi = {10.3847/1538-4357/ac695e},
archivePrefix = {arXiv},
       eprint = {2205.11536},
 primaryClass = {astro-ph.GA},
       adsurl = {https://ui.adsabs.harvard.edu/abs/2022ApJ...931...57S}
}

@ARTICLE{Sun2023,
       author = {{Sun}, Guochao and {Faucher-Gigu{\`e}re}, Claude-Andr{\'e} and {Hayward}, Christopher C. and {Shen}, Xuejian and {Wetzel}, Andrew and {Cochrane}, Rachel K.},
        title = "{Bursty Star Formation Naturally Explains the Abundance of Bright Galaxies at Cosmic Dawn}",
      journal = {\apjl},
         year = 2023,
        month = oct,
       volume = {955},
       number = {2},
          eid = {L35},
        pages = {L35},
          doi = {10.3847/2041-8213/acf85a},
archivePrefix = {arXiv},
       eprint = {2307.15305},
 primaryClass = {astro-ph.GA},
       adsurl = {https://ui.adsabs.harvard.edu/abs/2023ApJ...955L..35S}
}

@ARTICLE{Somerville2025,
       author = {{Somerville}, Rachel S. and {Yung}, L.~Y. Aaron and {Lancaster}, Lachlan and {Menon}, Shyam and {Sommovigo}, Laura and {Finkelstein}, Steven L.},
        title = "{Density-modulated star formation efficiency: implications for the observed abundance of ultraviolet luminous galaxies at z > 10}",
      journal = {\mnras},
         year = 2025,
        month = dec,
       volume = {544},
       number = {4},
        pages = {3774-3798},
          doi = {10.1093/mnras/staf1824},
archivePrefix = {arXiv},
       eprint = {2505.05442},
 primaryClass = {astro-ph.GA},
       adsurl = {https://ui.adsabs.harvard.edu/abs/2025MNRAS.544.3774S}
}

@ARTICLE{Subbarao1996,
       author = {{Subbarao}, M.~U. and {Connolly}, A.~J. and {Szalay}, A.~S. and {Koo}, D.~C.},
        title = "{Luminosity Functions From Photometric Redshifts. I. Techniques}",
      journal = {\aj},
         year = 1996,
        month = sep,
       volume = {112},
        pages = {929},
          doi = {10.1086/118066},
archivePrefix = {arXiv},
       eprint = {astro-ph/9606075},
 primaryClass = {astro-ph},
       adsurl = {https://ui.adsabs.harvard.edu/abs/1996AJ....112..929S}
}

@ARTICLE{Somerville1999,
       author = {{Somerville}, Rachel S. and {Primack}, Joel R.},
        title = "{Semi-analytic modelling of galaxy formation: the local Universe}",
      journal = {\mnras},
         year = 1999,
        month = dec,
       volume = {310},
       number = {4},
        pages = {1087-1110},
          doi = {10.1046/j.1365-8711.1999.03032.x},
archivePrefix = {arXiv},
       eprint = {astro-ph/9802268},
 primaryClass = {astro-ph},
       adsurl = {https://ui.adsabs.harvard.edu/abs/1999MNRAS.310.1087S}
}

@ARTICLE{Stanway2018,
       author = {{Stanway}, E.~R. and {Eldridge}, J.~J.},
        title = "{Re-evaluating old stellar populations}",
      journal = {\mnras},
         year = 2018,
        month = sep,
       volume = {479},
       number = {1},
        pages = {75-93},
          doi = {10.1093/mnras/sty1353},
archivePrefix = {arXiv},
       eprint = {1805.08784},
 primaryClass = {astro-ph.GA},
       adsurl = {https://ui.adsabs.harvard.edu/abs/2018MNRAS.479...75S}
}

@ARTICLE{Sheth2007,
       author = {{Sheth}, Ravi K.},
        title = "{On estimating redshift and luminosity distributions in photometric redshift surveys}",
      journal = {\mnras},
         year = 2007,
        month = jun,
       volume = {378},
       number = {2},
        pages = {709-715},
          doi = {10.1111/j.1365-2966.2007.11812.x},
archivePrefix = {arXiv},
       eprint = {astro-ph/0703537},
 primaryClass = {astro-ph},
       adsurl = {https://ui.adsabs.harvard.edu/abs/2007MNRAS.378..709S}
}

@ARTICLE{Stefanon2019,
       author = {{Stefanon}, Mauro and {Labb{\'e}}, Ivo and {Bouwens}, Rychard J. and {Oesch}, Pascal and {Ashby}, Matthew L.~N. and {Caputi}, Karina I. and {Franx}, Marijn and {Fynbo}, Johan P.~U. and {Illingworth}, Garth D. and {Le F{\`e}vre}, Olivier and {Marchesini}, Danilo and {McCracken}, Henry J. and {Milvang-Jensen}, Bo and {Muzzin}, Adam and {van Dokkum}, Pieter},
        title = "{The Brightest z {\ensuremath{\gtrsim}} 8 Galaxies over the COSMOS UltraVISTA Field}",
      journal = {\apj},
         year = 2019,
        month = sep,
       volume = {883},
       number = {1},
          eid = {99},
        pages = {99},
          doi = {10.3847/1538-4357/ab3792},
archivePrefix = {arXiv},
       eprint = {1902.10713},
 primaryClass = {astro-ph.GA},
       adsurl = {https://ui.adsabs.harvard.edu/abs/2019ApJ...883...99S}
}

@ARTICLE{Trayford2026,
       author = {{Trayford}, James W. and {Schaye}, Joop and {Correa}, Camila and {Ploeckinger}, Sylvia and {Richings}, Alexander J. and {Chaikin}, Evgenii and {Schaller}, Matthieu and {Ben{\'\i}tez-Llambay}, Alejandro and {Frenk}, Carlos and {Hu{\v{s}}ko}, Filip},
        title = "{Modelling the evolution and influence of dust in cosmological simulations that include the cold phase of the interstellar medium}",
      journal = {\mnras},
         year = 2026,
        month = feb,
       volume = {545},
       number = {4},
          eid = {staf2040},
        pages = {staf2040},
          doi = {10.1093/mnras/staf2040},
archivePrefix = {arXiv},
       eprint = {2505.13056},
 primaryClass = {astro-ph.GA},
       adsurl = {https://ui.adsabs.harvard.edu/abs/2026MNRAS.545f2040T}
}

@ARTICLE{Ventura2024,
       author = {{Ventura}, Emanuele M. and {Qin}, Yuxiang and {Balu}, Sreedhar and {Wyithe}, J. Stuart B.},
        title = "{Semi-analytic modelling of Pop. III star formation and metallicity evolution - I. Impact on the UV luminosity functions at z = 9-16}",
      journal = {\mnras},
         year = 2024,
        month = mar,
       volume = {529},
       number = {1},
        pages = {628-646},
          doi = {10.1093/mnras/stae567},
archivePrefix = {arXiv},
       eprint = {2401.07396},
 primaryClass = {astro-ph.GA},
       adsurl = {https://ui.adsabs.harvard.edu/abs/2024MNRAS.529..628V}
}

@ARTICLE{Willott2024,
       author = {{Willott}, Chris J. and {Desprez}, Guillaume and {Asada}, Yoshihisa and {Sarrouh}, Ghassan T.~E. and {Abraham}, Roberto and {Brada{\v{c}}}, Maru{\v{s}}a and {Brammer}, Gabe and {Estrada-Carpenter}, Vince and {Iyer}, Kartheik G. and {Martis}, Nicholas S. and {Matharu}, Jasleen and {Mowla}, Lamiya and {Muzzin}, Adam and {Noirot}, Ga{\"e}l and {Sawicki}, Marcin and {Strait}, Victoria and {Rihtar{\v{s}}i{\v{c}}}, Gregor and {Withers}, Sunna},
        title = "{A Steep Decline in the Galaxy Space Density beyond Redshift 9 in the CANUCS UV Luminosity Function}",
      journal = {\apj},
         year = 2024,
        month = may,
       volume = {966},
       number = {1},
          eid = {74},
        pages = {74},
          doi = {10.3847/1538-4357/ad35bc},
archivePrefix = {arXiv},
       eprint = {2311.12234},
 primaryClass = {astro-ph.GA},
       adsurl = {https://ui.adsabs.harvard.edu/abs/2024ApJ...966...74W}
}

@ARTICLE{Yan2017,
       author = {{Yan}, Zhiqiang and {Jerabkova}, Tereza and {Kroupa}, Pavel},
        title = "{The optimally sampled galaxy-wide stellar initial mass function. Observational tests and the publicly available GalIMF code}",
      journal = {\aap},
         year = 2017,
        month = nov,
       volume = {607},
          eid = {A126},
        pages = {A126},
          doi = {10.1051/0004-6361/201730987},
archivePrefix = {arXiv},
       eprint = {1707.04260},
 primaryClass = {astro-ph.GA},
       adsurl = {https://ui.adsabs.harvard.edu/abs/2017A&A...607A.126Y}
}

@ARTICLE{Yung2024,
       author = {{Yung}, L.~Y. Aaron and {Somerville}, Rachel S. and {Finkelstein}, Steven L. and {Wilkins}, Stephen M. and {Gardner}, Jonathan P.},
        title = "{Are the ultra-high-redshift galaxies at z > 10 surprising in the context of standard galaxy formation models?}",
      journal = {\mnras},
         year = 2024,
        month = jan,
       volume = {527},
       number = {3},
        pages = {5929-5948},
          doi = {10.1093/mnras/stad3484},
archivePrefix = {arXiv},
       eprint = {2304.04348},
 primaryClass = {astro-ph.GA},
       adsurl = {https://ui.adsabs.harvard.edu/abs/2024MNRAS.527.5929Y}
}

\appendix

\section{Effect of random sampling}
\label{sec:random_sampling}
\begin{figure*}
\centering
\includegraphics[width=\textwidth]{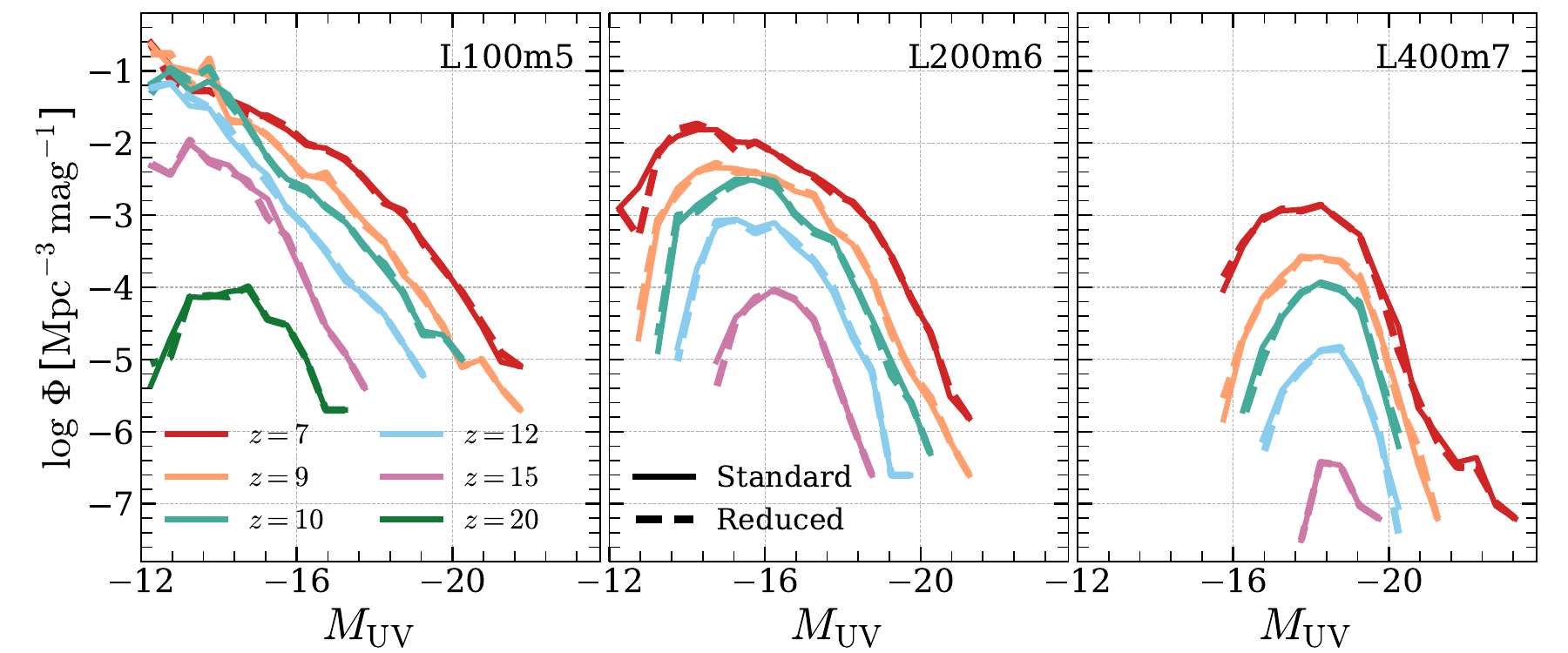}
\caption{The effect of random sampling on the UVLFs (dust-attenuated). Results for different \colibre simulations are shown in separate panels (from left to right: L100m5, L200m6 and L400m7). In each panel, the UVLFs at different redshifts are indicated by different colours. The solid curves represent the UVLFs of the {\it standard} sample (i.e., the originally selected sample; see \autoref{sec:sample}), while the dashed curves correspond to the UVLFs of the {\it reduced} sample (with smaller sample sizes; see Appendix~\ref{sec:random_sampling} for details).}
\label{fig:sampling_effect}
\end{figure*}

As mentioned in \autoref{sec:sample}, we do not perform radiative transfer on all galaxies in the \colibre simulations due to the computational cost, but instead model a representative subsample. To assess the impact of this random sampling strategy on UVLFs, we carried out the following test. We first construct a subset of the standard sample (i.e., the originally selected sample; see \autoref{sec:sample}) using the following procedure: for galaxies in the original subsample in a given stellar mass-SFR bin, if the number of galaxies exceeds 25 (half of the original target number; see \autoref{sec:sample} for details), we randomly select 25 galaxies; if it is fewer than 25, we include all (originally selected) galaxies. The sampling ratios are reduced accordingly. This new, smaller sample is referred to as the \textit{reduced sample}, relative to the \textit{standard sample} listed in \autoref{table:sample}. In \autoref{fig:sampling_effect}, we compare the UVLFs derived from the standard and reduced samples across the three \colibre simulations with different resolutions. As shown, across all redshifts from $z=7$ to $z=15$ (and also $z=20$ for L100m5), the UVLFs based on the reduced sample closely match those from the standard sample, indicating that even a smaller sample is sufficient to recover the UVLFs of the full galaxy population, highlighting the robustness and representativeness of our sampling method. This random sampling approach reduces the computational cost by approximately 99\% for L200m6 and L400m7 at $z=7$, while preserving the accuracy of the resulting luminosity functions.

\section{Sample distribution at various redshifts}
\label{sec:more_sample}
\begin{figure*}
\centering
\includegraphics[width=\textwidth]{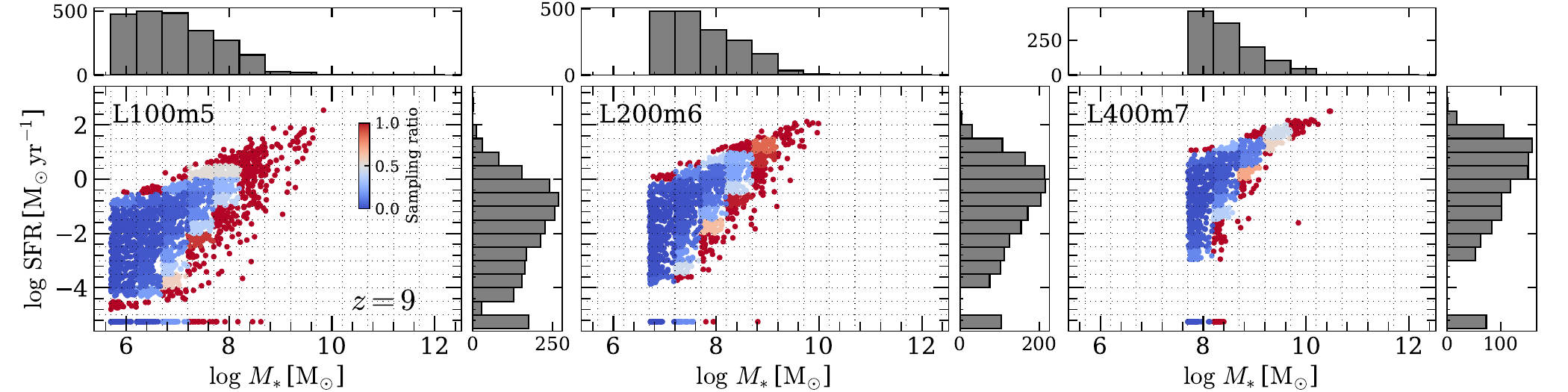}
\includegraphics[width=\textwidth]{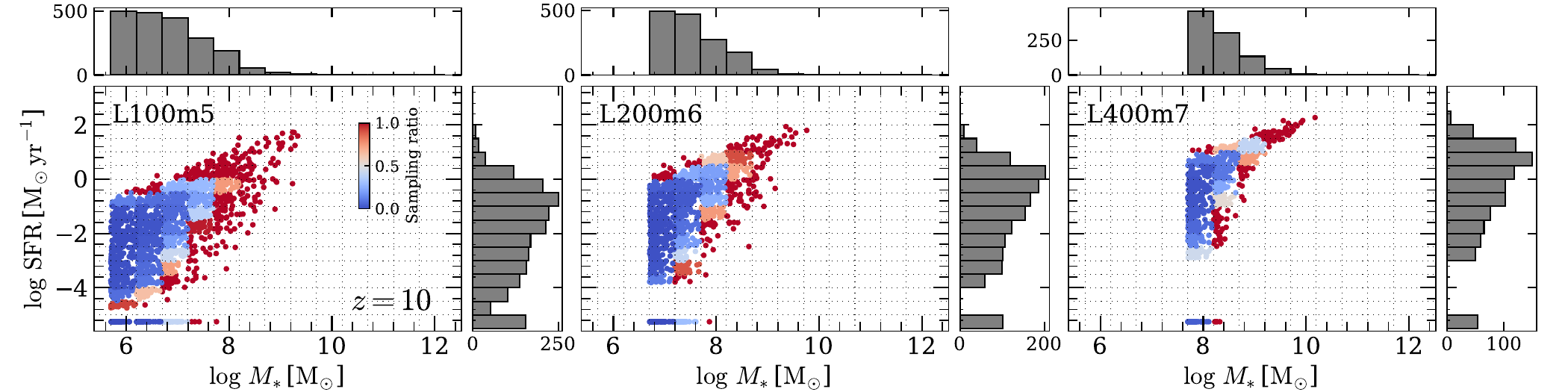}
\includegraphics[width=\textwidth]{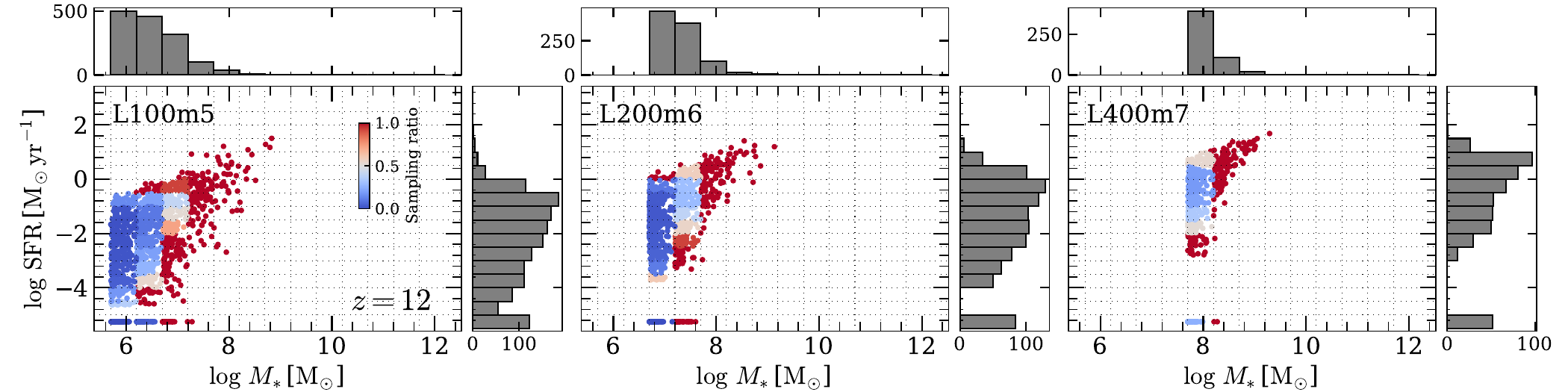}
\includegraphics[width=\textwidth]{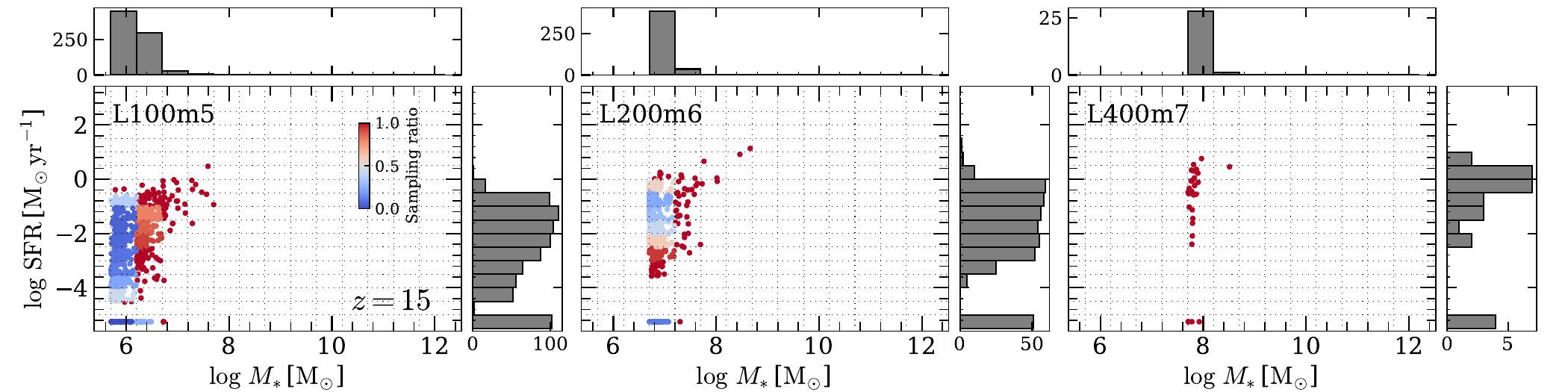}
\includegraphics[width=\textwidth]{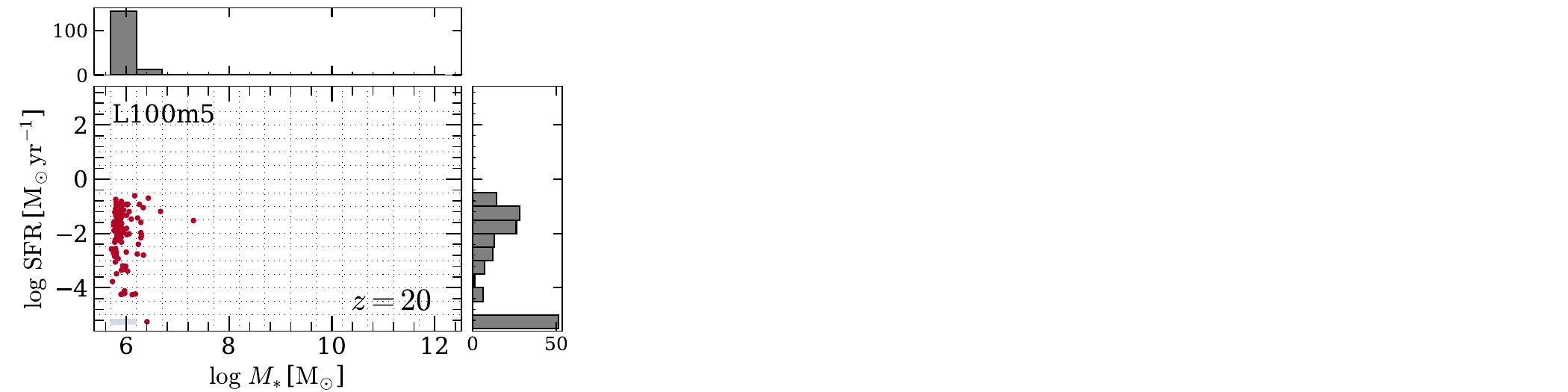}
\caption{The distribution of selected galaxies on the $\log\,M_{\ast}-\log\,\rm SFR$ plane from $z=9$ to $z=20$ (from top to bottom). The symbols are as in \autoref{fig:sample}.}
\label{fig:sample_more}
\end{figure*}

In \autoref{fig:sample_more}, we present the distribution of selected samples of galaxies from the \colibre simulations on the $\log\,M_{\ast}-\log\,\mathrm{SFR}$ plane (as in \autoref{fig:sample}) for redshifts $z=9$ to $z=20$. As shown, similarly to the trend observed for $z=7$ in \autoref{fig:sample}, rare objects with high stellar mass and high or low SFR have a higher selection probability.

\section{Exploring the Impact of \colibre and \skirt Setup Choices}
\label{sec:test}
\subsection{The effect of the feedback model}
\label{sec:feedback}
\begin{figure*}
\centering
\includegraphics[width=\textwidth]{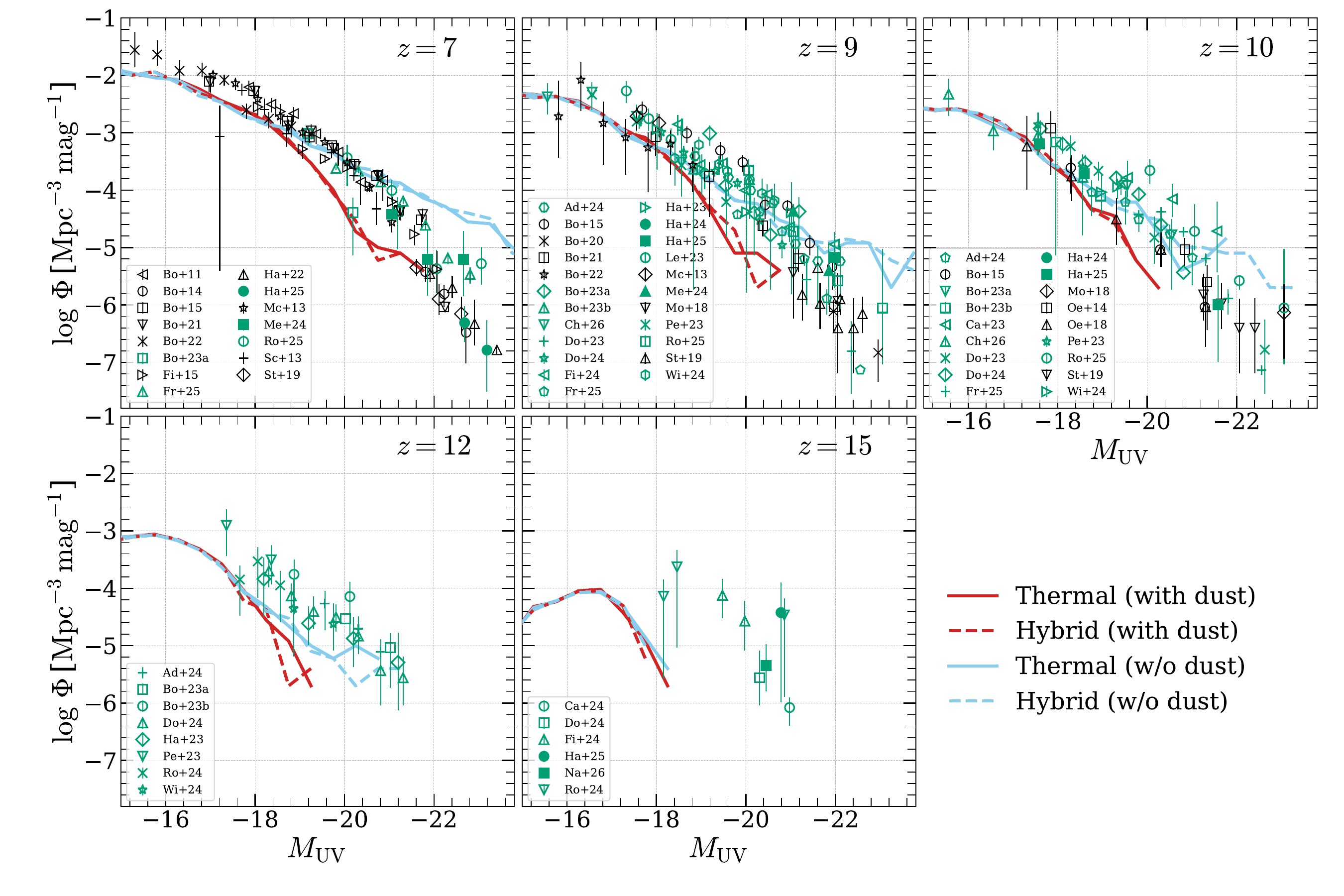}\vspace{-0.6cm}
\caption{Effect of the AGN feedback model on the UVLFs. In each panel, the UVLFs with the thermal AGN feedback model (the default in this paper) and the hybrid (thermal+jet) feedback model (see \citealt{Husko2026} for details) are shown with solid and dashed curves, respectively. Dust-attenuated and dust-free results are shown in red and blue, respectively. The \colibre L100m6 and L100m6h simulations are used for this test. The symbols for the observational data are the same as in \autoref{fig:uvlf}.}
\label{fig:uvlf_agn}
\end{figure*}

In \autoref{fig:uvlf_agn}, we investigate the effect of the AGN feedback model on the UVLFs. We present results from two \colibre L100m6 simulations that employ different AGN feedback models: the purely thermal model (solid; default in this work) and the hybrid (thermal+kinetic jet) model (dashed; see \citealt{Husko2026} for details). The UVLFs, both dust-attenuated and dust-free, produced by the two models show only minor differences, with the slight discrepancy at the bright end likely due to the small number of galaxies in the simulation volume.

\subsection{The effect of star-forming region resampling}
\label{sec:resampling}
As mentioned in \autoref{sec:skirt}, we follow the fiducial \skirt setup for \colibre described in \citet{Gebek2026} and resample the young stellar particles using the star-forming gas (both the currently existing star-forming gas and the parent gas of the young stellar particles) to mitigate the limited resolution of the simulations. Including the parent gas particles of recently formed stars ensures that the resampling conserves the mass of young stars and captures the full star formation activity over the past 10~Myr. This approach enforces a constant star formation history over the past 10~Myr, such that the resulting stellar population includes stars with ages from nearly 0 to 10~Myr. This is important because the youngest stars make a disproportionate contribution to the ionizing UV luminosity. The impact of this resampling on the UV luminosities decreases with increasing numerical resolution, as the young stellar populations are then better directly resolved. Thus, one of the primary motivations for adopting this treatment is also to improve convergence across different resolutions. Another motivation is to interface with the {\sc toddlers} model for star-forming regions. However, a limitation of this star-forming region resampling scheme is that it does not maintain full self-consistency with the dust modelling. In particular, the inferred UV flux is computed from resampled stellar populations that are, by construction, too young to have contributed dust through core-collapse supernovae, while the dust content is not adjusted accordingly. In practice, however, this inconsistency is unlikely to be important for most galaxies, since their dust content is expected to be dominated by dust growth in the ISM enriched by earlier generations of stars. It may nevertheless be more relevant for the youngest systems and should therefore be borne in mind when interpreting the UV luminosities derived with this method.

\begin{figure*}
\centering
\includegraphics[width=\textwidth]{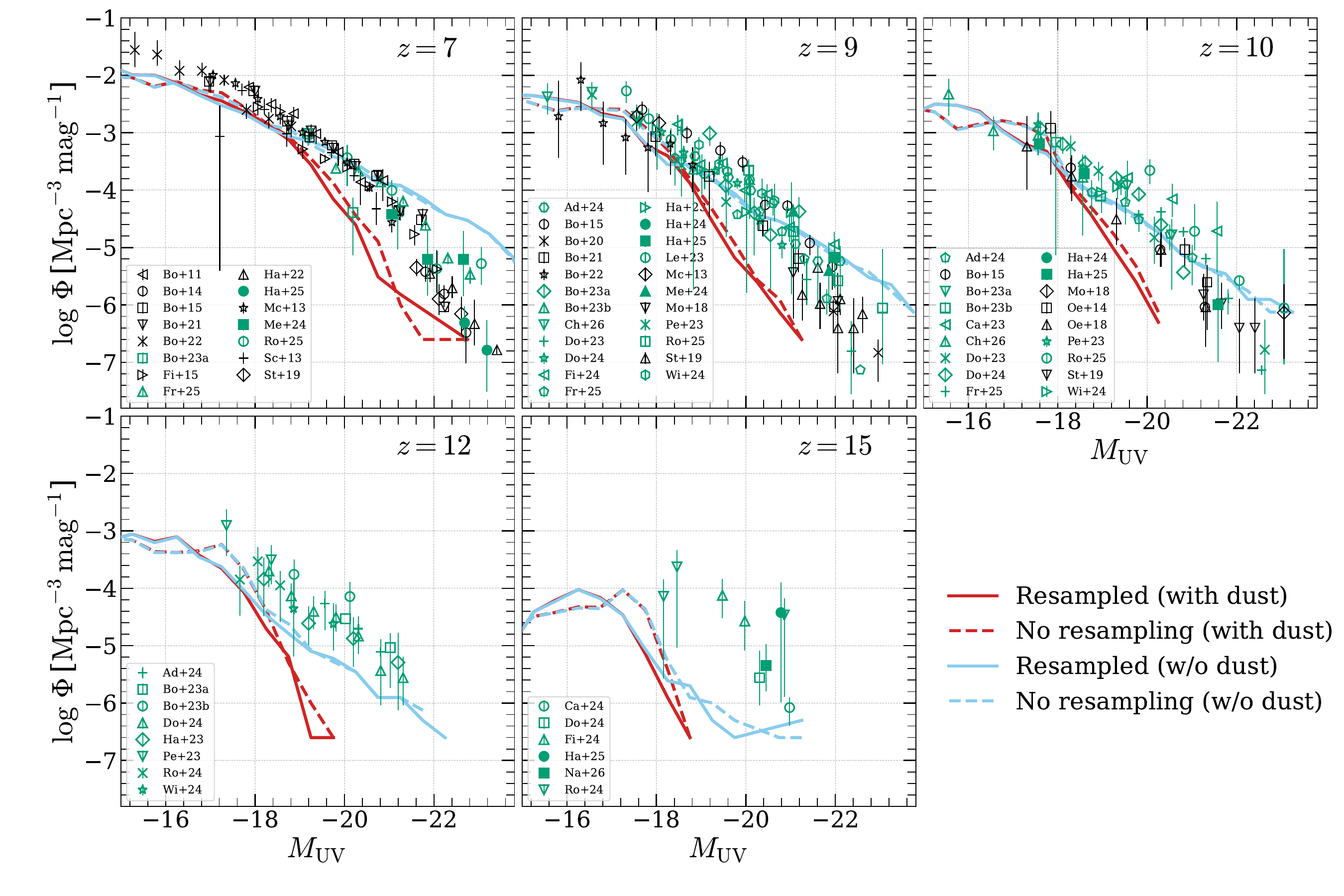}\vspace{-0.6cm}
\caption{Effect of star-forming region resampling on the (combined) UVLFs. In each panel, the solid curve represents results with resampling (default in this work), while dashed curves represent results without resampling. Dust-attenuated and dust-free results are shown in red and blue, respectively. The L200m6 \colibre simulation is used for this test. Observational data points are the same as in \autoref{fig:uvlf}.}
\label{fig:uvlf_resampling}
\end{figure*}

In \autoref{fig:uvlf_resampling}, we investigate the effect of this SFH resampling by comparing the UVLFs with resampling (the default in this paper) and those without resampling (i.e. including stellar particles of all ages and excluding the star-forming gas). As can be seen, at $z\leqslant 10$, the resampling procedure slightly reduces the dust-attenuated UV luminosities of galaxies at a given number density, except at the brightest end at $z=7$, while the dust-free UV luminosities are nearly unchanged. At higher redshifts ($z\geqslant 12$), the effect of SFH resampling becomes larger for the dust-attenuated results and the dust-free results also show measurable differences. These differences in the UV luminosities are expected to propagate into the UV luminosity density, with SFH resampling lowering $\rho_{\rm UV}$, particularly at $z\geqslant 12$. Nevertheless, the differences introduced by SFH resampling are not sufficiently large to account for the discrepancy between the predicted and observed UVLFs. Moreover, the increasingly close agreement between the dust-attenuated and dust-free results towards higher redshift indicates that the low dust content of these galaxies limits the impact of dust-related choices in the \skirt radiative transfer modelling.

\subsection{The effect of the projected aperture size}
\label{sec:aperture}
\begin{figure*}
\centering
\includegraphics[width=\textwidth]{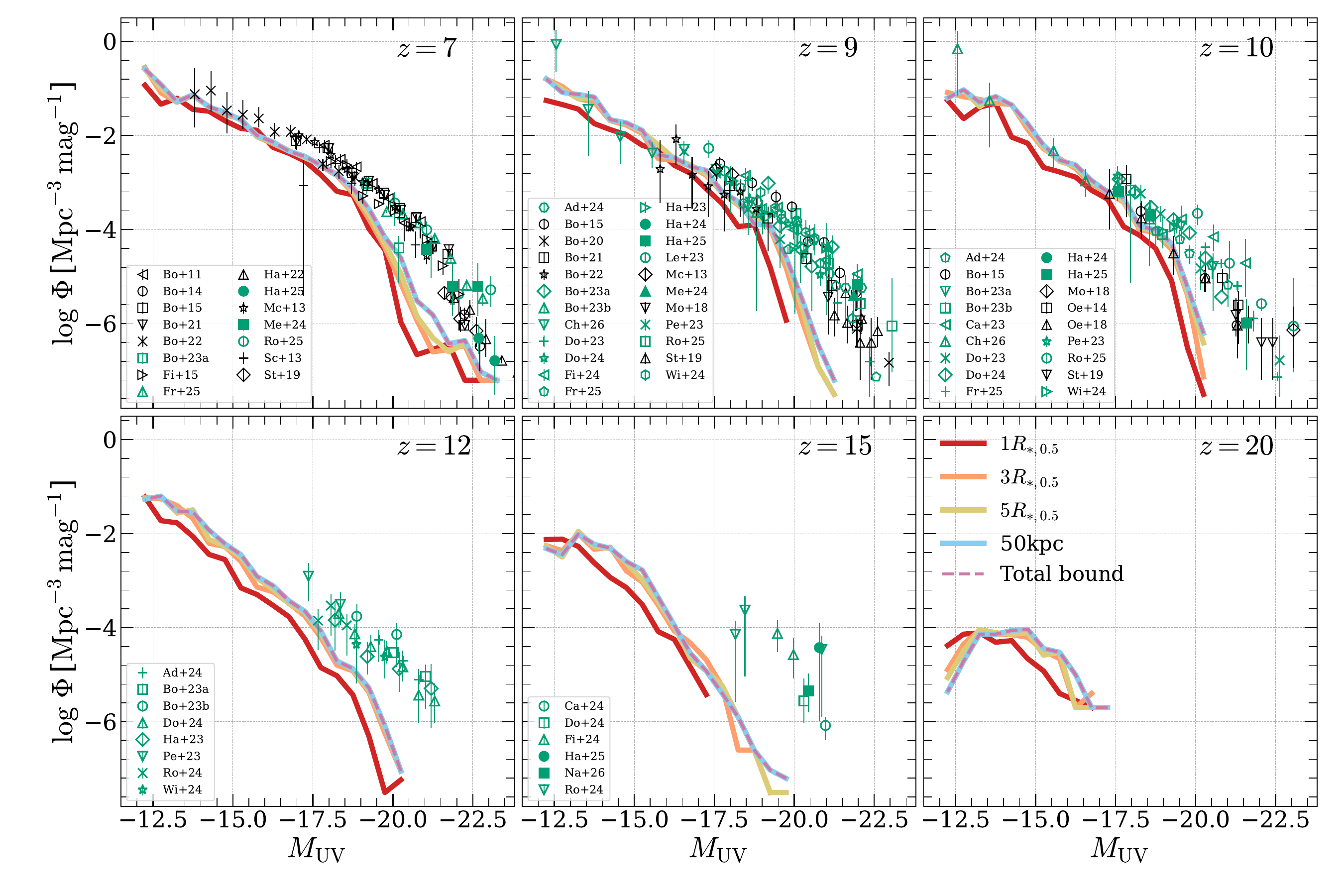}\vspace{-0.6cm}
\caption{Comparisons of \colibre UVLFs (the dust-attenuated combined UVLFs for $z=7-15$ and UVLF from L100m5 for $z=20$) with different projected aperture sizes. In this work, we adopt a proper projected aperture of 50 kpc (blue) as the default. The symbols for observational data are the same as in \autoref{fig:uvlf}.}
\label{fig:uvlf_aperture}
\end{figure*}

In this section, we investigate the influence of the (projected) aperture size on the UVLFs. As mentioned in \autoref{sec:skirt}, we adopt a projected circular aperture with a radius of $50\,\rm kpc$ for calculating the galaxy luminosity, to be consistent with other studies from the \colibre project (although a spherical aperture is typically used in some other studies). In \autoref{fig:uvlf_aperture}, we present the combined, dust-attenuated UVLFs using different projected aperture sizes ($1\,R_{\ast,0.5}$, $3\,R_{\ast,0.5}$, $5\,R_{\ast,0.5}$, $50\,\rm kpc$ and an aperture including all gravitationally bound particles; here, $R_{\ast,0.5}$ is the half-stellar-mass radius provided by the SOAP catalogue).

As expected, UVLFs measured within larger apertures exhibit higher amplitudes, since larger apertures contain more particles (including both stellar particles and star-forming gas). The differences in UVLFs across aperture sizes can be as large as $\sim 1\,\rm mag$ in the horizontal direction, highlighting the importance of aperture choice when comparing UVLFs. The UVLFs obtained with a $50\,\rm kpc$ aperture are nearly identical to those including all gravitationally bound particles, indicating that at $z>7$, a $50\,\rm kpc$ aperture is sufficiently large to enclose essentially all particles associated with a given galaxy.

{\it JWST} photometric measurements of high-redshift galaxies do not rely on a single fixed physical aperture. Instead, they typically use small apertures to optimize the signal-to-noise ratio and then apply aperture corrections to estimate total fluxes. For example, \citet{Harikane2023} measured colours in $0.3^{\prime\prime}$-diameter circular apertures on PSF-matched images and estimated total magnitudes by applying an aperture correction defined relative to {\sc mag\_auto}. \citet{Finkelstein2023}, by contrast, adopted small elliptical Kron apertures (Kron factor 1.1; minimum radius 1.6) and derived total fluxes using a two-step, simulation-based aperture-correction procedure. Thus, although the observational implementations are different, both studies ultimately aim to recover {\it total} magnitudes rather than the raw fluxes measured within the initial apertures. In this context, our use of a 50 kpc aperture is intended to encompass the full UV-emitting extent of each simulated galaxy and thereby provide an effectively total magnitude, making the comparison with these {\it JWST}-based measurements reasonable.

\label{lastpage}
\end{document}